\documentclass[aj,onecolumn]{emulateapj}

\makeatletter

\newcommand{\Rmnum}[1]{\expandafter\@slowromancap\romannumeral #1@}
\makeatother

\newcommand{\fA}{\ensuremath{A_{21}}}
\newcommand{\fP}{\ensuremath{\varphi_{21}}}
\newcommand{\M}{\mathrm}

\newcommand{\gps}{\ensuremath{g_{\rm P1}}}
\newcommand{\rps}{\ensuremath{r_{\rm P1}}}
\newcommand{\ips}{\ensuremath{i_{\rm P1}}}
\newcommand{\nrps}{\ensuremath{r_{{\rm P1},0}}}
\newcommand{\nips}{\ensuremath{i_{{\rm P1},0}}}
\newcommand{\ngps}{\ensuremath{g_{{\rm P1},0}}}
\newcommand{\zps}{\ensuremath{z_{\rm P1}}}
\newcommand{\yps}{\ensuremath{y_{\rm P1}}}

\newcommand{\rsdss}{\ensuremath{r_{\rm SDSS}}}
\newcommand{\isdss}{\ensuremath{i_{\rm SDSS}}}
\newcommand{\sig}{{\mathrm{sig}}}

\newcommand{\sdss}{{\mathrm{SDSS}}}
\newcommand{\rf}{{\mathrm{ref}}}
\newcommand{\diff}{{\mathrm{diff}}}
\newcommand{\psf}{{\mathrm{PSF}}}
\newcommand{\zp}{{\mathrm{ZP}}}

\begin{document}

\slugcomment{{\sc Accepted to AJ:} January 15, 2013} 
\title{Properties of M31. II: A Cepheid disk sample derived 
from the first year of PS1 PAndromeda data}

\author{Mihael Kodric\altaffilmark{1,2}, Arno
  Riffeser\altaffilmark{1,2}, Ulrich Hopp\altaffilmark{1,2}, Stella
  Seitz\altaffilmark{1,2}, Johannes Koppenhoefer\altaffilmark{1,2},
  Ralf Bender\altaffilmark{2,1}, Claus Goessl\altaffilmark{1,2}, Jan
  Snigula\altaffilmark{2,1}, Chien-Hsiu Lee\altaffilmark{3,1},
  Chow-Choong Ngeow\altaffilmark{3}, K. C. Chambers\altaffilmark{4},
  E. A. Magnier\altaffilmark{4}, P. A. Price\altaffilmark{4},
  W. S. Burgett\altaffilmark{4}, K. W. Hodapp\altaffilmark{4},
  N. Kaiser\altaffilmark{4}, R.-P. Kudritzki\altaffilmark{4}}

\altaffiltext{1}{University Observatory Munich, Scheinerstrasse 1, 81679 Munich, Germany}
\altaffiltext{2}{Max Planck Institute for Extraterrestrial Physics, Giessenbachstrasse, 85748 Garching, Germany}
\altaffiltext{3}{Graduate Institute of Astronomy, National Central University, Jhongli City, 32001, Taiwan}
\altaffiltext{4}{Institute for Astronomy, University of Hawaii at Manoa, Honolulu, HI 96822, USA}

\begin{abstract}

  We present a sample of Cepheid variable stars towards M31 based on
  the first year of regular M31 observations of the PS1 survey in the
  $\rps$ and $\ips$ filters.  We describe the selection procedure for
  Cepheid variable stars from the overall variable source sample and
  develop an automatic classification scheme using Fourier
  decomposition and the location of the instability strip. We find
  1440 fundamental mode (classical $\delta$) Cep stars, 126 Cepheids
  in the first overtone mode, and 147 belonging to the Population II
  types. 296 Cepheids could not be assigned to one of these classes
  and 354 Cepheids were found in other surveys. These 2009 Cepheids
  constitute the largest Cepheid sample in M31 known so far and the
  full catalog is presented in this paper. We briefly describe the
  properties of our sample in its spatial distribution throughout the
  M31 galaxy, in its age properties, and we derive an apparent
  period-luminosity relation (PLR) in our two bands. The Population I
  Cepheids nicely follow the dust pattern of the M31 disk, whereas the
  147 Type II Cepheids are distributed throughout the halo of M31.  We
  outline the time evolution of the star formation in the major ring
  found previously and find an age gradient. A comparison of our PLR
  to previous results indicates a curvature term in the PLR.

\end{abstract}


\section{Introduction}
\label{chapterintro}
Classical $\delta$~Cep variable stars are young massive stars of
Population I in their post main sequence evolution. Due to their
period-luminosity relation (\citealt{1908AnHar..60...87L} and
\citealt{1912HarCi.173....1L}), they are the anchor of the
extragalactic distance scale. As such, they are a central topic of
many investigations in nearby galaxies (see
e.g. \citealt{2001ApJ...553...47F}, for a historical summary including
the early HST projects, more recent work is discussed e.g. by
\citealt{2011ApJ...730..119R} and \citealt{2011ApJ...743..176G}).
$\delta$~Cep stars can provide distance estimates out to several 10
Mpc (e.g. \citealt{2011ApJ...730..119R}), but the major uncertainty of
these distance values is still related to the local calibration which
until recently totally rested on the LMC distance. As there is an
ongoing and unsolved debate to which extent the period-luminosity
relation (PLR) of $\delta$~Cep stars depends on stellar properties
like e.g. the metallicity \citep{2011ApJ...741L..36M}, the LMC is not
an optimal choice given its significantly reduced metallicity compared
to the large spirals to which the PLR is normally
applied. \cite{2011ApJ...741L..36M} summarize the recent discussion
with the conclusion that the period Wesenheit
(\citealt{1976MNRAS.177..215M}, \citealt{1982ApJ...253..575M},
\citealt{1983IBVS.2425....1O}) relation (from colors $V$ and $I$) does
not depend on metallicity while \cite{2011ApJ...733..124S}, based on a
large sample of $\delta$~Cepheid stars observed with HST in two fields
of the spiral galaxy M~101, conclude the opposite and in addition
report supporting evidence from other studies
\citep{2011ApJ...743..176G}. On the theoretical side there have been
predictions that the Wesenheit PLR should depend on metallicity if
based on photometric bands sensitive to metal line blanketing (like
the combination of Johnson $B$ and $V$) while the red band
combinations of the Johnson-Cousins system should not show this
dependence \citep{2007A&A...476..863F}.  The discussion of the PLR
properties - in individual photometric bands - has become even more
complicated as some authors reported a non-universal slope for the PLR
(see \citealt{2008ApJ...686..779S} and references therein), which has
not been confirmed by others and still is under debate.

The recent discussion on the calibration of the cosmological distance
scale and the subsequent determination of the Hubble constant has
somehow shadowed that these relatively young and massive stars can be
also used as excellent tracers of the recent evolution of stellar
populations as they allow number counts in precisely known time bins.

The LMC has been the historical place for the establishment of the PLR
\citep{1912HarCi.173....1L} as it is close by and can be well resolved
from the ground while already distant enough that the member stars are
essentially at the same distance.  Thus, the LMC (and SMC) were
intensively studied for variable stars and transient phenomena by
various projects yielding also extensive and well observed samples of
Cepheid stars in fundamental, first and second overtone pulsation as
well as Population II pulsators (W Vir, RV Tau etc). The most
prominent example is probably the OGLE project which provided a well
established PLR based on $B, V$, and $I$ observations of $\sim$~3400
$\delta$~Cep stars in LMC and SMC
(\citealt{1999AcA....49..223U},\citealt{1999AcA....49..437U},\citealt{1999AcA....49..201U}).

The nearest example of a Cepheid population in a typical large spiral
galaxy of roughly solar metallicity and outside our own Milky Way
galaxy is the one in the Andromeda galaxy M31. The first detections
go back to \cite{1929ApJ....69..103H}.  \cite{1965AJ.....70..212B}
summarize the early observational results which already comprised a
sample size of $\sim$~260 Cepheid variables.

Wide field CCD imagers time series studies brought M31 again
into the focus of several projects aiming to detect and study
transients and variables stars (e.g. \citealt{2004ASPC..310...33M},
the DIRECT project; \citealt{2006A&A...445..423F},
\citealt{2007A&A...473..847V}).

The \cite{2006A&A...445..423F} sample (based on the WeCAPP micro
lensing project \citep{2001A&A...379..362R} was observed over almost 8
years with ground based CCD imaging centered on the M31
bulge. Therefore, the $\delta$~Cep stars as typical Population I
objects are only a minor contribution (33) to the total sample of
variable stars and they are even outnumbered by other types of
Cepheids (93).

The \cite{2007A&A...473..847V} sample was obtained from observations
covering the north-west quarter of the M31 disk. Spanning almost 5
years, the $\sim$~250 observations in $B$ and $V$ allowed to obtain a
sample of 416 Cepheid stars, to study their light curves in detail and
to derive distance estimates to M31. A shortcoming of the DIRECT data
is that the 250 individual observations were obtained in 21 nights
only.  The window function (inherent to the observations) implies a
strong bias against finding objects within certain period ranges. This
can distort results like e.g. the period-number distribution of the
sample.

Ground-based M31 observations have been supplemented with single epoch
high resolution HST-imaging (PHAT, \citealt{2012ApJ...752...95J}),
especially in the NIR. This allowed to improve on some of the
short-comings of ground-based data for this galaxy and paved the way
for a precise distance determination of M31
\citep{2012ApJ...745..156R}.

In this paper we expand the studies of the Cepheid population in M31
with our wide field ground based data taken within the first year of
the Pan-STARRS 1 project \citep{2002SPIE.4836..154K}. Among others,
the two major advantages of our data are that, firstly, they cover the
complete M31 disk, and, secondly, they are much more evenly
distributed in time using observations taken in up to 102 nights
($\sim 80 \%$ of the nights have two epochs, giving a total of 183
epochs).

Throughout this paper we will refer to Fundamental mode (FM) Cepheids
(classical $\delta$~Cep variable stars), First overtone (FO) Cepheids
(classical $\delta$~Cep variable stars pulsating in the excited mode
of the first overtone), Type II Cepheids (W Virginis) and use the term
Cepheids for all three subgroups. BL Herculi stars are too faint to be
detected with our survey and RV Tauri are excluded manually
(c.f. Section \ref{sec.period}), so that our Type II Cepheids only
consist of W Virginis stars.

This paper is organized as follows: In Section \ref{sec.observations}
we present the PAndromeda survey, our observations in 2010 and 2011,
and a description of our data reduction, including the determination
of the light curves for the resolved sources. Our period determination
is described in Section \ref{sec.period}. The Cepheids detected in the
PAndromeda data-set are presented in Section \ref{sec.detection}. This
Section also describes the automatic Cepheid detection and
classification. An overview of the catalog is shown in Section
\ref{sec.catalog}. We discuss our results, such as (among others) the
PLR, the spatial distribution of the Cepheids and the spatial age
distribution, in Section \ref{sec.results}. We conclude the paper in
Section \ref{sec.conclusion}.

\section{Observations \& Data reduction}
\label{sec.observations}

\subsection{PAndromeda}

The 1.8~m Panoramic Survey Telescope and Rapid Response System
(Pan-STARRS, \citealt{2004AN....325..636H},
\citealt{2002SPIE.4836..154K}) on Haleakala, Maui, began its regular
survey in spring 2010. Pan-STARRS 1 (PS1) uses the Giga Pixel Camera
(GPC, \citealt{2009amos.confE..40T}) which is currently one of the
largest cameras in the world. It consists of 60 OTAs (orthogonal
transfer arrays), each of them segmented into an array of $8\times8$
cells. The total number of pixels is $1.4 \cdot 10^9 $, with a pixel
size of 0.258 arcsec. The FOV of one OTA is
$21\arcmin\times21\arcmin$. The aims of PS1 are to map 3$\pi$ of the
sky in $\gps$, $\rps$, $\ips$, $\zps$, $\yps$ bands. In addition to
the large 3$\pi$ area some fields are visited more frequently and have
deeper exposure times. One of these so called medium deep fields
targets the Andromeda galaxy (M31). With the $\sim$7~deg$^2$ field of
view of Pan-STARRS, a complete monitoring of the disk and large parts
of the halo is possible with one single exposure.  The PAndromeda
survey observes M31 for 0.5 h per night (including overhead) during a
period of 5 months per year, equivalent to 2\% of the overall PS1
time. We split the integration time into two visits per night
(separated by $3 \, \mathrm{h} \pm 0.5 \, \mathrm {h} $), for the
whole season\footnote{Per visit 7 exposures of $60 \, \mathrm{s}$ in
  the $\rps$ band and 5 exposures of $60 \, \mathrm{s}$ in the $\ips$
  band are taken. The total integration time per night is therefore
  $1440 \, \mathrm{s}$. The remaining time is spent on readout,
  focusing and filter change.}. During the first two seasons we
focused on the filters $\rps$ and $\ips$ \citep{2012arXiv1203.0297T}.
Observing M31 two times per night with the $\rps$ and $\ips$ filters
is a consequence of the PAndromeda survey goal to find microlensing
events as described in \cite{2012AJ....143...89L}.

In the first season PAndromeda monitored M31 from 2010-07-23 to
2011-12-27. In this work we used in addition part of the second
season from 2011-07-25 to 2011-08-12. This results in a gap of about
half a year in the light curves and gives us a total of 183
epochs\footnote{Due to masking the total number of epochs for a light
  curve can be less than 183.}.

\subsection{Data reduction}

The details of the data reduction for PAndromeda are described in
\cite{2012AJ....143...89L}. In the following section we summarize the data
reduction steps and point out differences to the approach taken in
this work.

The GPC1 exposures are de-biased, flat-fielded, and astrometrically
calibrated using the Pan-STARRS Image Processing Pipeline (IPP;
\citealt{2006amos.confE..50M}). During this process the images are
remapped to a common grid, the so called skycells. These warped images
have a pixel scale of 0.200 arcsec/pixel, which is smaller than the
natural pixel size of the GPC1 (0.258 arcsec/pixel). For better data
management we integrated our pipeline \citep{2011ExA...tmp..141K} into
the AstroWISE data management system \citep{2007ASPC..376..491V}.

\begin{figure}[h!]
  \epsscale{1.0}
  \plotone{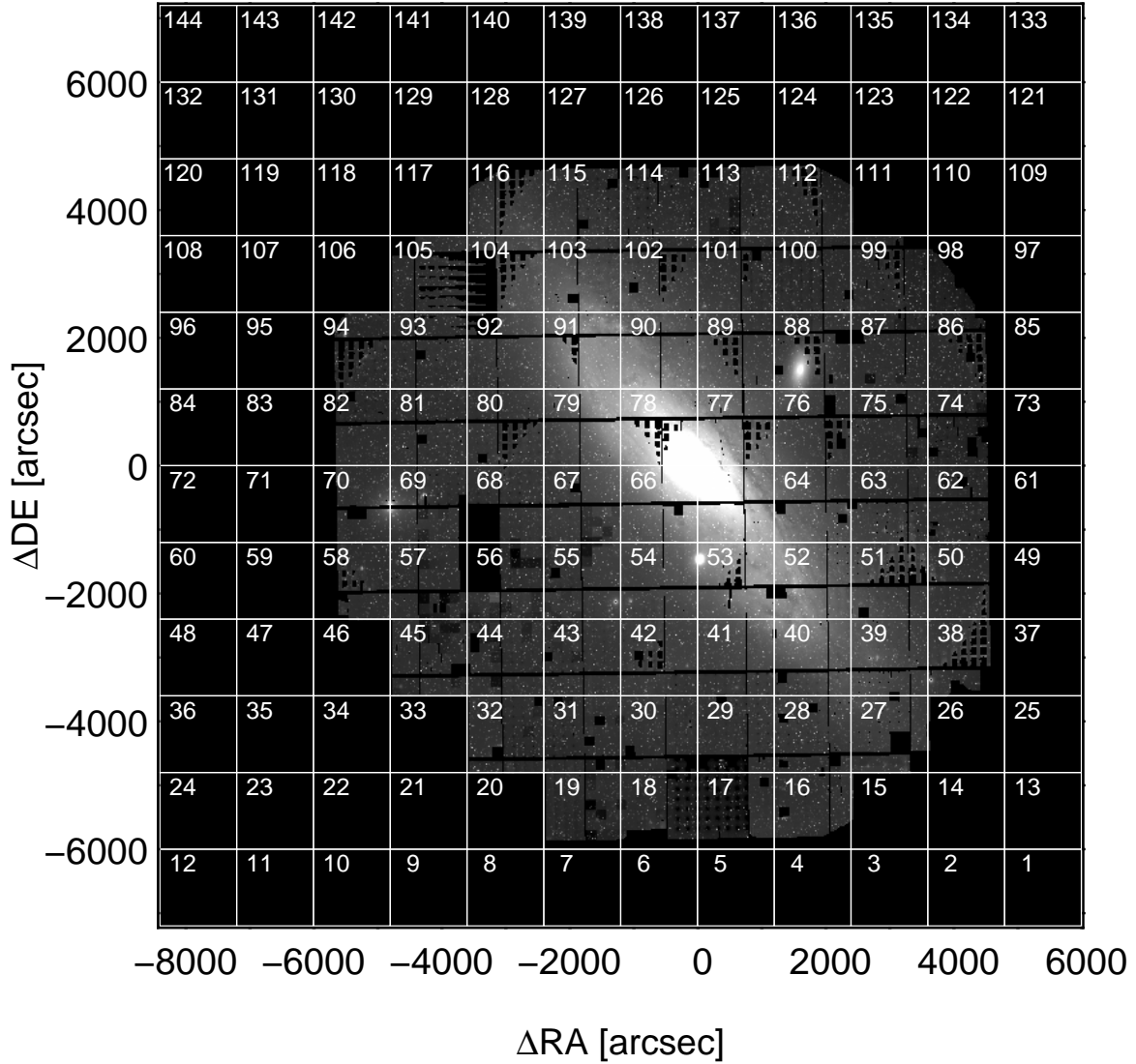}
  \label{Fig.skycelllayout}
  \caption{Skycell layout for the first season. As can be seen the
    entire disk of the Andromeda galaxy, M~32 and NGC~205 can be imaged with one
    pointing. The underlying M31 image is a single $\rps$-band frame,
    which was chosen to have relatively small masked area. The masked
    areas (in black) are e.g ill-functioning chips, video guide star
    cells and areas on the focal plane with sub-standard imaging
    performance.}
\end{figure}

In this work we use only data from 24 skycells, namely 039-041,
051-054, 064-068, 076-079, 088-092, 102-104. As shown in
Fig. \ref{Fig.skycelllayout}, the skycell layout is not optimized in
the sense that it does not follow the detector boundaries.  We plan to
optimize the skycell layout and use an intrinsic pixel scale close to
0.258 arcsec/pixel in the future. The following data reduction steps
are applied to each skycell and both filters $\rps$ and $\ips$.

Since we use difference image photometry \citep{1996AJ....112.2872T},
we have to produce a stack of images with the best seeing, the so
called reference image.  We photometrically align the 70 best seeing
images, so that the images have the same zero point and the same (non
zero) sky background. In the next step we replace masked areas in the
photometrically aligned images with values from another image that has
the most similar PSF (\citealt{ediss5984}, p. 134). The weighted stack
of the aligned images is called the reference image.

We use SExtractor to detect sources on the reference image
\citep{1996A&AS..117..393B}.  For the data utilized in this work,
SExtractor is used in a mode that modifies the detection threshold
depending on the background. While this mode is very useful it hampers
completeness tests.

For all sources in the reference image, we perform PSF photometry on a
background (sky and M31) subtracted reference image. In order to
obtain the background in the reference image, we fit a bicubic spline
model. To suppress blending effects we iteratively subtract
neighboring stars. The fluxes measured in the reference image are
added later to the fluxes in the difference images, so as to create
the light curves. The flux error in the reference image determined by
the PSF photometry does not take into account how well our constant
PSF model matches the real sources in the reference frame which can
show slight PSF-variation over the field.  This can result in wrong
flux errors. To obtain reasonable errors we rescale the error of each
source with the $\sqrt{\chi^2}$ of its PSF fit.

To minimize masking and to obtain a better signal-to-noise ratio we
stack the images of each visit. The procedure is the same as for the
reference images. The visit stacks are photometrically aligned to the
reference image, so that the visit stacks and the reference image have
the same zero point and sky background. At this point they only
deviate in signal-to-noise and PSF-shape.

\begin{figure}[h!]
  \epsscale{0.75}
  \plotone{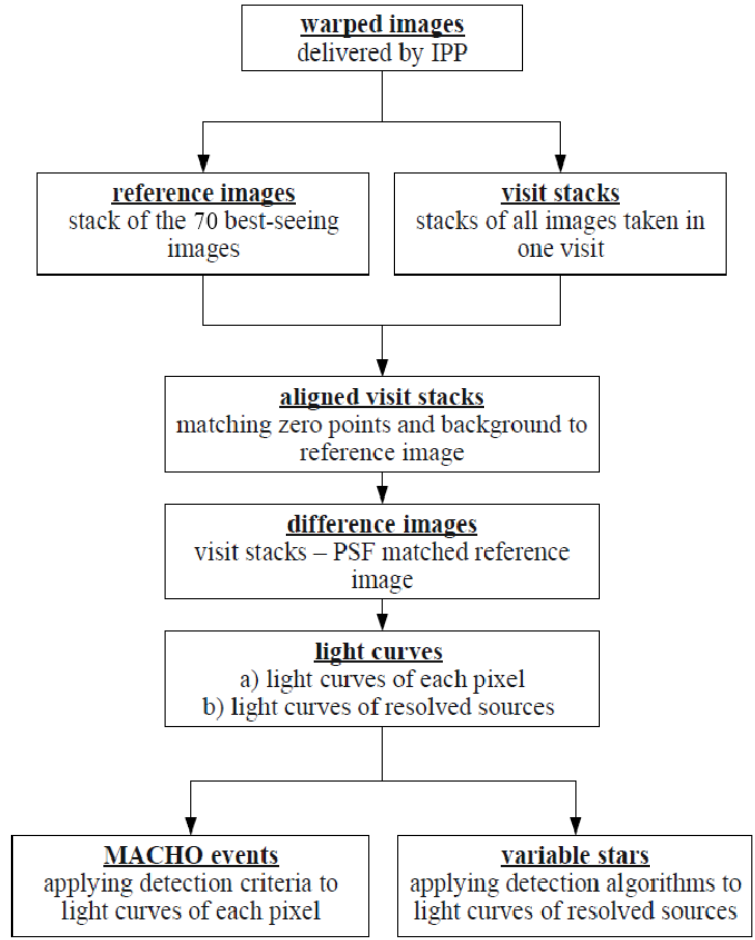}
  \caption{Summary of the PAndromeda data reduction pipeline (taken
    from \citealt{2012AJ....143...89L} (modified)). We use 70 images
    for the reference image stacking, instead of the 30 that
    \cite{2012AJ....143...89L} use.}
  \label{Fig.pipeline}
\end{figure}

To create the difference images we first align the PSF of the
reference image to match the PSF of the visit stack by calculating a
constant convolution kernel by least-squares minimization from all
pixels in subareas of $75\arcsec\times75\arcsec$ (see
\citealt{1998ApJ...503..325A}).  This PSF aligned reference image is
then subtracted from the visit stack. On this difference image we
perform PSF photometry to obtain light curves. The PSF is constructed
with sub-pixel precision from the convolved reference image using
isolated stars. Since we only measure the flux difference we add the
flux measured in the reference image to this flux difference to obtain
the light curves for the resolved sources. Fig. \ref{Fig.pipeline}
summarizes the PAndromeda data reduction pipeline.

Note that the astrometric precision and photometric stability have
been discussed in \cite{2012AJ....143...89L}. For the flux calibration
we make use of the fact, that the $\rps$ and $\ips$ bands are rather
similar to the Sloan Digital Sky Survey (SDSS;
\citealt{2009ApJS..182..543A}) $\rsdss$ and $\isdss$ bands and thus the
flux calibration should be insensitive to color terms
\citep{2012arXiv1203.0297T}. To calibrate our magnitude to the SDSS
magnitude system with stars from SDSS we use:

\begin{equation}
  m_\sdss = -2.5 \log\left[ F_\rf(\psf_\rf) + \Delta F_\diff(\psf_\diff) +\zp_\psf \right]
\end{equation}
$F_\rf(\psf_\rf)$ is the flux of a source determined by PSF photometry
in the reference image and $\Delta F_\diff(\psf_\diff)$ the flux
determined by PSF photometry in the difference
image. $\zp_\psf$ accounts for the difference between SDSS
magnitudes and our instrumental magnitudes and does also consider the
PSF construction.
 
Our pipeline does not take into account the correlated noise that
arises from the change of the pixel scale.  To quantify this we
constructed light curves of all pixels, normalize each single
deviation from the fitted mean by its error.  The distribution of
these normalized deviations is clearly dominated by the non-varying
pixels. Then the correction factor can easily be estimated by
comparing this distribution with a standard Gaussian-distribution
($\sig=1$). This rescaling factor is the same for every skycell and
filter.

Throughout this paper, we will use the AB magnitude system.

\section{Period determination}
\label{sec.period}

The periodicity of the light curves is determined with the software
{\tt SigSpec} \citep{2007A&A...467.1353R}. {\tt SigSpec} calculates a
quantity called \textit{spectral significance} (sig) for each peak in
the discrete Fourier transform (DFT) amplitude spectrum of a light
curve. Similar to the false alarm probability\footnote{The false alarm
  probability $\Phi_\M{FA}(A)$ is the integrated probability density
  function and describes the probability that an amplitude exceeds a
  given limit $A$ as described in \cite{2007A&A...467.1353R}.}
$\Phi_\M{FA}(A)$
\begin{equation}
  \sig(A) := - \log \left[\Phi_\M{FA}(A)\right]
\end{equation}
is a measure for the probability that a peak arises from noise. This
value does not only depend on the amplitude and frequency, but also on
the phase and the sampling of the light curve. This fact gives {\tt
  SigSpec} an advantage over the use of other Fourier based algorithms
such as the Lomb-Scargle algorithm (\citealt{1976Ap&SS..39..447L},
\citealt{1982ApJ...263..835S}). While both methods produce comparable
results concerning the period, we found that {\tt SigSpec} finds more
periodic light curves over a given false alarm probability threshold
than the Lomb-Scargle algorithm. This is due to the fact that certain
periods have to be excluded in the Lomb-Scargle algorithm, depending
on the sampling of the light curve in order to avoid aliasing.

In the following paragraph we outline the parameters with which we run
{\tt SigSpec}. As statistical weight we use the inverse variance ($w_i
= \sigma_i^{-2}$) of the PSF flux measurement and we use
$[\mathrm{JD}-2450000]$ for the epoch.  We apply a $\sig$-threshold of
5 ({\tt siglimit 5}) to determine the periodicity of the light
curves. This means that in less than one out of $10^5$ cases pure
noise could produce the measured period. With a lower threshold the
number of detections would increase (e.g. by a factor of $\sim 10$
with a threshold of 3), but most of the detections would be low
signal-to-noise detections. With the parameter {\tt ufreq 1} we define
the lowest period to be considered as 1 day. We exclude smaller
periods because they can hardly be detected with a sampling of two
visits per night. An upper limit for the period has not been defined,
but a period cut is performed later on.  The last two parameters ({\tt
  multisine:lock} and {\tt iterations 1}) force {\tt SigSpec} to
consider only the most significant peak\footnote{The {\tt SigSpec}
  default value would perform an iterative sequence, the so called
  prewhitening sequence. After the most significant peak is
  identified, a fit to all signals that are identified so far is
  performed. This fit is then prewhitend (subtracted) from the
  spectrum and the next iteration starts until the maximum peak is
  under the threshold.}  in the DFT amplitude spectrum ensuring that
noise has no impact on the determined period. We use no variability
index since {\tt SigSpec} only determines periods for variable
sources.

Inspection by eye of the folded light curves, that {\tt SigSpec}
detects, showed the good performance in the period determination of
{\tt SigSpec}. One problem, however is the lack of an upper period
limit that permits {\tt SigSpec} to determine periods even if the
signal is not periodic\footnote{{\tt SigSpec} can determine a period
  for almost every signal that is not constant. If the signal is not
  periodic and not constant the program usually detects the total
  observation time as the period.}. The inspection showed that for
large periods some of the folded light curves look like Cepheid light
curves. We did not exclude those light curves at this stage, but
rather applied a period cut later on. As our data set also contains 10
days of observations of season 2011, the data show a gap of half a
year between these data points and the 2010 data points. In some cases
the 2011 data points have a small flux shift from the 2010 data points
in the folded light curve. This may have different reasons. Firstly
the shift could be due to flux variations. Secondly it could be caused
by a small deviation from the true period, which can be observed in
the folded light curve if the period is small enough. And lastly the
frequency spacing used by {\tt SigSpec} could be too low. Also in the
case of RV Tauri variables {\tt SigSpec} detects the wrong period, as
it identifies the time span between the primary and secondary minimum
as the period. We identified and rejected those light curves
manually\footnote{RV Tauri light curves are rather easy to distinguish
  from the Cepheid light curves, since the RV Tauri light curves
  appear as an overlay of two shifted light curves.}. From 724894
sources that were detected in the $\rps$ band and which have light
curves in the $\rps$ and $\ips$ band\footnote{We detect the variable
  sources in the $\rps$ band and therefore there is always a light
  curve in the $\rps$ band for each variable source. But the same
  source can be masked in the $\ips$ band so that there is no light
  curve in the $\ips$ band.}, {\tt SigSpec} finds 75362 periodic
variable light curves with $\mathrm{sig} > 5$.

\begin{figure}[h!]
  \epsscale{0.5}
  \plotone{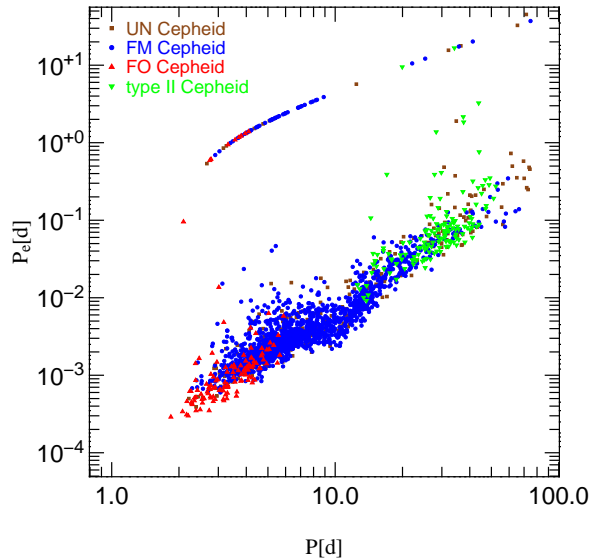}
  \caption{Period errors $P_e$ determined through the bootstrap method
    for the 3-dimensional parameter space classified Cepheid
    catalog. The large period errors of $P_e \gtrsim 0.5$ days are
    caused by aliasing produced by $\sim 3.5 \%$ of the bootstrap
    light curves.}
  \label{Fig.perError}
\end{figure}

Since {\tt SigSpec} does not determine the period error, we use the
bootstrap method to estimate the period uncertainties. A random
sampling is drawn from the light curve (one epoch can be drawn
multiple times) and the period for that sampling is determined with
{\tt SigSpec} (with a $\sig$-threshold of 1). This procedure is
performed 1000 times for each light curve, so that the $1 \sigma$
error can be determined from the resulting distribution. The obtained
period errors for the Cepheids of the 3-dimensional parameter space
classified Cepheid catalog (see Section \ref{2.6}) are shown in
Fig. \ref{Fig.perError}. The classification of the different Cepheid
types shown in Fig. \ref{Fig.perError}, as well as the meaning of the
3-dimensional parameter space classified Cepheid catalog, is described
in the next section.

In general the period errors are small, except for some light curves
where the distribution has an additional peak at $\sim 1.5 \,
\mathrm{d}$ due to aliasing. For those light curves the large error is
only caused by this additional peak. It is not a broader distribution
that causes the larger error and therefore the period error would be
small if there is no aliasing. Since we find no aliasing in the final
Cepheid light curves and the systematics have a larger effect than the
period errors, we disregard the period errors, which means that the
period errors shown in the plots do not contribute to any selection
criteria or fit.
\bigskip
\section{Cepheid detection and classification}
\label{sec.detection}

As described in the previous section we narrowed the number of
variable sources down to those that have periodic light
curves. In this sample of variable sources we want to identify the
Cepheids.

The classical approach to find the Cepheids
(e.g. \citealt{1999AcA....49..223U}, \citealt{2007A&A...473..847V})
is to preselect sources in a certain brightness and color range and to
visually inspect them. But the position of the instability strip spans
an area depending on brightness and color
(\citealt{2002ApJ...576..402F} and \citealt{2008A&A...482..883M},
including the updates by \citealt{2010ApJ...724.1030G}) and is
incorrectly described by a rectangle in the color-magnitude diagram
(see Fig. \ref{Fig.Wesenheit}).  For our color cut we therefore use
the instability strip of the Cepheids translated into our filter
system (\citealt{2002ApJ...576..402F}, \citealt{2008A&A...482..883M},
\citealt{2010ApJ...724.1030G}). We choose a different approach than
the visual inspection and establish an automatic Cepheid
detection. The motivation for this is that the selection based purely
on light curves is always biased, in particular if the light curves
become more noisy. We define limits in a 3-dimensional parameter
space which define our Cepheids. In order to define the parameter
space we select a subsample of the periodic light curves and visually
select the Cepheids. The manual selection is of course also biased,
but the advantage of this method is that the selection of the
remaining Cepheids only depends on the defined parameter space. The
parameters used are derived from a Fourier decomposition of the light
curves. With this method we were also able to identify the Cepheid
type. The fundamental mode Cepheids and the first overtone Cepheids
can be distinguished this way as e.g. shown by
\cite{1999AcA....49..223U} and \cite{2007A&A...473..847V}. We can
additionally also distinguish the Type II Cepheids from the other two
mentioned Cepheid types.
\\
Additionally we impose certain requirements on the remaining variable
sources detected by {\tt SigSpec}, so as to exclude bad light curves.

In the following Sections we describe our approach that we outlined
here.

\subsection{Fourier decomposition}
\label{Fdecomp}

A Fourier decomposition of the folded light curves can be used to
distinguish between fundamental mode (FM) Cepheids and first overtone
(FO) Cepheids (\citealt{1999AcA....49..223U},
\citealt{2006A&A...459..321V}). We also use Fourier decomposition to
identify Type II Cepheids and to define the 3-dimensional parameter
space in which we detect the Cepheids.

We fit a Fourier series to the light curve of the form
\begin{equation}
  \label{sincos}
  m(t) = a_0 + \sum_{n=-N}^{-1} a_{n}\cdot \cos(n  \omega  t) + \sum_{n=1}^{N} a_{n}\cdot \sin(n  \omega  t)  
\end{equation}
with $P$ being the period of the light curve determined by {\tt
  SigSpec}), $\omega = \frac{2 \pi}{P}$ and $t \in [0,P]$.  Prior to
the fit we convert the flux to magnitudes and fold the light curves,
so that they show the variation of the magnitude over the phase
$\theta$ times the period\footnote{That is why $t \in [0,P]$ and not
  $t \in [0,1]$, what would be the case for folded light curves that
  vary over the phase.} $P$. With this fit we determine the minimum
and maximum of the light curve and the coefficients $a_n$. We also
calculate the magnitude of the mean flux, needed later for the
PLR. This is different to a mean magnitude, which we would get if we
take the mean of the light curve. For this reason we transform the
Fourier series fit back to flux and calculate the mean flux and then
convert this to a mean magnitude.

For an analysis of the Fourier decomposition we transform the Fourier
series to the form:
\begin{equation}
  m(t) = b_0 + \sum_{k=1}^{N} b_{k}\cdot \cos(k \omega t + \varphi_k)
\end{equation}
with $b_k = \sqrt{a_{-k}^2+a_k^2}$ and $\varphi_k = \arctan\frac{-a_k}{a_{-k}}$.
This allows us to calculate the amplitude ratio
\begin{equation}
  A_{21} \equiv \frac{b_2}{b_1}
\end{equation}
and the phase difference
\begin{equation}
  \varphi_{21} \equiv \varphi_2 - 2 \cdot \varphi_1
\end{equation}
for each light curve \citep{2007A&A...473..847V}. The period $P$ and
the coefficients $\fA$ and $\fP$ are used to define the 3-dimensional
parameter space ($P$, $\fA$, $\fP$) and to identify the Cepheid type.

We chose $N=5$ for the Fourier series fit. Other degrees of freedom
(d.o.f.) would result in slightly different amplitudes $a_n$. In most
cases the deviations of the $a_n$ from other d.o.f., compared to the
$a_n$ of $N=5$, is small, but there are also light curves with
significant deviations for higher d.o.f.. While $N=5$ is sufficient to
fit bumps (c.f. left panel Fig. \ref{bsplcFM2}) in the light curves,
higher d.o.f. are more susceptible to `overshoots` and tend to be more
influenced by outliers. This is why we do not determine the optimal
d.o.f. for each light curve as e.g. described in
\cite{2009A&A...507.1729D}. The principal component analysis (PCA)
technique demonstrated in \cite{2009A&A...507.1729D} does not offer
any advantage in the form of saving computation time, since we use a
linear least squares fit and not a non linear fit for which the PCA is
faster.

Additionally we derive a polynomial fit with two boundary conditions
of the form:
\begin{equation}
  p(t) = a_0 + \sum_{n=1}^{K} a_n t^n \quad \M{with} \quad  p(0) = p(P) \;\wedge\;  p'(0) = p'(P) \;\wedge\;  t \in [0,P]
\end{equation}
to the folded light curves (in magnitudes). For this fit a change in
the number of d.o.f. also changes the $a_n$, so we chose $K = 10$ in
order to have the same number of d.o.f. as in the Fourier series
fit. With this polynomial fit we have an alternative light curve
description to the Fourier series. The best fitting Fourier and
polynomial fits are affected by data gaps in a different way and they
produce different `overshoots`. We compare the Fourier and polynomial
fits to identify and exclude badly fitted light curves, as we will
describe in the next section.

\subsection{Wesenheit-color cuts}
\label{2.3}

Young stars are located in spiral arms which are also populated by
interstellar dust. The dust in the arms show a clumpy distribution and
Cepheids may be either located before, within, or behind this dust.

Although Population II Cepheids are not located in the M31 disk it is
not a priori known whether they stay in front of or behind the
disk. Therefore, even for the majority of the Population II objects an
understanding of the internal reddening is required for each single
object.

As already mentioned the preselection of sources in a certain
brightness and color range depends on the instability strip. But it
also depends on the reddening. By using the Wesenheit magnitude the
brightness is independent of reddening and only the color is affected
by extinction. For this reason we use the Wesenheit-color plane and
select Cepheids according to the location of the instability strip.

The Wesenheit magnitude is defined to be independent of reddening
(\citealt{1976MNRAS.177..215M}, \citealt{1982ApJ...253..575M}, \citealt{1983IBVS.2425....1O})
\begin{equation}
  W = \rps-R (\rps- \ips) = {\nrps} -R ({\nrps} - {\nips}).
\end{equation}
With $\rps = {\nrps} + A_{\rps}$, $\ips = {\nips} + A_{\ips}$ and
$A_{\rps}$ and $A_{\ips}$ being the extinction corrections in the
$\rps$ and $\ips$ bands we can derive the coefficient:
\begin{equation}
\begin{array}{l}
 {\nrps} + A_{\rps} - R ({\nrps} - {\nips}) -R (A_{\rps} - A_{\ips}) = {\nrps} -R ({\nrps} - {\nips}) \\
\Rightarrow R = \frac{A_{\rps}}{A_{\rps} - A_{\ips}}
\end{array}
\end{equation}

The galactic extinction coefficients $\frac{A_{\rps}}{E(B-V)}$ and
$\frac{A_{\ips}}{E(B-V)}$ are \citep{2012arXiv1203.0297T}:
\begin{equation}
  \begin{array}{l}
    \frac{A_{\rps}}{E(B-V)} = 2.585 - 0.0315 \cdot (\ngps-\nips)\\
    \frac{A_{\ips}}{E(B-V)} = 1.908 - 0.0152 \cdot (\ngps-\nips)
  \end{array}
  \label{eq.extinction}
\end{equation}

\begin{figure}[t!]
  \epsscale{1.0}
  \plottwo{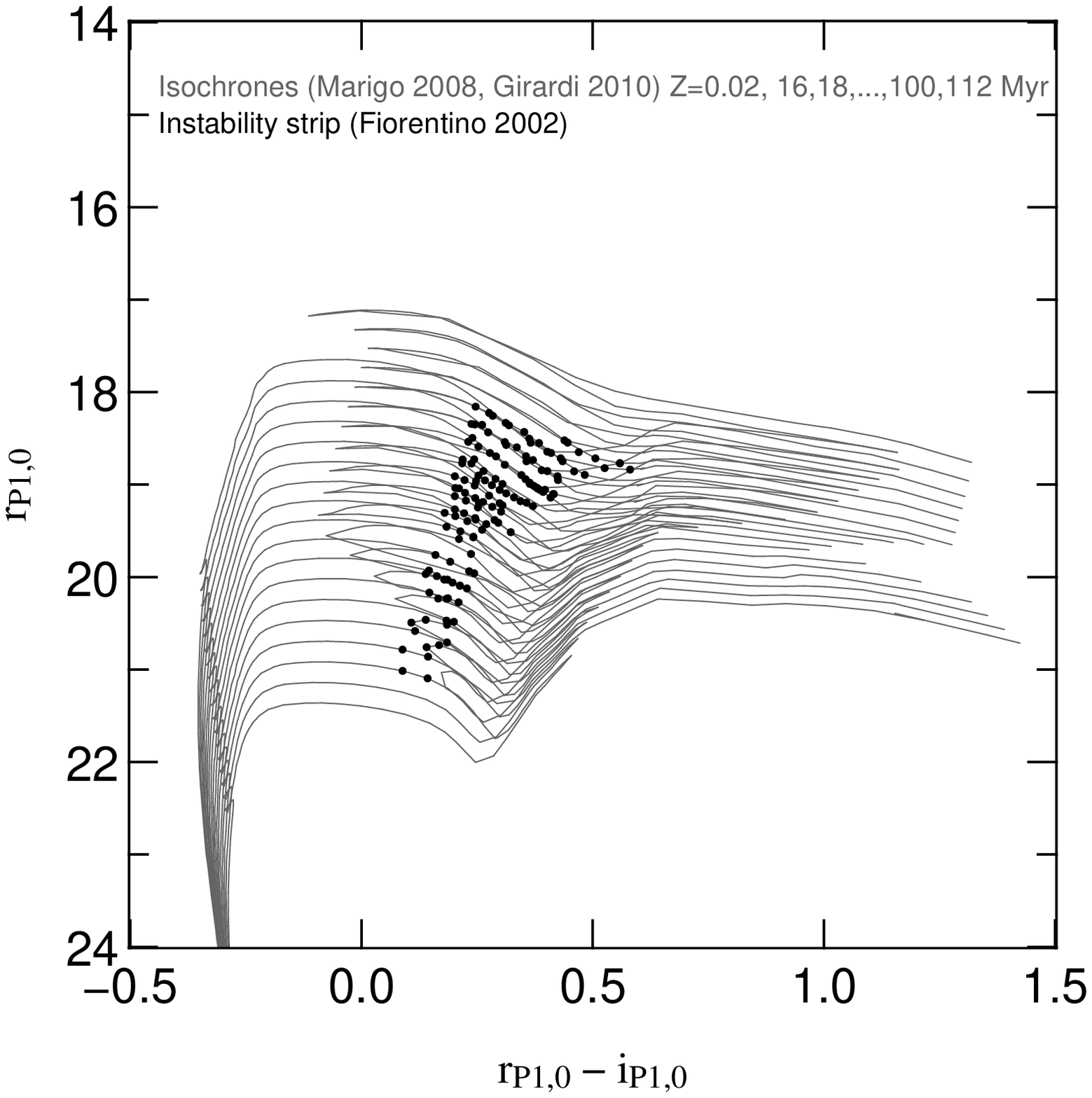}{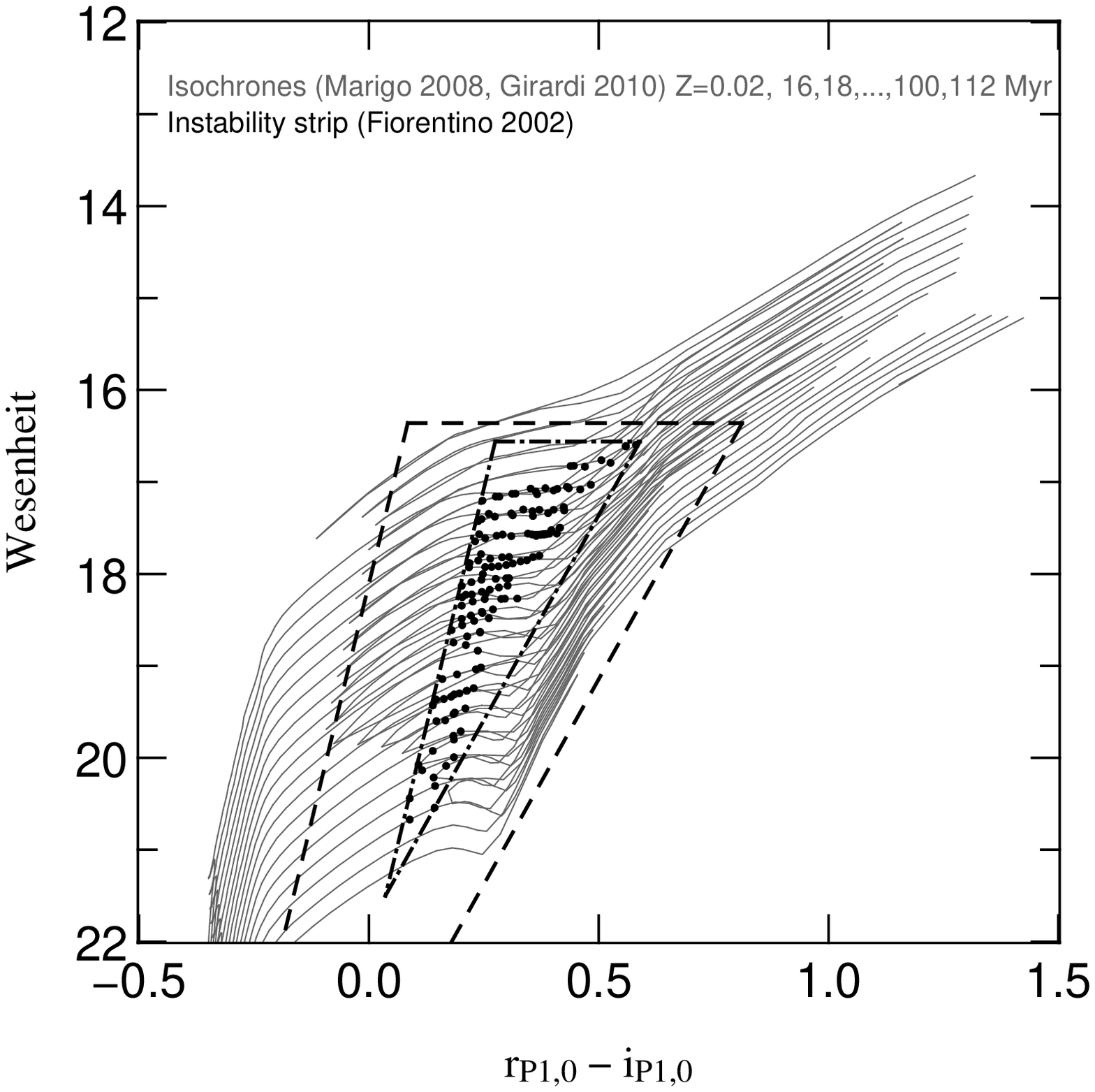}
  \label{Fig.Wesenheit}
  \caption{Domain of the instability strip for Cepheids used for the
    Wesenheit-color cut. A distance modulus of 24.36
    \citep{2010ASPC..435..375V} has been applied to the magnitudes.
    Left panel: Location of the instability strip for Cepheids in the
    magnitude-color plane according to instability theory. Right
    panel: Location of the instability strip for Cepheids in the
    Wesenheit-color plane according to instability theory. The area
    inside the dashdotted lines encloses the instability strip and the
    area within the dashed lines is where we require the mean
    magnitude of the light curve to be in, within their error
    margin. To define this domain we enlarged the boundaries of the
    dashdotted region by $0.2 ~ mag$ in both color directions and by
    $0.2~mag$ in direction of brighter Wesenheit magnitudes.}
\end{figure}

\begin{figure}[h!]
\epsscale{0.5}
  \plotone{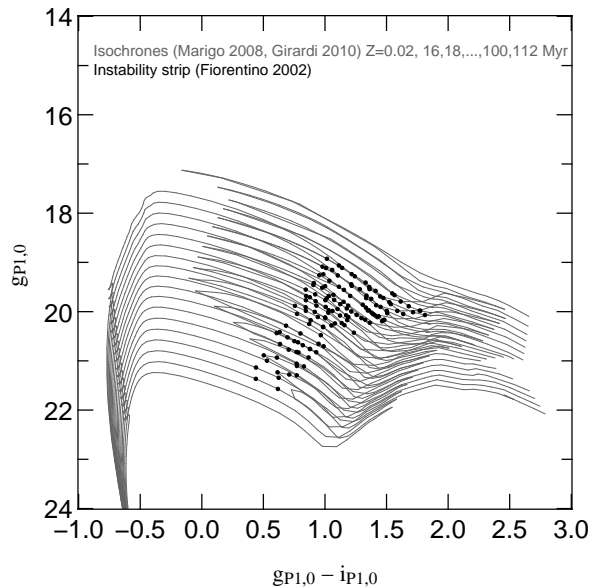}
  \caption{Domain of the instability strip for Cepheids used to
    determine the $(\ngps-\nips)$  correction in the galactic extinction
    coefficients. Since the correction is small, $(\ngps-\nips) = 1.0$ is a
    reasonable choice. }
  \label{Fig.gi}
\end{figure}

\cite{2002ApJ...576..402F} provides the position of the instability
strip for Cepheids depending on $\frac{L}{L_\odot}$ and
$T_{\mathrm{eff}}$ ($L_\odot$ is the solar luminosity and
  $T_{\mathrm{eff}}$ the effective Temperature). We combine this with
the isocrones from \cite{2008A&A...482..883M} with the corrections by
\cite{2010ApJ...724.1030G} with $Z=0.019$, for different
$\frac{L}{L_\odot}$ and $T_{\mathrm{eff}}$, thus allowing us to
determine the theoretical location of the Cepheids in the
color-magnitude diagram, as shown in the left panel of
Fig. \ref{Fig.Wesenheit}. The right panel of Fig. \ref{Fig.Wesenheit}
shows the domain defined by this theoretical prediction in the
color-Wesenheit diagram (defined by a triangle with the corners
(0.036/21.50), (0.275/16.56) and (0.59/16.56)). A candidate is
qualified as a Cepheid variable star if its mean magnitude is located
inside this triangle (including its $1\sigma$ error). Similar to
Fig. \ref{Fig.Wesenheit} we also obtained the location of the Cepheids
in the $\ngps$ vs. $\ngps - \nips$ plane, see Fig. \ref{Fig.gi}.  For
$(\ngps-\nips) = 1.0$ and from Eq.~\ref{eq.extinction} we get
$\frac{A_{\rps}}{E(B-V)} = 2.5535$, $\frac{A_{\ips}}{E(B-V)} = 1.8928$
and $R = 3.86$. Since the $(\ngps-\nips)$ correction is small, the
min-max error in $R$ is $\Delta R = 0.04$, which makes $(\ngps-\nips)
= 1.0$ a reasonable choice.

While the Wesenheit magnitude is independent of extinction, the color
is not. 
Therefore we apply an extinction correction to the colors.  We use the
$E(B-V)$ map from \cite{2009A&A...507..283M} and the foreground
extinction of M31 derived by \cite{1998ApJ...500..525S} with
$E_{\mathrm{fg}}(B-V)=0.062$ and apply the recalibration factor of 0.86
\citep{2011ApJ...737..103S}, to determine a mean extinction for each
light curve depending on its (projected) position:
\begin{eqnarray}
\overline{A_{\rps}} =  \frac{E(B-V) + 0.86 \cdot 0.062}{2} \cdot \frac{A_{\rps}}{E(B-V)}\\
\overline{A_{\ips}} =  \frac{E(B-V) + 0.86 \cdot 0.062}{2} \cdot \frac{A_{\ips}}{E(B-V)}
\end{eqnarray}

Note that for Cepheids that are outside the Montalto et al. map 
we only corrected for the foreground extinction.

Obviously we could use an extinction corrected magnitude and use it
instead of the Wesenheit, but since the correction is associated with
a large systematic uncertainty it is better to use the Wesenheit,
which does not require an extinction correction. The Wesenheit thus
has a higher statistical error, but a lower systematic error relative
to the extinction corrected magnitudes.

With the Wesenheit and the extinction corrected $({\nrps} -{\nips})$
color we are able to select only those light curves belonging to the
instability strip. This Wesenheit-color, applied to the 75362 periodic
light curves detected by {\tt SigSpec}, excludes 70\% of the light
curves.

\subsection{Manual classification}
\label{2.4}
 
In order to find out in which part of the $P$, $\fA$, $\fP$
3-dimensional parameter space the Cepheids are located, we need a
'training' set of Cepheids which we obtain from a manual
classification, i.e. by visual inspection of the preselected light
curves.

The visual inspection of a subsample of Cepheid candidates and
admission or rejection of Cepheids based on the light curve can lead
to a bias if the Cepheid subsample itself does not represent all
Cepheid types. The most unbiased way to select a Cepheid candidate
subsample for manual classification is to use the \textit{spectral
  significance} (sig). On the one hand the significance threshold has
to be chosen low enough to contain a fair amount of Cepheids, so that
there are enough Cepheids for the exploration of the 3 dimensional
parameter space. On the other hand, we do not want the Cepheid
candidate subsample to contain so many Cepheids already, that the
classification with the parameter space does not detect significantly
more Cepheids than the subsample.

For the manual classification we thus selected all light curves that
have $sig > 25$ in the \rps band. As reference light curves for the
manual classification we used the Cepheid light curves from
\cite{sterken2005light}.

The resulting manual classified subsample contains 1020 Cepheids.
This subsample makes up $\sim$ 50\% of the final Cepheid sample.

\subsection{Type classification}
\label{2.5}

\begin{figure}[h!]
  \epsscale{1.0}
  \plottwo{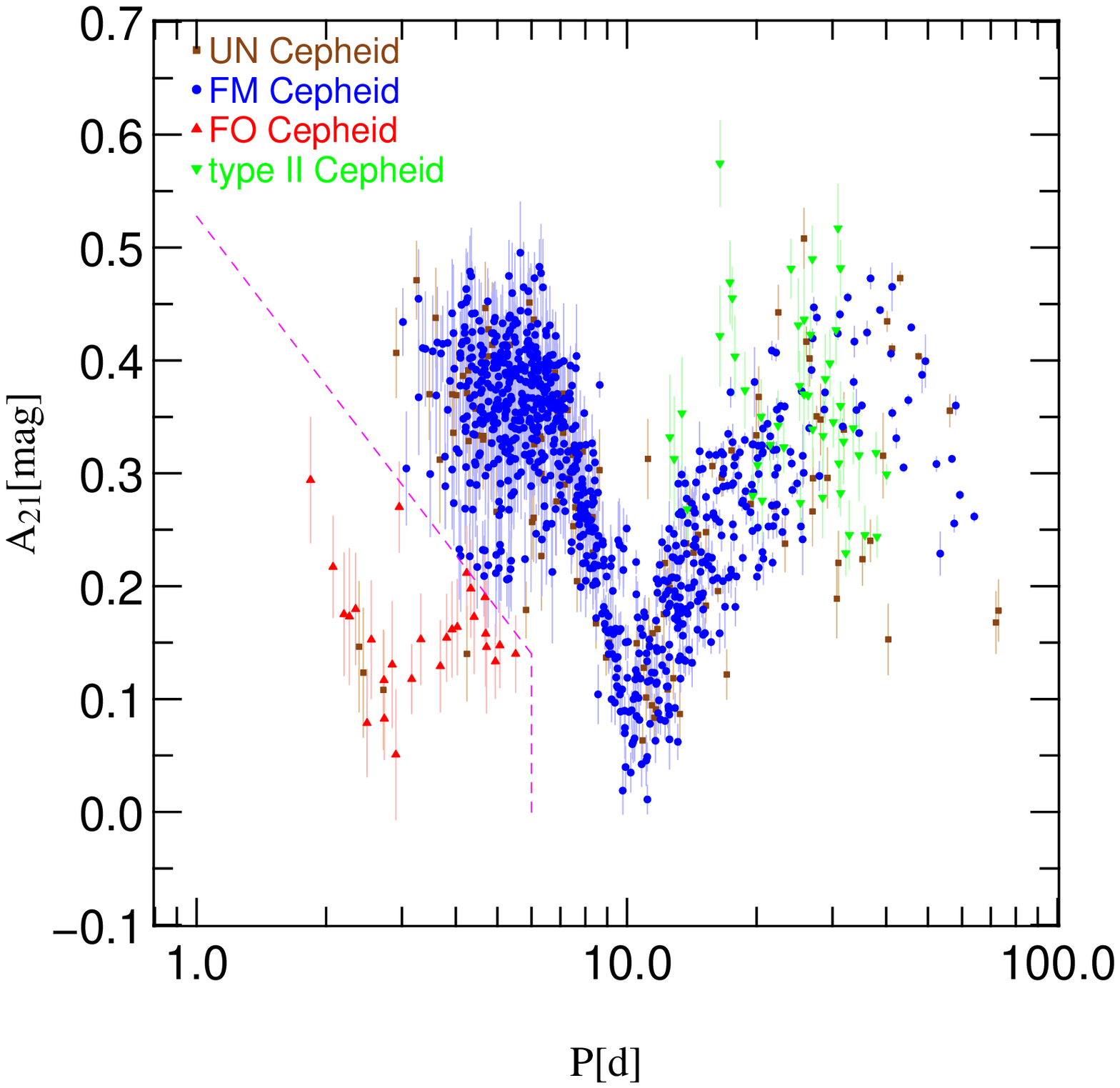}{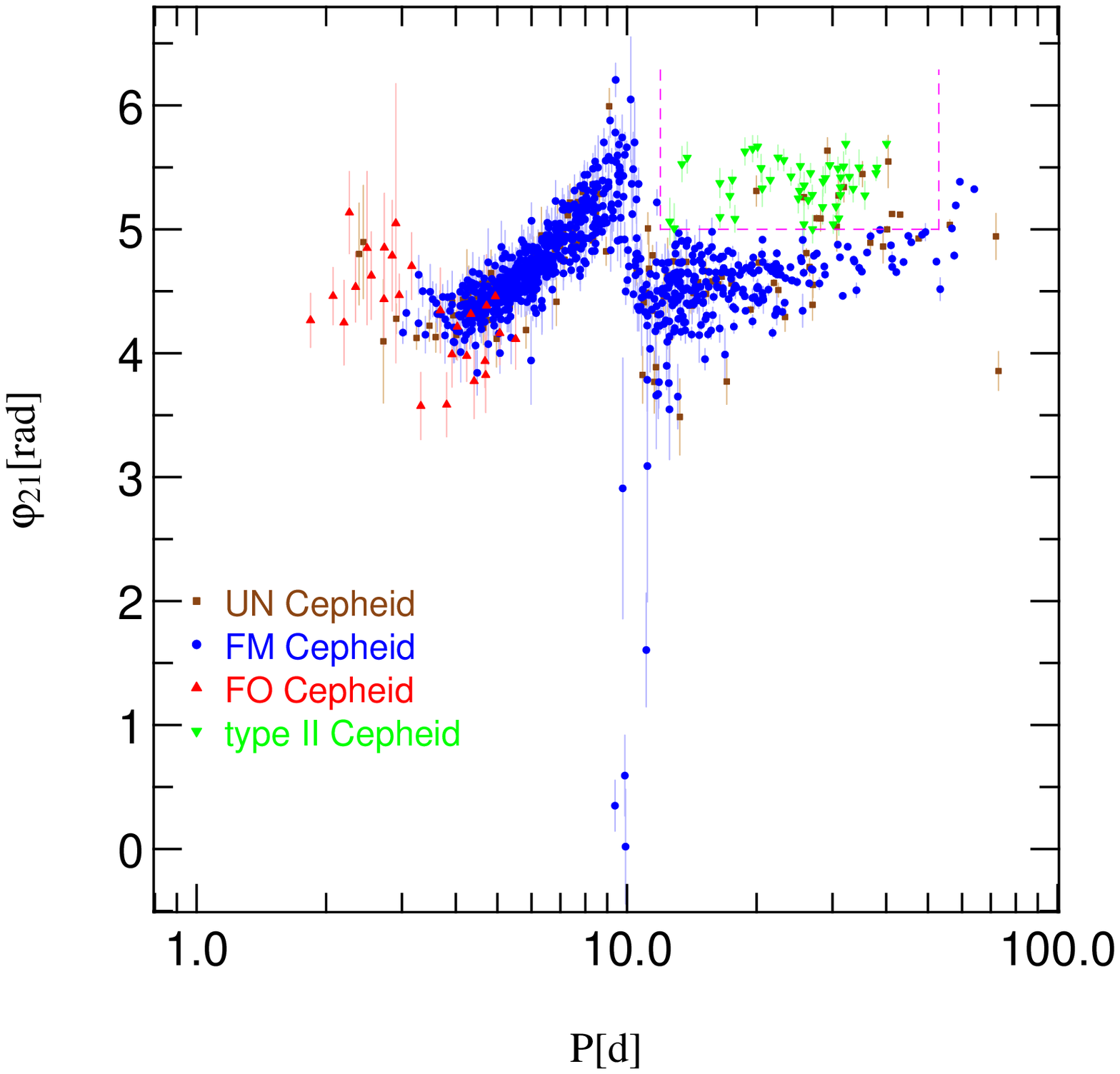}
  \caption{3-dimensional parameter space for the manually classified
    Cepheid catalog in the $\rps$ band. The scope defined by the
    magenta dashed lines defines the parameter space in which most of
    the FO Cepheids ($P < 6.0 \, \mathrm{d}$ and $\fA <
    -0.498289*log(P)+0.527744$) or rather most of the Type II Cepheids
    ($11.95 \, \mathrm{d} < P < 53.0 \, \mathrm{d}$ and $\fP > 5 \,
    \mathrm{rad}$) reside in, while the FM Cepheids occupy the
    remaining parameter space. Since there are transitions between the
    different scopes, we perform a $3 \sigma$ clipping in the Period -
    Wesenheit relation (cf. Fig. \ref{Fig.PWesenheitMan}) for each
    Cepheid type. All the clipped Cepheids and all those Cepheids with
    a large error in the Wesenheit ($\Delta W > 0.9 $) are classified
    as unclassified (UN) Cepheids. Left panel: Amplitude ratio $\fA$ in
    the \rps band. Right panel: Phase difference $\fP$ in the \rps band.
  }
  \label{Fig.FourierMan}
\end{figure}

As already mentioned in Section \ref{Fdecomp} the Fourier
decomposition can be used to distinguish between FM Cepheids and FO
Cepheids. This is done by identifying different locations of the
Cepheids in the \fA-P and \fP-P planes
(e.g. \citealt{1999AcA....49..223U},
\citealt{2007A&A...473..847V}). Fig. \ref{Fig.FourierMan} shows these
projections for the manually classified Cepheid catalog. When looking
at the Cepheids in a 3-dimensional plot in \fA-\fP-P space, it becomes
clear that there are indeed three different relatively well separated
distributions with only small overlap areas.

The FM Cepheids correspond to the `V-shaped` sequence that can be seen
in in the left panel of Fig. \ref{Fig.FourierMan}. In a 3-dimensional
plot the FM sequence appears spiral shaped and the two branches are
shifted slightly from one another. The FO Cepheid sequence is
perpendicular to the FM sequence, but there is also an overlap regime
between them where FM and FO objects can be hardly distinguished.  For
objects in or near this overlap area the classification into FM or FO
is therefore rather uncertain as a real parameter overlap seems to be
present.

We additionally find a third sequence not present in
\cite{1999AcA....49..223U} and \cite{2007A&A...473..847V} and
corresponding to Type II Cepheids. This sequence is more distinct from
the FM Cepheids than the FO Cepheids, nevertheless there is also a
transition between the two.

\begin{figure}[h!]
  \epsscale{0.5}
  \plotone{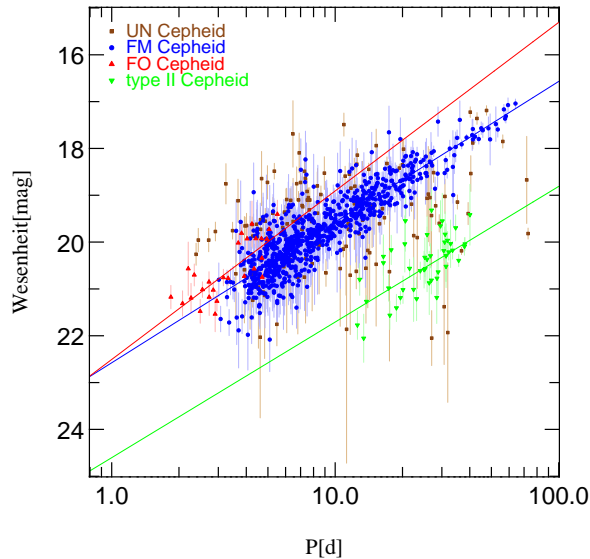}
  \caption{Period - Wesenheit relation for the manually classified
    Cepheid catalog. The Period - Wesenheit relation is used to
    identify those Cepheids that reside between the type
    classification scopes (cf. Fig. \ref{Fig.FourierMan}). This is
    done by performing $3 \sigma$ clipping for a linear relation ($W =
    a \cdot log(P) + b$, c.f. Section \ref{chapterpl}) for each
    Cepheid type (FM Cepheids: blue symbols and line; FO Cepheids: red
    symbols and line; Type II Cepheids: green symbols and line). All
    the clipped Cepheids and all those Cepheids with a large error in
    the Wesenheit ($\Delta W > 0.9 $) are classified as unclassified
    (UN) Cepheids.}
  \label{Fig.PWesenheitMan}
\end{figure}

In a first step we define the parameter space where most of the FO
Cepheids and the Type II Cepheids are located and assign the
according type to the Cepheid. We use
\begin{equation}
 P < 6.0~\mathrm{d}  \quad\mathrm{and} \quad   \fA <-0.498 \log(P)+0.528
\end{equation}
for the FO parameter space.\\
For the Type II
space we use 
\begin{equation}
  11.95~\mathrm{d} < P < 53.0~\mathrm{d}   \quad\mathrm{and} \quad \fP > 5~\mathrm{rad.}
\end{equation}
The remaining Cepheids are FM Cepheids. Fig. \ref{Fig.FourierMan}
shows the three different sequences. Nevertheless there are
transitions between the sequences. There are FM Cepheids in the FO and
Type II parameter space and FO Cepheids or rather Type II Cepheids in
the FM sequence. In order to account for these ambiguous cases we
identify these Cepheids in a second step by using the Period -
Wesenheit relation. By using the Wesenheit and not the $\rps$ band we
get a classification independent of reddening. We perform iterative $3
\sigma$ clipping for a linear relation ($W = a \cdot \log(P) + b$,
c.f. Section \ref{chapterpl}) for each Cepheid type and classify all
clipped Cepheids and all those Cepheids with a large error in the
Wesenheit ($\Delta W > 0.9 $) as unclassified (UN)
Cepheids. Fig. \ref{Fig.PWesenheitMan} shows that this classification
works well, but also indicates the problem with this
approach. Cepheids that are e.g. outside the defined Type II parameter
space will be classified as UN Cepheids although it is apparent from
Fig. \ref{Fig.PWesenheitMan} that they are most likely Type II
Cepheids.
 
\subsection{Selection criteria}
\label{2.2}
With a visual inspection of the light curves only a color cut is
needed to identify the Cepheids. Since we want to identify Cepheids
using the $P$, $\fA$, $\fP$ 3-dimensional parameter space cut we have
to apply additional selection criteria to exclude bad light
curves. Note that the following criteria can be applied in any order. Table \ref{tabcuts} summarizes the
selection cuts.

The first two criteria (\Rmnum{1} \& \Rmnum{2}) concern the period of
the light curve. As already mentioned in Section \ref{sec.period} we have not
applied a period cut yet. Our data covers approximately 150 days in
2010 and a few days only in 2011. While inspecting the light curves by
eye, these few days in 2011 can help to confirm that the detected
period is valid, even if the period is $\sim 100 \,
\mathrm{d}$. However since there is a gap of approximately half a year
between the 2010 and 2011 data with only few days of
observations in 2011, the light curve could be irregular in the time
span of the gap and the data points in 2011 could fit to the folded
light curve by chance. We decided therefore to select only light
curves with periods below 75 days (\Rmnum{1}) , so that the whole period cycle is
sampled at least twice in the 2010 data. We note that there could very
well be Cepheids with larger periods than that, but we will be able to
confirm those only after the release of the full 2011 data. The second
criterion (\Rmnum{2}) is that the period determined in the $\ips$
filter differs from that in the $\rps$ filter by less than one
percent. This criterion excludes most of the light curves (84 \%).

\begin{figure}[h!]
  \epsscale{1.0}
  \plottwo{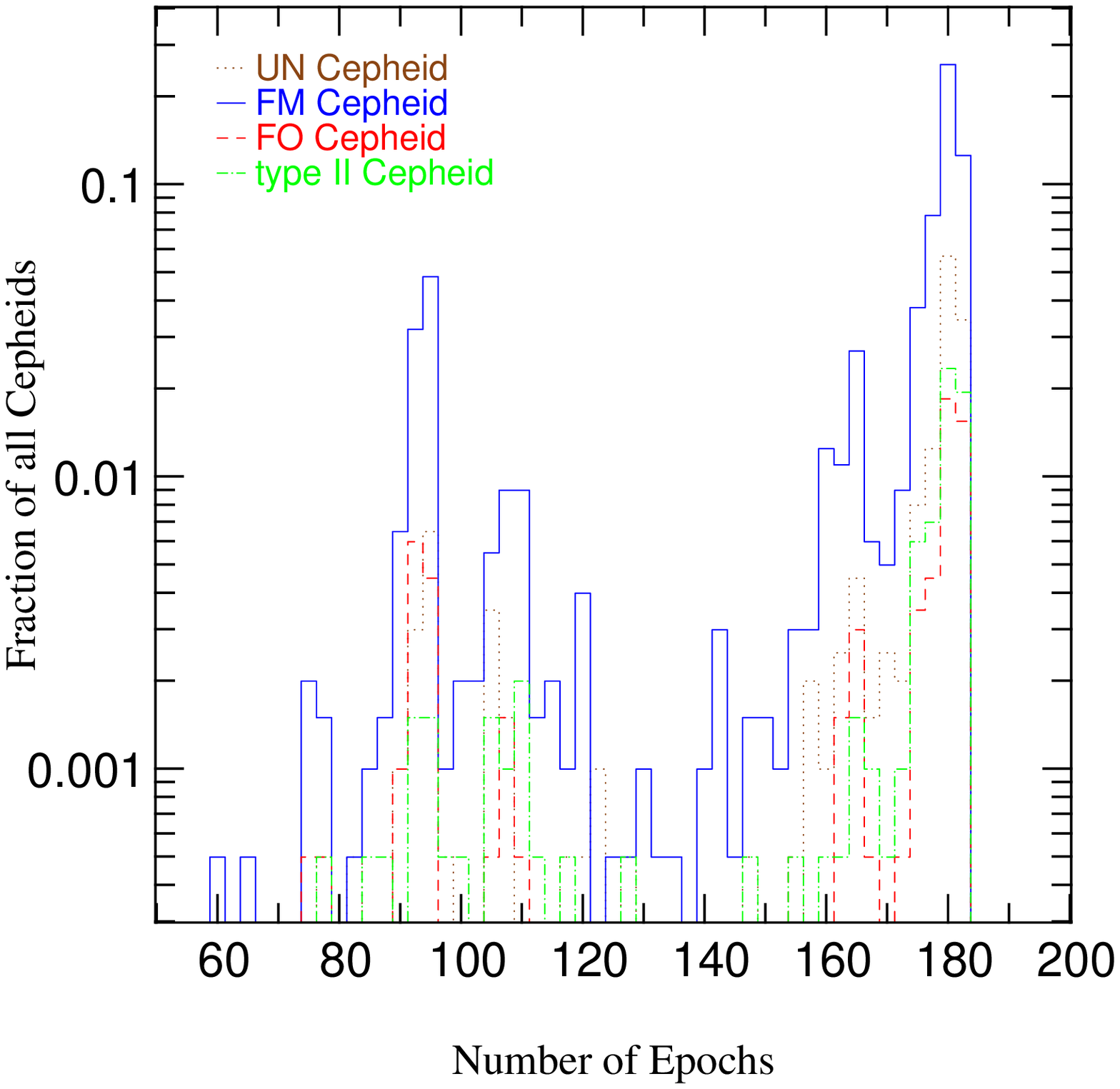}{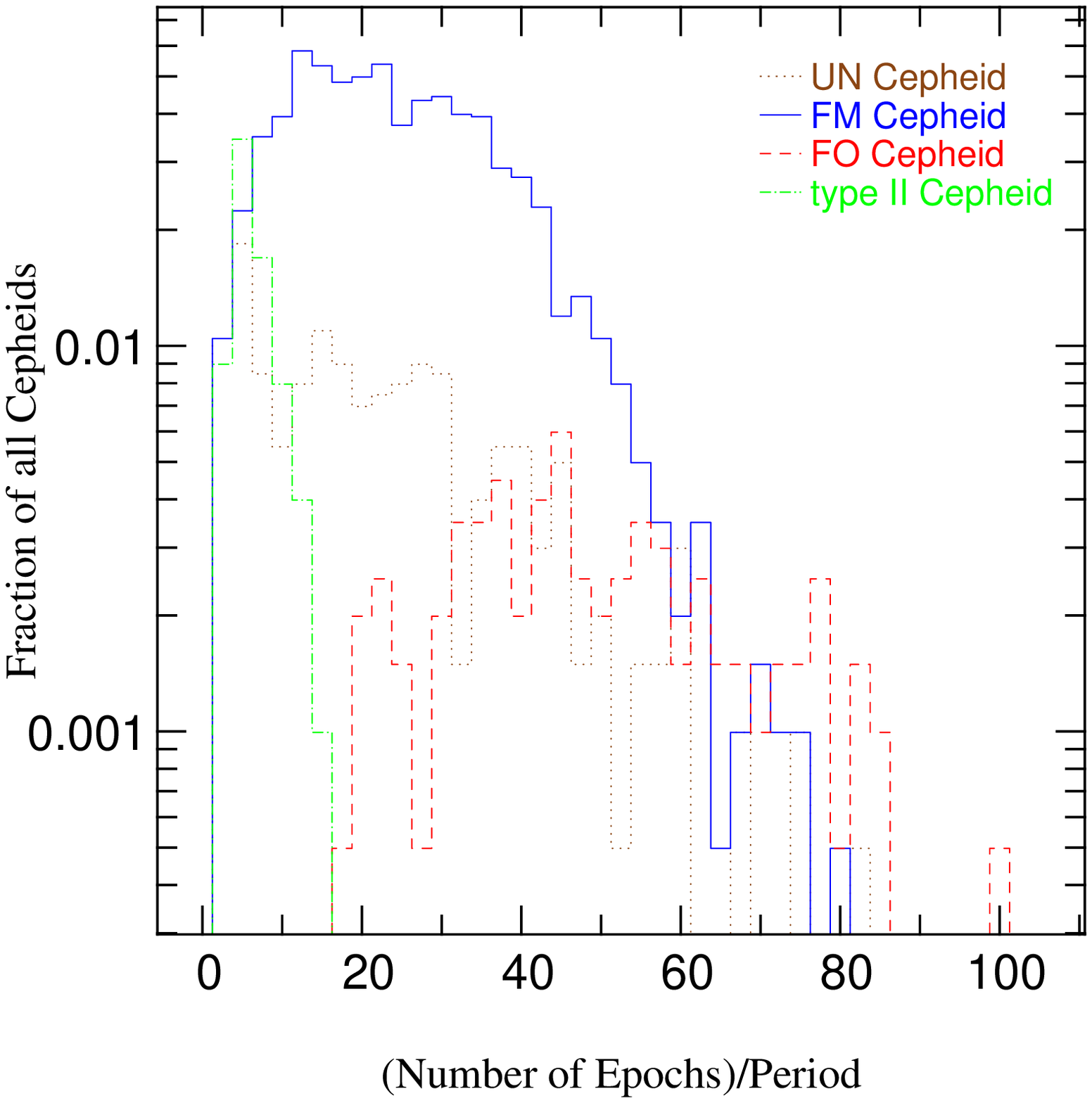}
  \caption{Epoch distribution for the 3-dimensional parameter space
    classified Cepheid catalog in the \rps~band. Left panel:
    Distribution of the number of epochs. Right panel: Distribution of
    the number of epochs divided by the period.}
  \label{Fig.epoch3d}
\end{figure}

We do not require a light curve to contain more than a certain number
of epochs. The reason for this is that we use the Fourier series fit
and the polynomial fit to account for gaps in the data that might be
problematic and produce `overshoots`. Most of our light curves have
$\sim 180$ epochs, but there can also be light curves with as few as 60
epochs\footnote{60 is the lowest number of epochs for a light curve
  associated to a Cepheid in the 3-dimensional parameter space
  classified Cepheid catalog.}.  The left panel of
Fig. \ref{Fig.epoch3d} shows the distribution of the number of epochs
for the 3-dimensional parameter space classified Cepheid catalog. The
right panel of Fig. \ref{Fig.epoch3d} shows that there are only a few
Cepheids with a small number of epochs when normalized to the period
of the light curve in the 3-dimensional parameter space classified
Cepheid catalog. A visual inspection shows that these light curves
constitute Cepheids. This indicates that we need no criterion
requiring a certain number of epochs when applying the three
`overshoot` criteria (The classification of the different Cepheid
types shown in Fig. \ref{Fig.epoch3d} is described in the next
Sections).

The criteria \Rmnum{3} \& \Rmnum{4} exclude extreme `overshoots` by
selecting only those light curves, where the magnitudes $m(t)$ of the
fitted light curve are reasonable: $m(t) < 30~\mathrm{mag}$ and $m(t) >
15~\mathrm{mag}$. 

Note that criteria \Rmnum{1} to \Rmnum{4} are applied to both
  $\rps$ and $\ips$ band. All other criteria only concern the $\rps$
band.

In order to also exclude less extreme `overshoots`, we apply criterion
\Rmnum{5}, which selects only light curves with similar Fourier series
fit and polynomial fit. We check for similarity by calculating the
area between both fits and demand that the area $\int\limits_0^P
(m(t)-p(t))^2 dt$ is below a threshold. The median of this area is
$\sim 0.2 ~ \mathrm{mag} \cdot \mathrm{d}$ and the largest area is $\sim
13$ for the 3-dimensional parameter space classified Cepheid
catalog. So our threshold of 30 is quite high when compared to those
numbers, but the `overshoots` usually produce orders of magnitude larger
values than this threshold. The choice of the threshold is arbitrary
to some degree, although the threshold of 30 has proven to work well.

The last two selection criteria (\Rmnum{6} \& \Rmnum{7}) concern the
noise in the folded light curves. If the maximum and minimum of the
Fourier series fit are similar, we cannot exclude that the light curve
is constant and the variation is caused by noise. Additionally light
curves affected by blending have smaller amplitudes. Therefore we
select only those light curves with: $\max(m(t))-\min(m(t)) \geq 0.1 ~
\mathrm{mag}$. The $\chi^2$ of a fit can be low, although the scatter
is large, if the errors in the data points are large enough. Therefore
we introduce a last selection criterion (\Rmnum{7}), which concerns
the scatter of the light curve. We want to allow for a few points to
scatter, that is why we compare the median scatter of the Fourier
series fit with the total change in magnitude:
$\mathrm{median}\left[|m(t)-y(t)|\right] \leq 0.2
\left(\max(m(t))-\min(m(t))\right)$ ($m(t)$ is the magnitude of the
Fourier series fit at time $t$ and $y(t)$ is the measured magnitude at
time $t$).  The threshold of $0.2$ was optimized by a visual
inspection.

\begin{table}[h]
  \centering
  \caption{Selection criteria for the light curves. $P$ is the period, 
    $m(t)$ the magnitude of the Fourier series fit at time $t$, $p(t)$ the polynomial
    fit at time $t$ and $y(t)$ is the measured magnitude at time $t$. The last column
    is the percentage excluded by the criterion (if each criterion is applied individually)
    from the 75362 periodic light curves detected by SigSpec.}
  \renewcommand{\arraystretch}{1.4}
  \begin{tabular}{c|ccc|c}
  \quad & Selection criterion & \quad & \quad & percentage excluded\\[2ex]
  \tableline
  \Rmnum{1} & $P_{\rps}$ & $\leq$ & $75 \, \mathrm{d}$ & 66\%\\[2ex]
  \tableline
  \Rmnum{2} & $| \frac{P_{\rps} - P_{\ips}}{P_{\rps}} |$ & $<$ & $0.01$ & 84\%\\[2ex]
  \tableline
  \Rmnum{3} & $m(t)$ & $<$ & $30 \, \mathrm{mag}$ & 19\%\\[2ex]
  \tableline
  \Rmnum{4} & $m(t)$ & $>$ & $15 \, \mathrm{mag}$ & 20\%\\[2ex]
  \tableline
  \Rmnum{5} & $\int\limits_0^P (m(t)-p(t))^2 dt$ & $<$ & $30$ & 31\%\\[2ex]
  \tableline
  \Rmnum{6} & $\max(m(t))-\min(m(t))$ & $\geq$ & $0.1 \, \mathrm{mag}$ & 4\%\\[2ex]
  \tableline
  \Rmnum{7} & $\frac{median(|m(t)-y(t)|)}{max(m(t))-min(m(t))}$ & $\leq$ & $0.2 \, \mathrm{mag}$ & 42\%\\[2ex]
  \tableline
  \end{tabular}
  \label{tabcuts}
\end{table}

\subsection{3-dimensional parameter space classification}
\label{2.6} 

After the selection criteria are applied (c.f. Section \ref{2.2}) we
are left with some objects with well defined light curves that are not
Cepheids. This becomes obvious if we compare their location in the
3-dimensional parameter space (\fA-\fP-P) with that of our manually
classified Cepheids. Therefore we use the manually classified
subsample to define a `3-dimensional parameter space` occupied by
Cepheids. All variables surviving the selection criteria in Section
\ref{2.2} which at the same time reside in the 3-dimensional Cepheid
subspace are considered Cepheids as well. The complexity of this 3d
subspace is so high that there are no 2d projections that could
classify Cepheids as effectively.

We span a $10 \times 10 \times 10$ grid\footnote{A grid with more
  grid-points could leave gaps, which we want to avoid.} in \fA-\fP-P
space with $1 \leq \log(P) \leq 75$, $0 \leq \fA \leq 0.75$ and $0
\leq \fP \leq 2 \pi$. We then identify those boxes around the
grid-points where the Cepheids of the manually classified subsample
reside in. The space defined by those grid-points that have Cepheids
around them, defines the 3-dimensional parameter space. The subsample
size and the rather large grid span force us to accept grid-points
that only have one data point around it. While this is not ideal, the
alternative would be to accept only grid-points that have a number of
Cepheids above a certain threshold and this would drastically decrease
the number of Cepheids at large periods, Type II Cepheids and FO
Cepheids.

After the Wesenheit-color cut and the selection criteria are applied
there are 2030 light curves within this parameter space, including
also those already identified with the manual classification. As
stated in Section \ref{sec.period}, these light curves can contain RV
Tauri light curves. We inspect those 2030 light curves manually and
find 21 RV Tauri light curves, which we exclude from the 3-dimensional
parameter space classified Cepheid catalog. The final 3-dimensional
parameter space classified Cepheid catalog contains 2009 Cepheids
(including the 1020 manually classified Cepheids).

\begin{figure}[h!]
  \epsscale{1.0}
  \plottwo{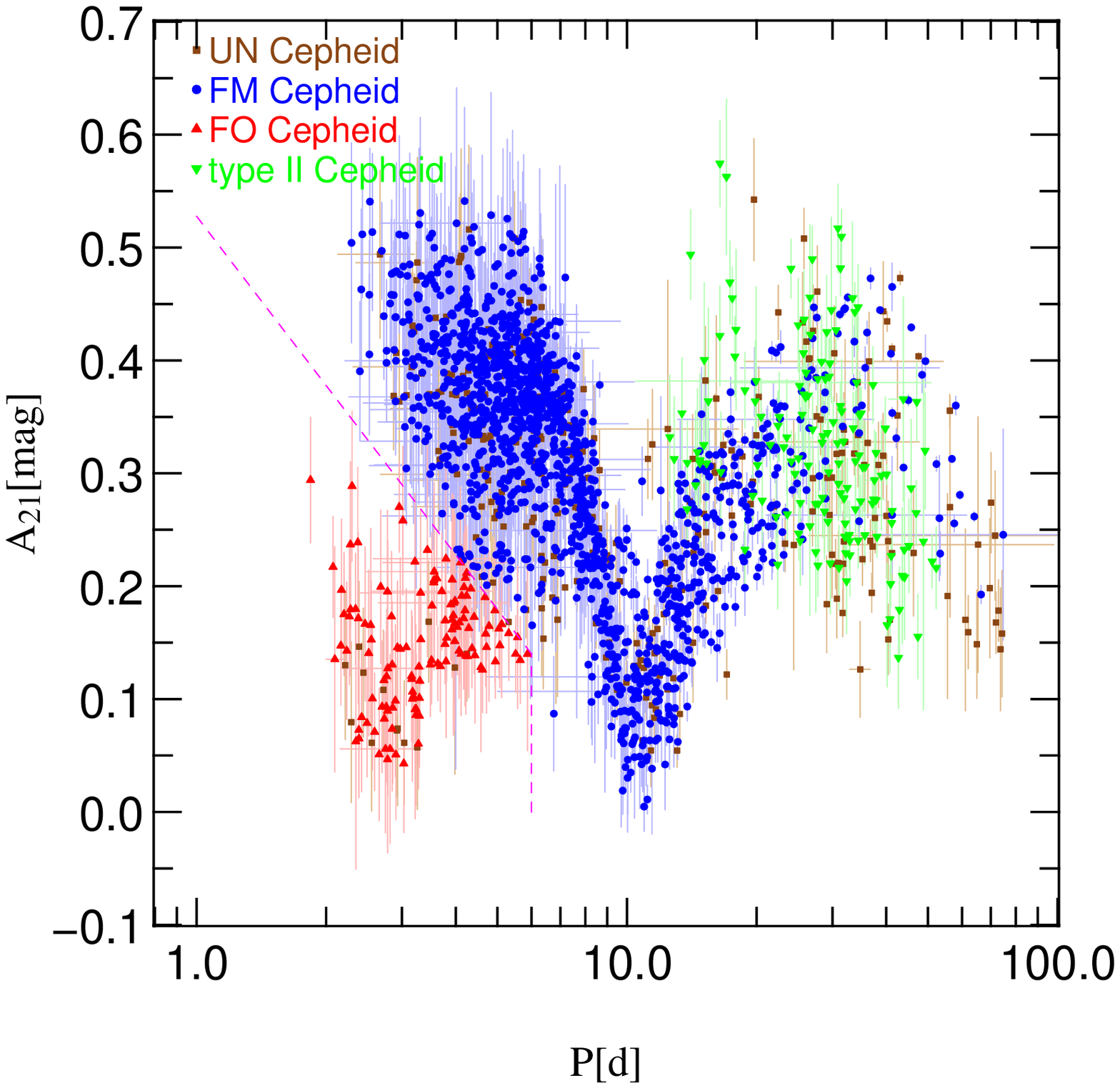}{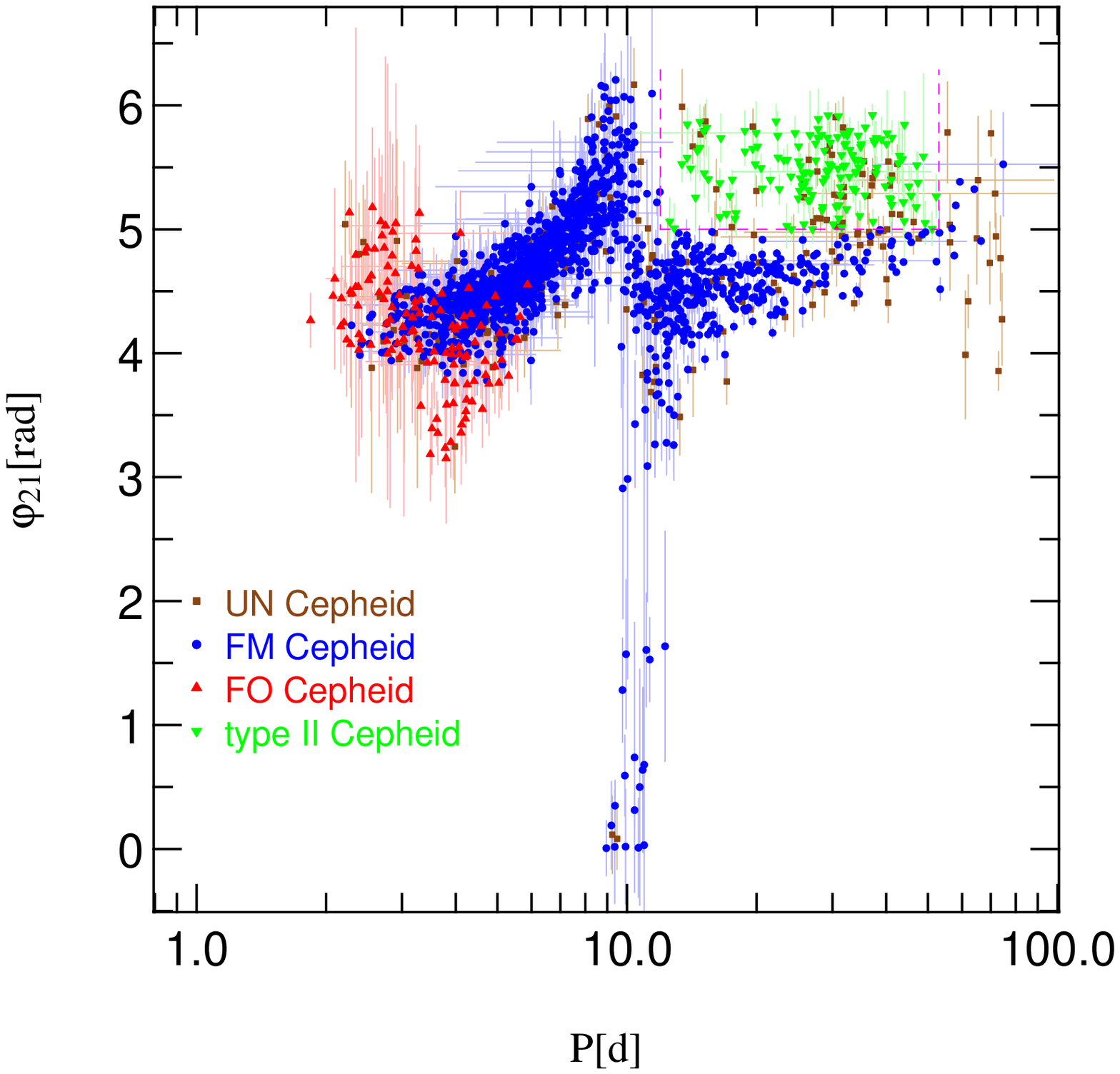}
  \caption{3-dimensional parameter space for the $P$, $\fA$, $\fP$ 3
    dimensional parameter space classified Cepheid catalog in the
    $\rps$ band.  The scopes for the type classification are the same
    as in Fig. \ref{Fig.FourierMan}. The transitions between the
    different sequences become more pronounced when compared to those
    in the manually classified Cepheid catalog, while the shape of the
    sequences is preserved (c.f. Fig. \ref{Fig.FourierMan}). Left
    panel: Amplitude ratio $\fA$ for the 3-dimensional parameter space
    classified Cepheid catalog. Right panel: Phase difference $\fP$
    for the 3 dimensional parameter space classification.}
  \label{Fig.A21P213d}
\end{figure} 

\begin{figure}[h!]
  \epsscale{0.5} \plotone{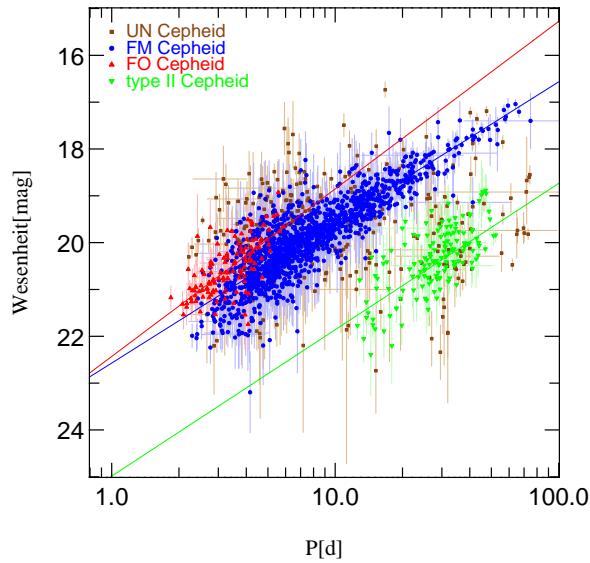}
  \caption{Same as Fig. \ref{Fig.PWesenheitMan}, but for the 3-dimensional parameter
      space classified Cepheid catalog.}
  \label{Fig.PWesenheit3d}
\end{figure}

As can be seen in Fig. \ref{Fig.A21P213d} the transition region between the
different Cepheid types becomes more populated, in comparison to the
manual classified Cepheid catalog
(c.f. Fig. \ref{Fig.FourierMan}). This makes the classification of the
UN Cepheids with the Period - Wesenheit relation
(Fig. \ref{Fig.PWesenheit3d}) even more important. Of the 75362
periodic light curves detected by SigSpec the 3 dimensional parameter
space cut alone excludes 94\%.
 
It is important to note that while this procedure works well, there
are Cepheids that are excluded by this strict 3-dimensional parameter
space classification cut. This is due to the grid size and the
sampling of the parameter space through the manually classified
subsample.

\section{The Catalog}
\label{sec.catalog}

The 3-dimensional parameter space classified Cepheid catalog contains
a total of 2009 Cepheids, consisting of 1440 FM Cepheids, 126 FO
Cepheids, 147 Type II Cepheids and 296 UN Cepheids.

\begin{figure}[h!]
  \epsscale{1.0}
  \plottwo{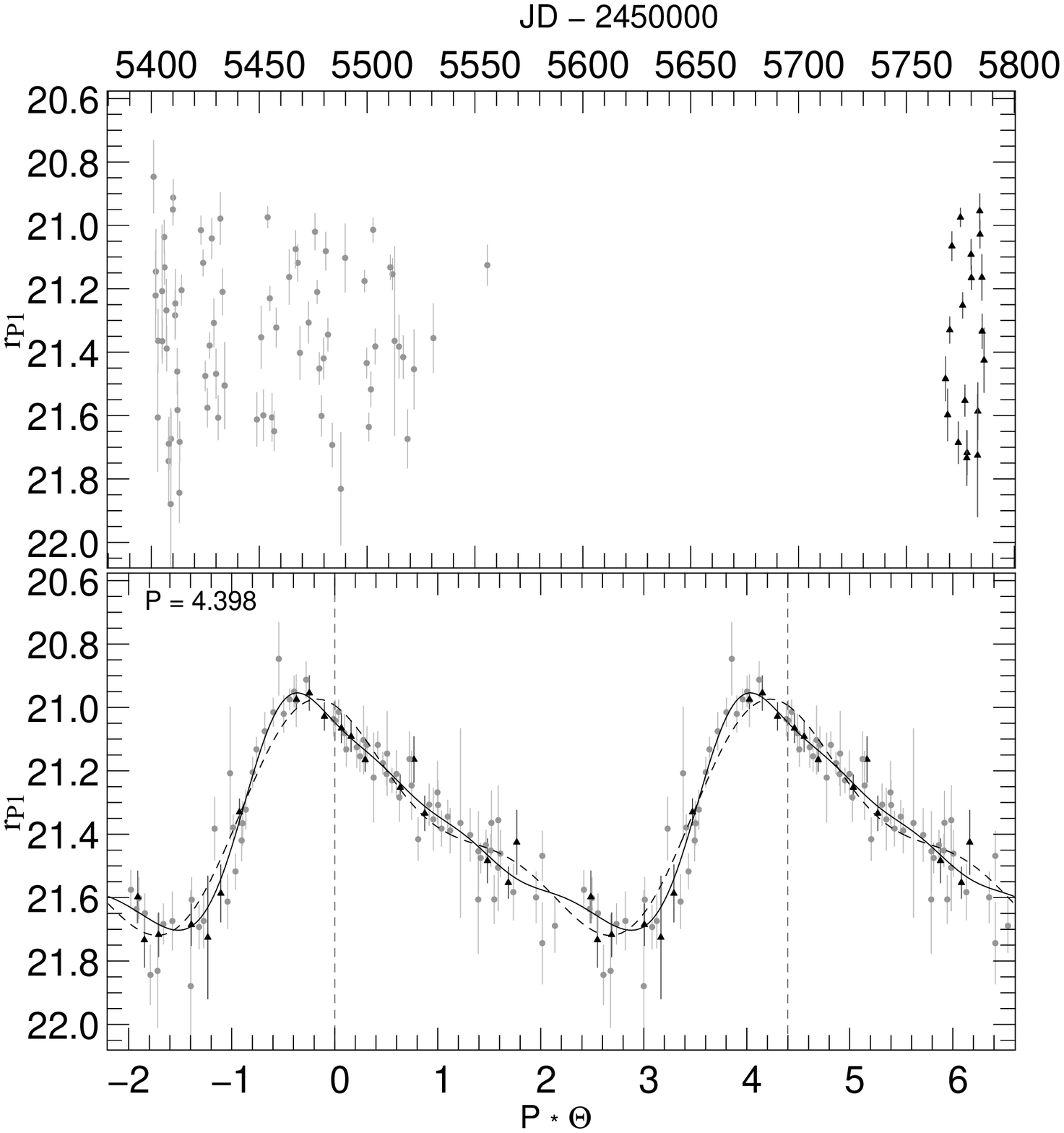}{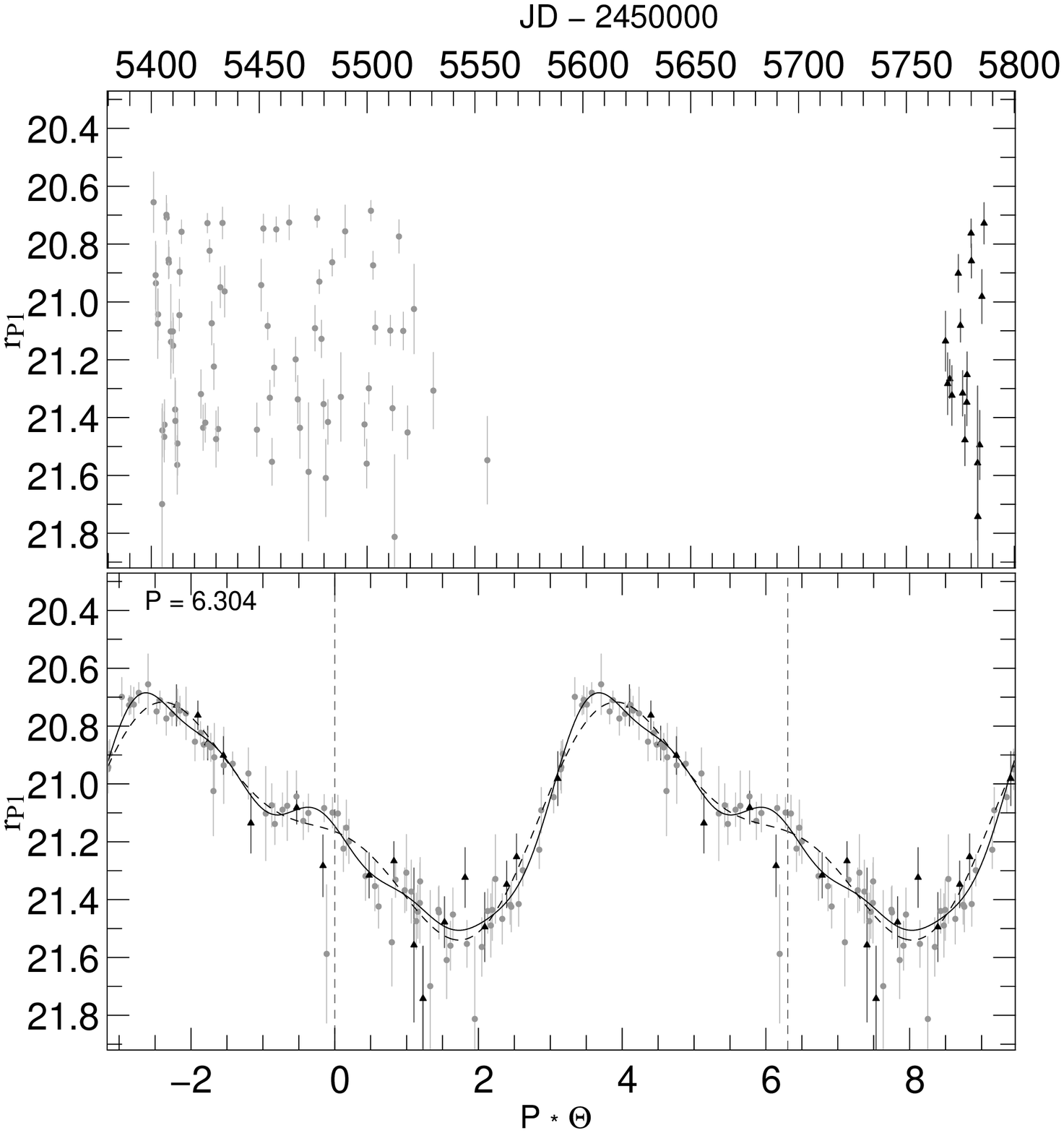} 
  \caption{Light curves in the \rps~band for a part of the excerpt of
    the 3-dimensional parameter space classified Cepheid catalog. Each
    top panel shows the unfolded light curve, while each bottom panel
    shows the variation of the folded light curve over the period
    times the phase $\Theta$. The folded light curve has been
    continued in both directions for better illustration, only the
    middle part between the dashed lines shows the real folded light
    curve. The grey circle data points represent the 2010 season,
    while the gray triangle data points are within the 2011
    season. The Fourier series fit is the black solid line and the
    black dashed line represents the light curve that is only defined
    by $\fA$, $\fP$ and the mean magnitude $r_{P1}$ (c.f. Section
    \ref{Fdecomp}). Left panel: Light curve of a FM Cepheid with
    $P=4.398~d$. Right panel: Light curve of a FM Cepheid with
    $P=6.304~d$.}
\label{bsplcFM1}
\end{figure}

\begin{figure}[h!]
  \epsscale{1.0}
  \plottwo{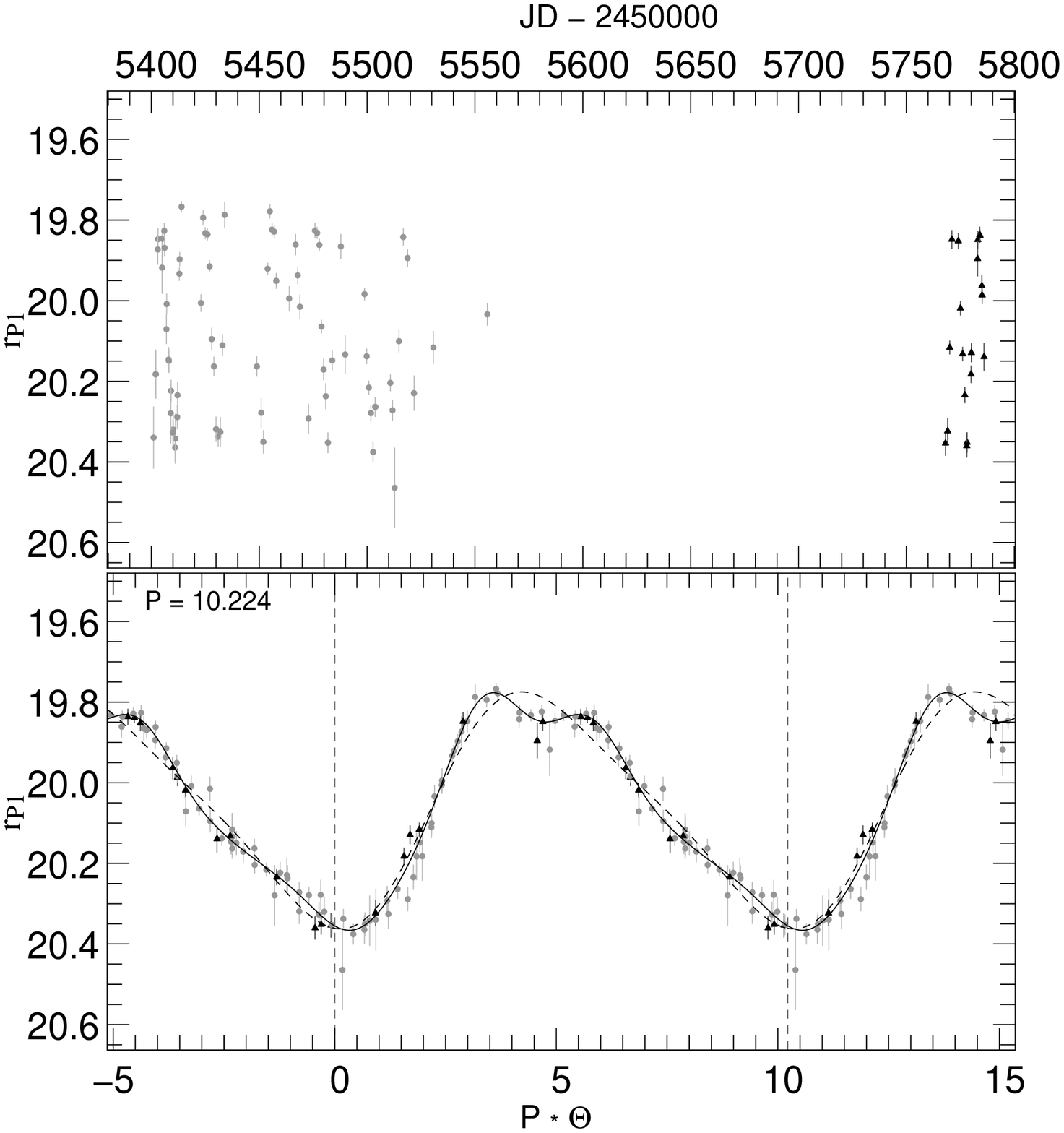}{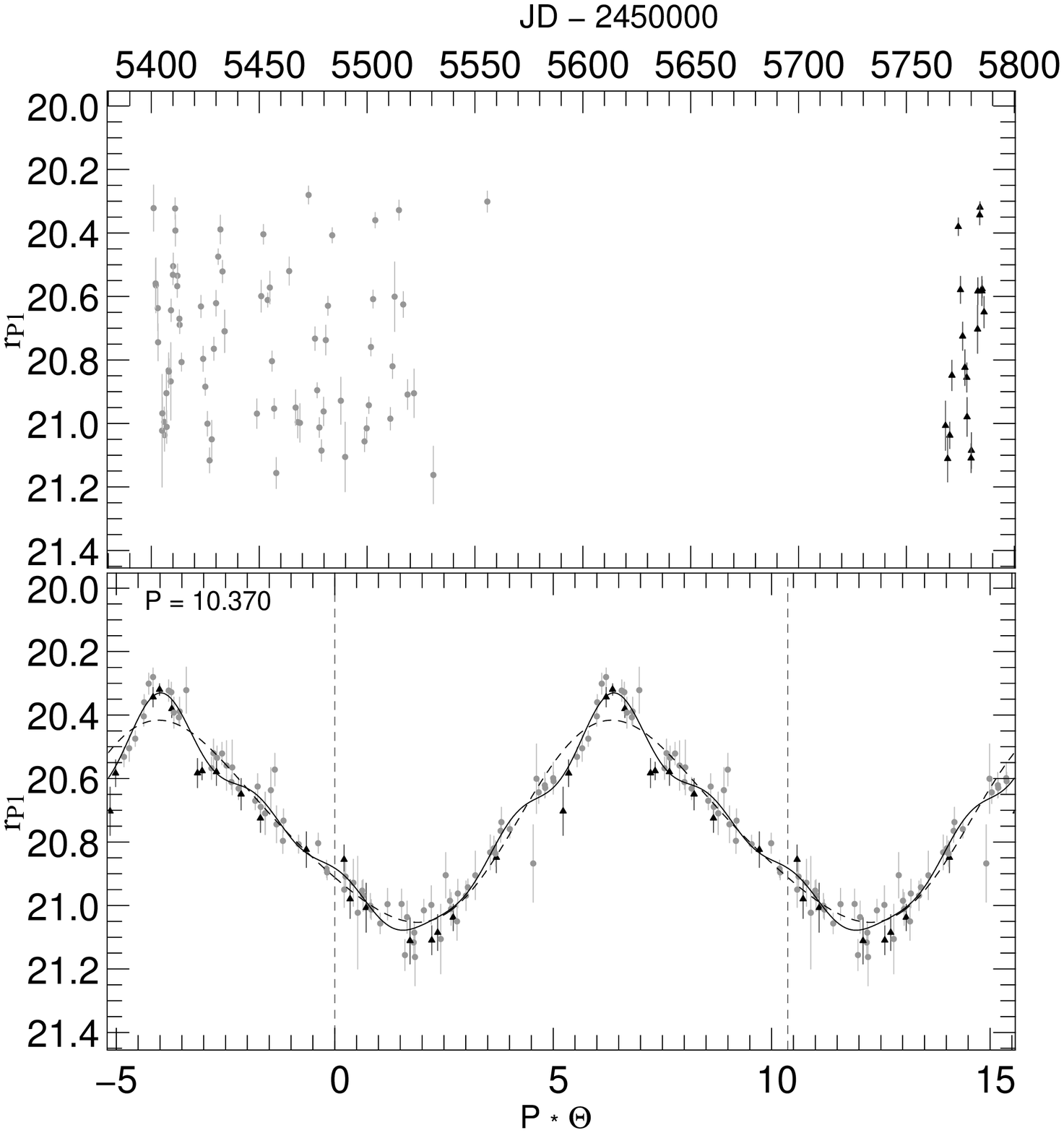} 
  \caption{Same as Fig. \ref{bsplcFM1} for the next two entries
    in the excerpt of the 3-dimensional parameter space classified
    Cepheid catalog. Left panel: Light curve of a FM Cepheid with
    $P=10.224~d$. Right panel: Light curve of a FM Cepheid with
    $P=10.370~d$.}
\label{bsplcFM2}
\end{figure}

\begin{figure}[h!]
  \epsscale{1.0}
  \plottwo{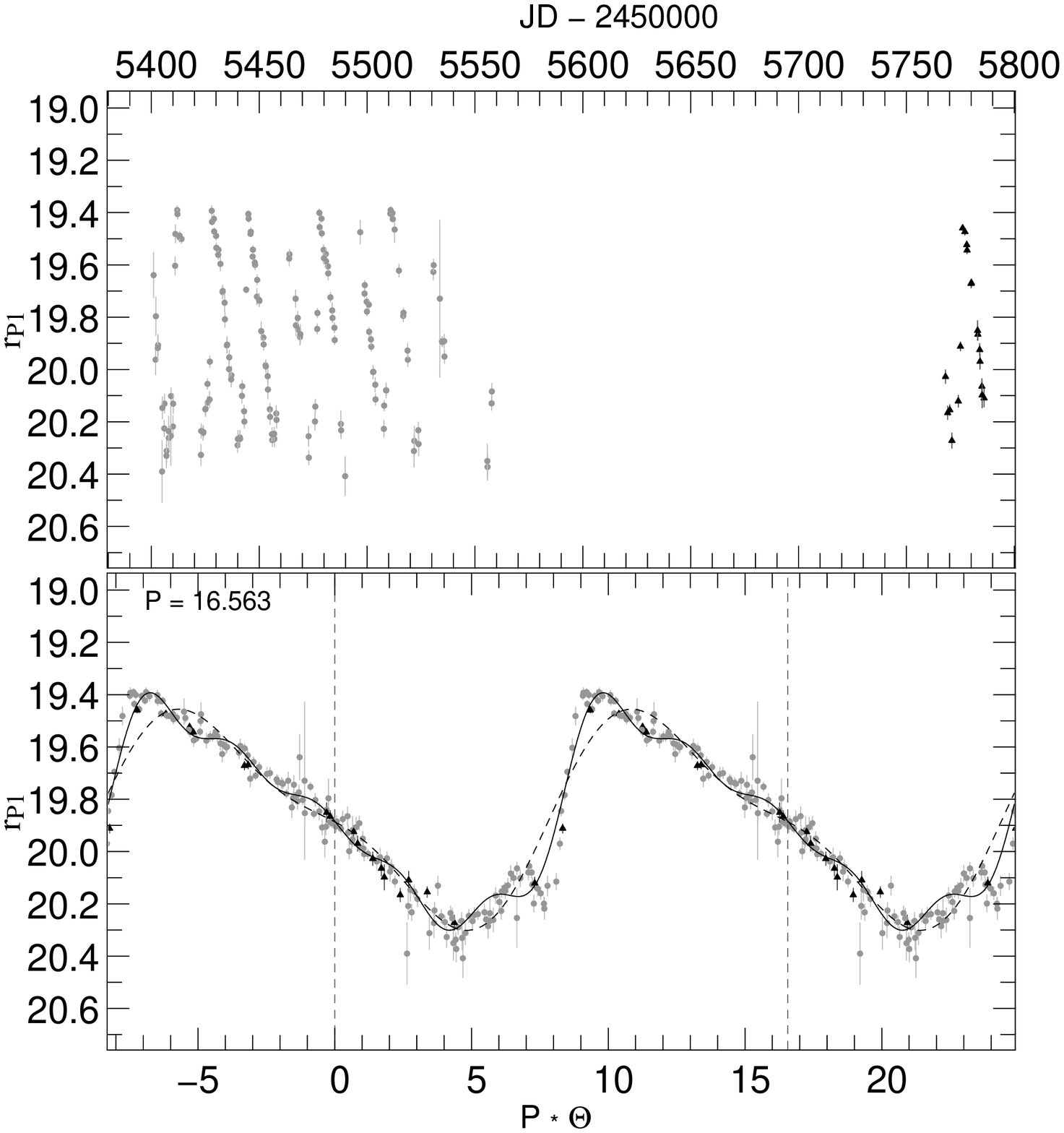}{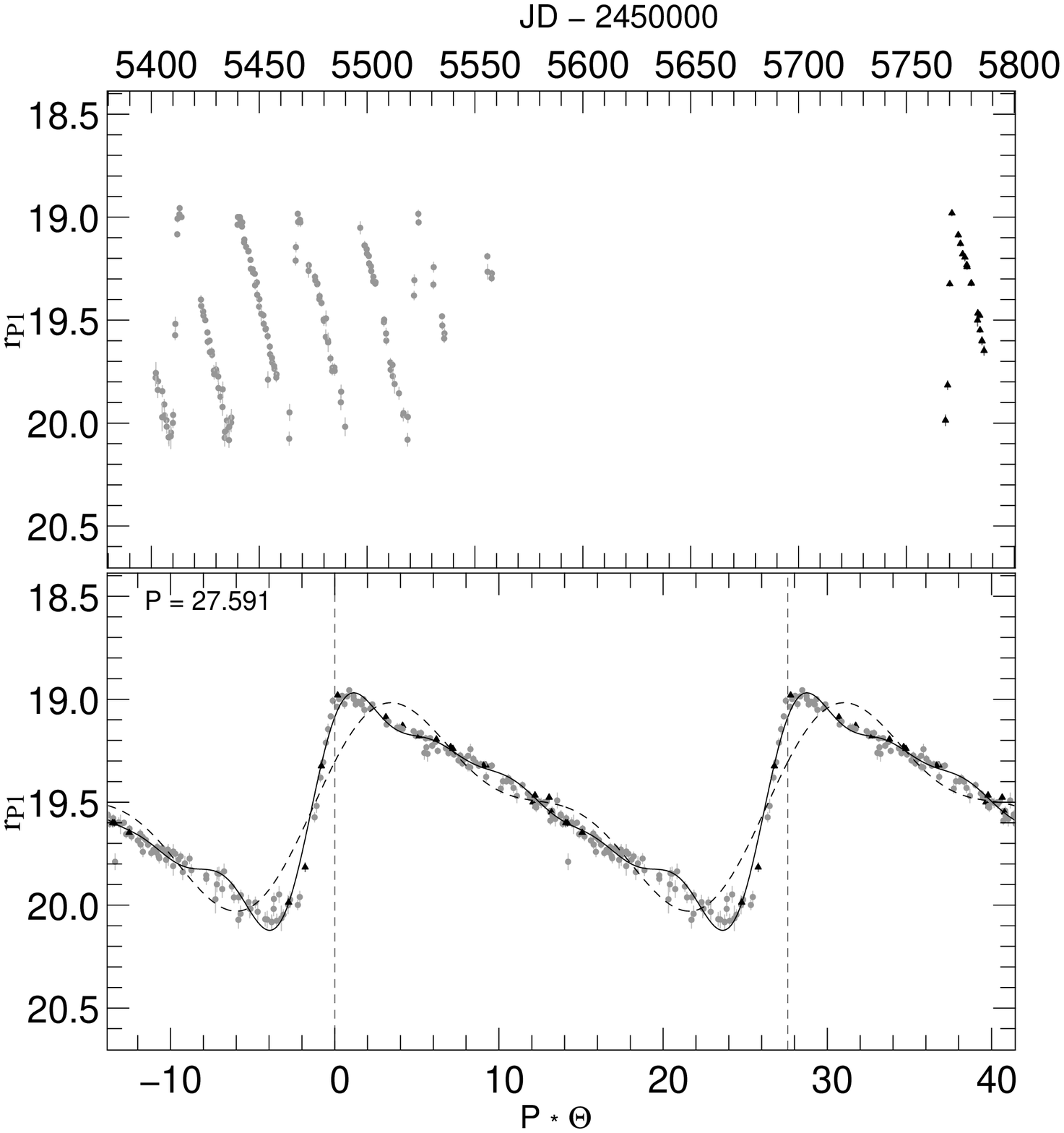} 
  \caption{Same as Fig. \ref{bsplcFM1} for the next two entries in
    the excerpt of the 3-dimensional parameter space classified
    Cepheid catalog. Left panel: Light curve of a FM Cepheid with
    $P=16.563~d$. Right panel: Light curve of a FM Cepheid with
    $P=27.591~d$.}
\label{bsplcFM3}
\end{figure}

\begin{figure}[h!]
  \epsscale{1.0}
  \plottwo{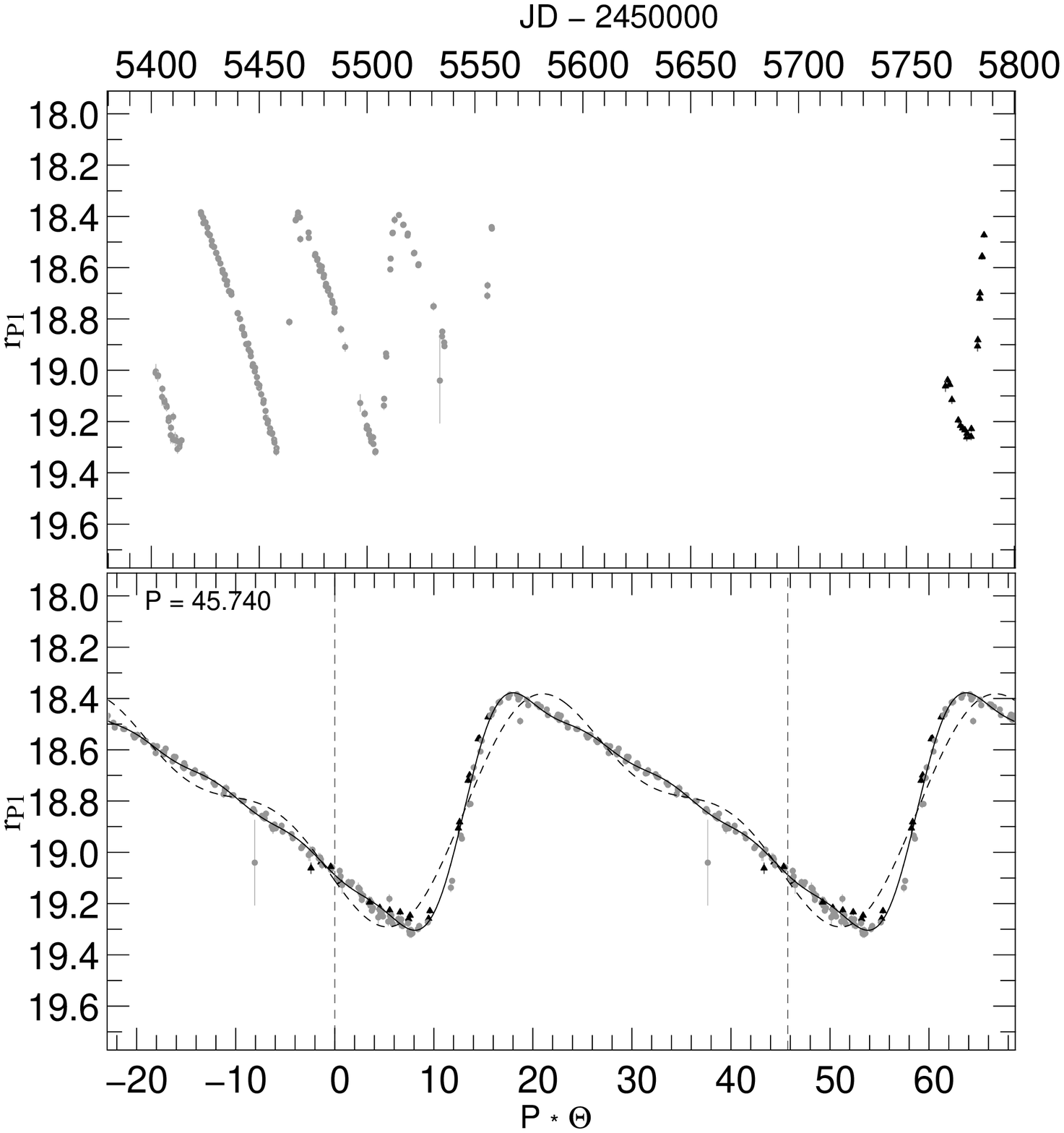}{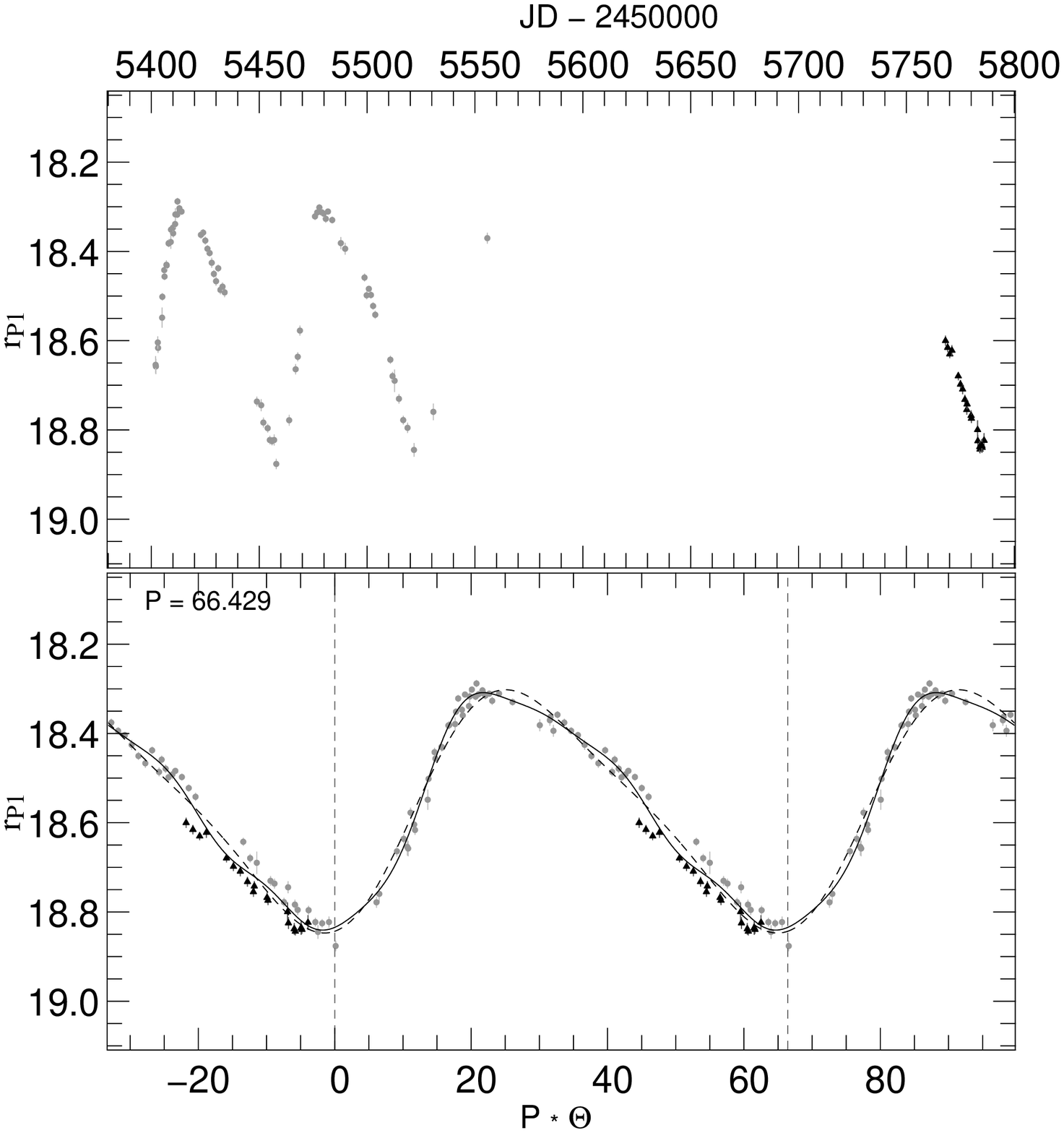} 
  \caption{Same as Fig. \ref{bsplcFM1} for the next two entries in
    the excerpt of the 3-dimensional parameter space classified
    Cepheid catalog. Left panel: Light curve of a FM Cepheid with
    $P=45.739~d$. Right panel: Light curve of a FM Cepheid with
    $P=66.429~d$.}
\label{bsplcFM4}
\end{figure}

\begin{figure}[h!]
  \epsscale{1.0}
  \plottwo{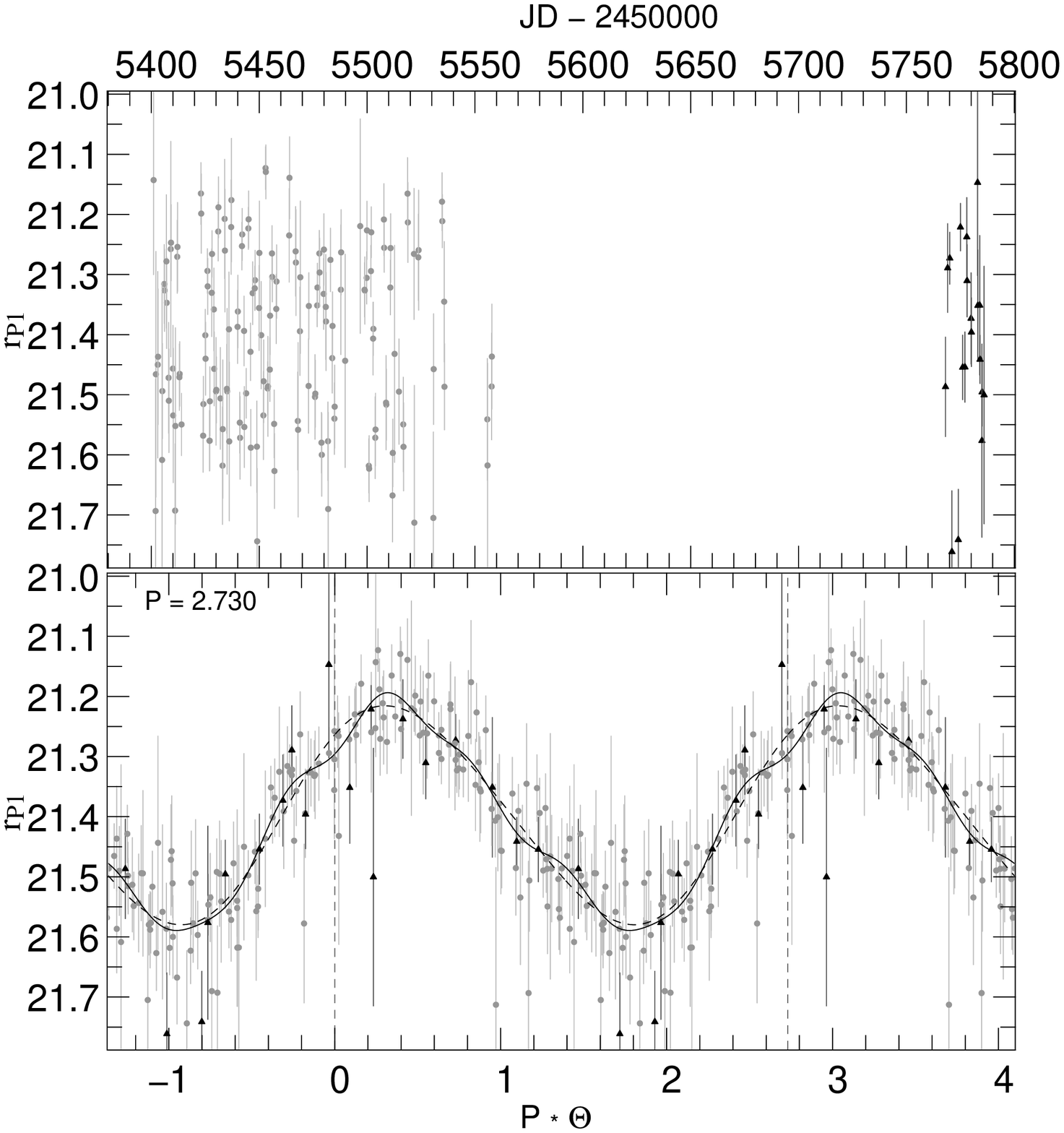}{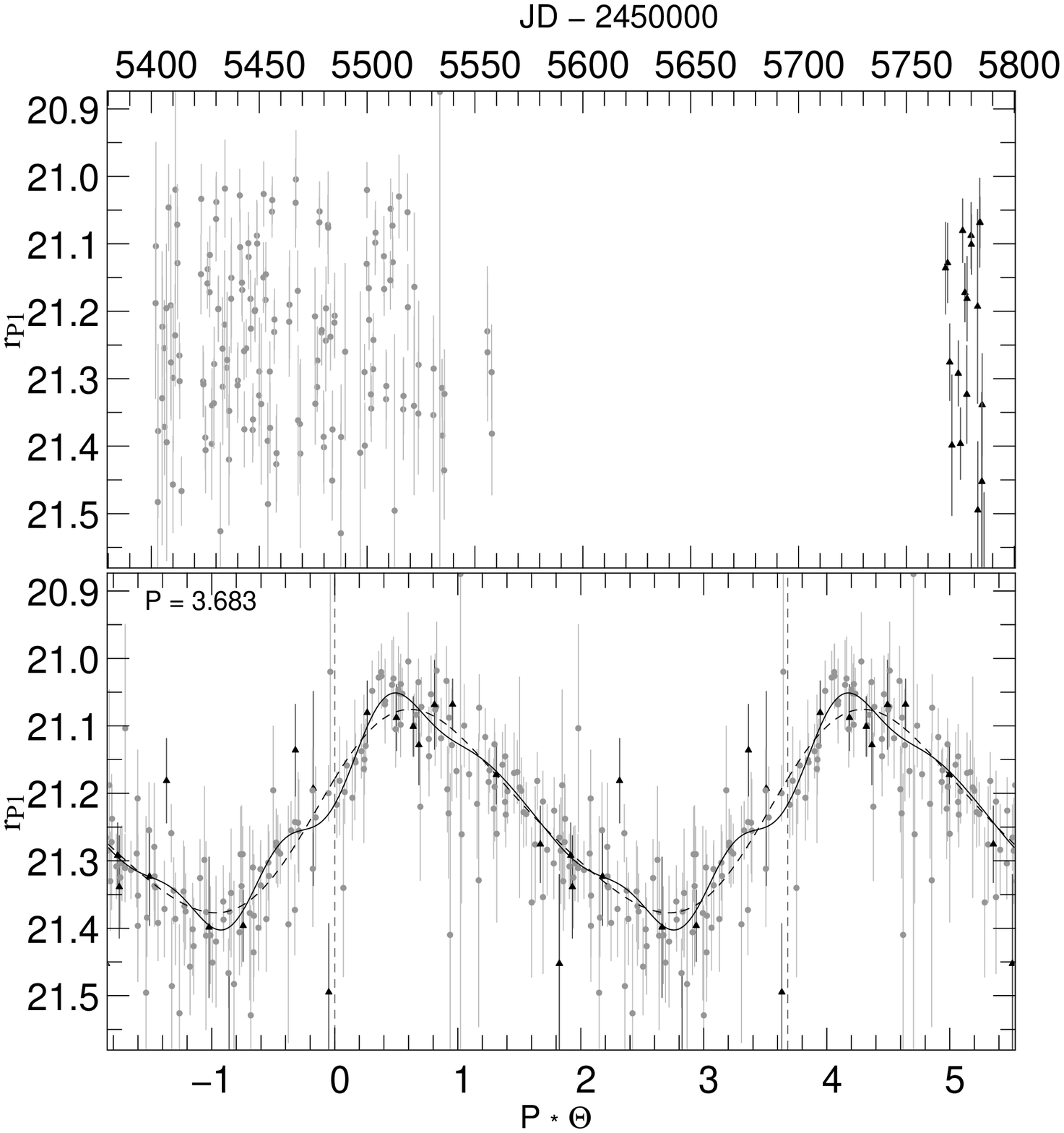} 
  \caption{Same as Fig. \ref{bsplcFM1} for the next two entries in
    the excerpt of the 3-dimensional parameter space classified
    Cepheid catalog. Left panel: Light curve of a FO Cepheid with
    $P=2.730~d$. Right panel: Light curve of a FO Cepheid with
    $P=3.683~d$.}
\label{bsplcFO}
\end{figure}

\begin{figure}[h!]
  \epsscale{1.0}
  \plottwo{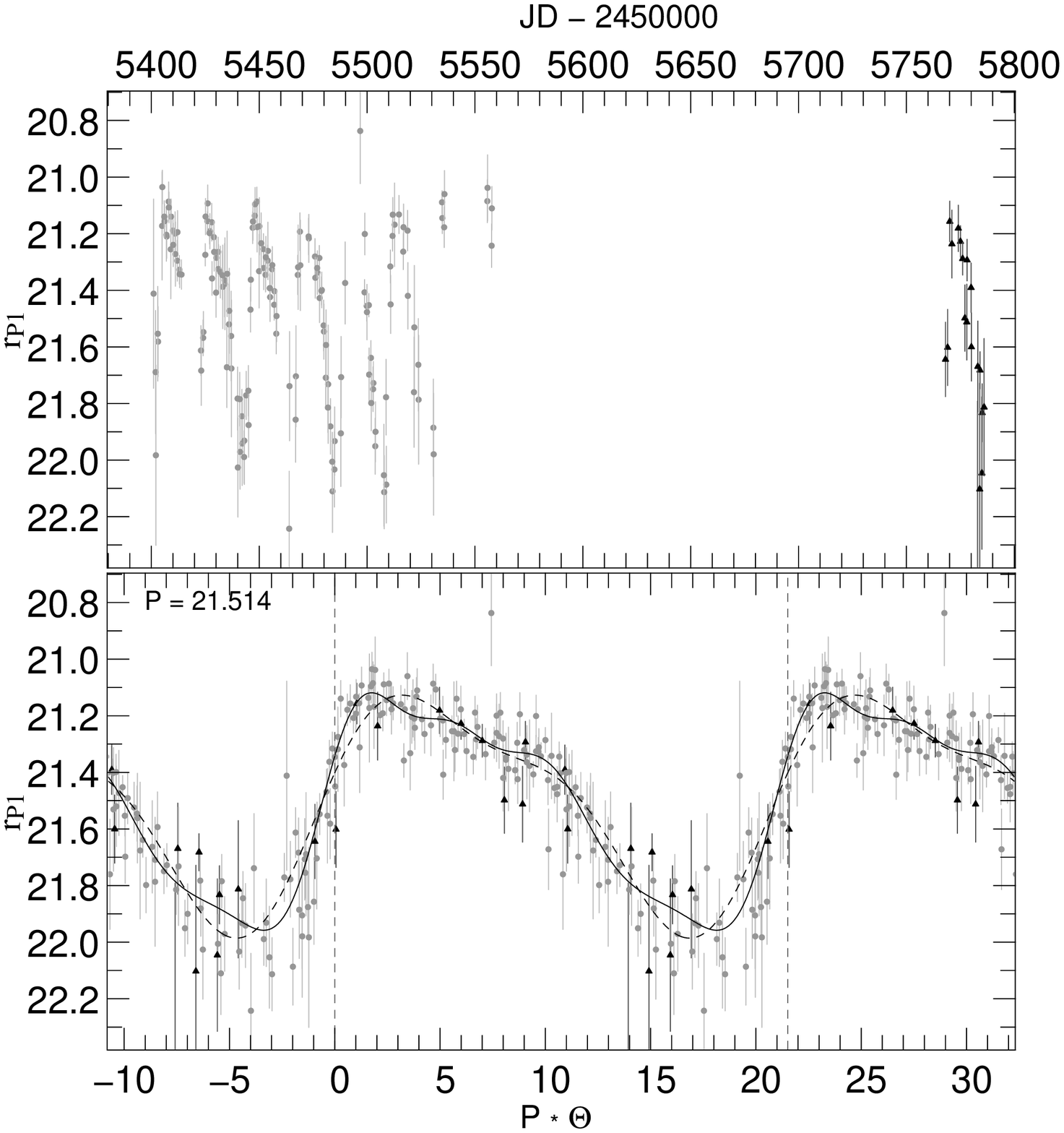}{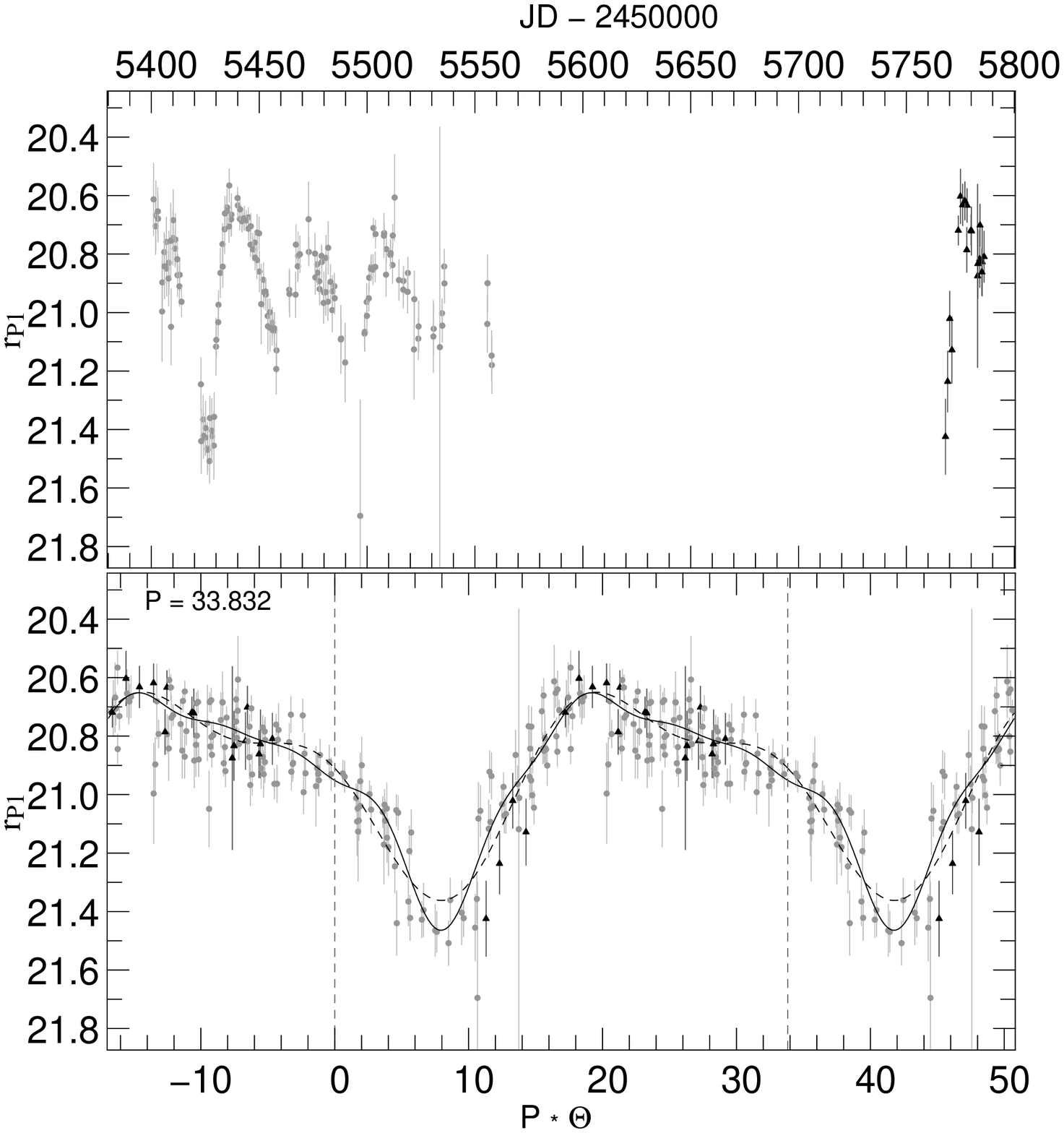} 
  \caption{Same as Fig. \ref{bsplcFM1} for the next two entries
    in the excerpt of the 3-dimensional parameter space classified
    Cepheid catalog. Left panel: Light curve of a Type II Cepheid with
    $P=21.513~d$. Right panel: Light curve of a Type II Cepheid with
    $P=33.831~d$.}
\label{bsplcT2}
\end{figure}

\begin{figure}[h!]
  \epsscale{1.0}
  \plottwo{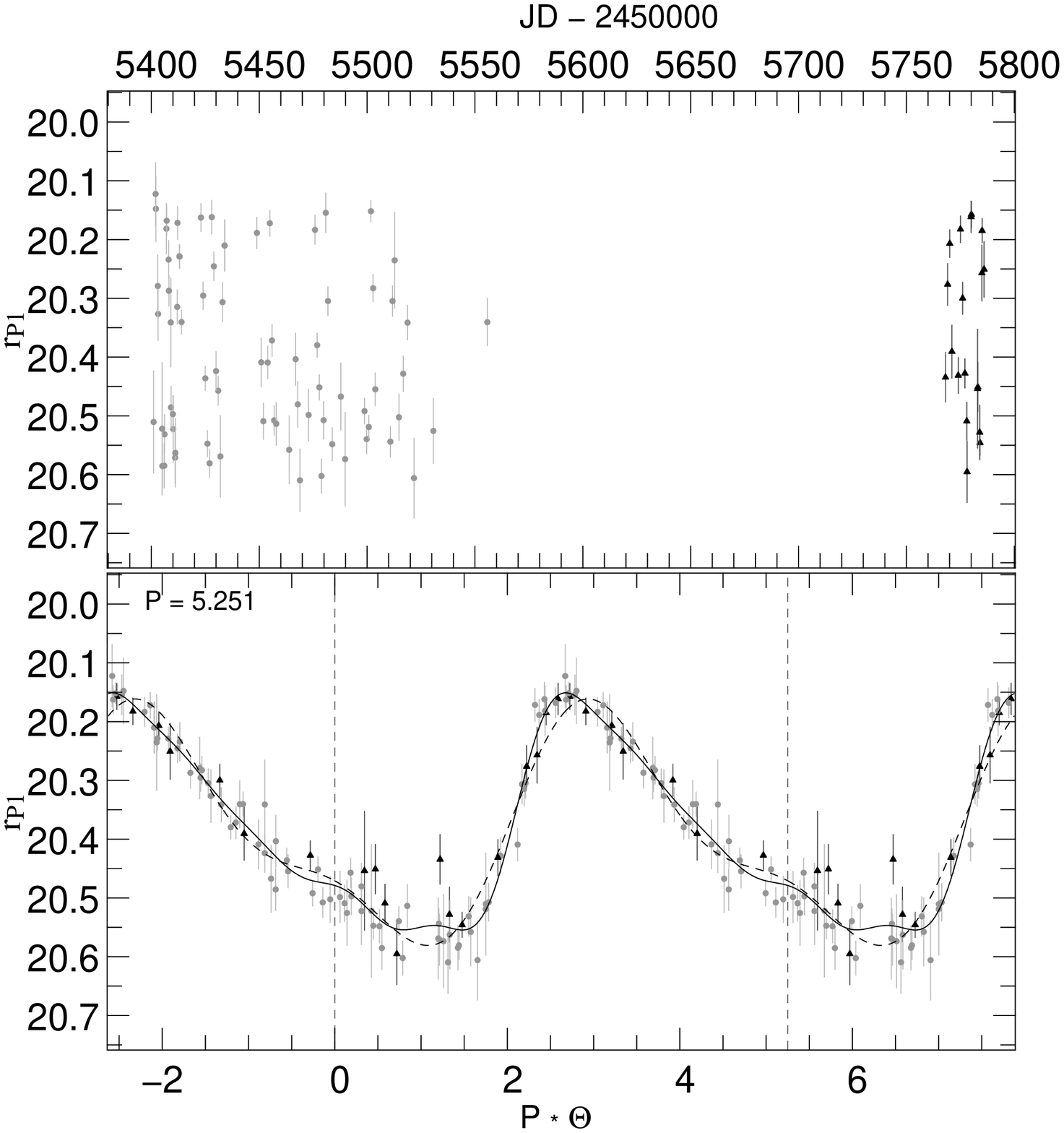}{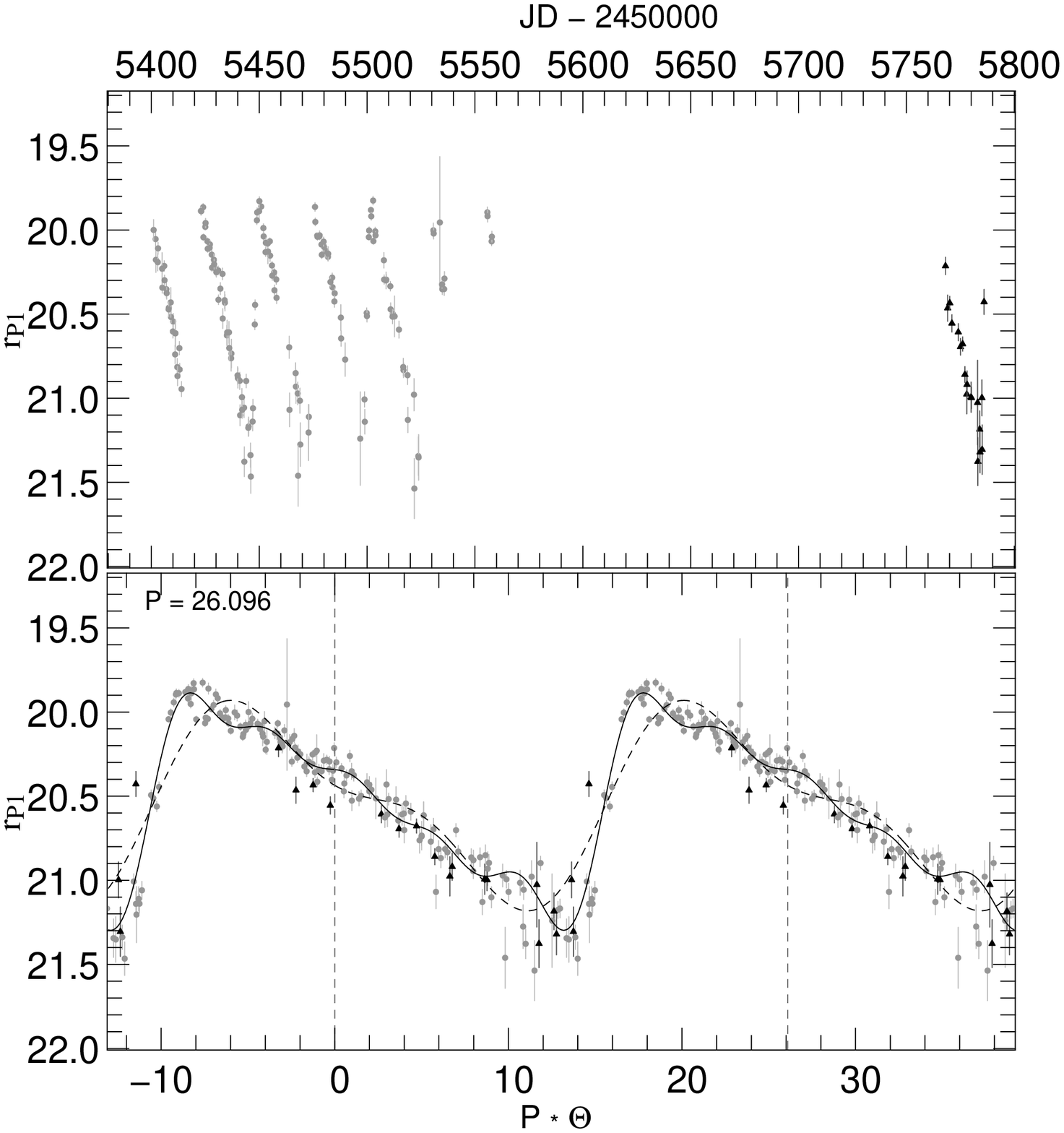} 
  \caption{Same as Fig. \ref{bsplcFM1} for the next two entries
    in the excerpt of the 3-dimensional parameter space classified
    Cepheid catalog. Left panel: Light curve of a UN Cepheid with
    $P=5.251~d$. Right panel: Light curve of a UN Cepheid with
    $P=26.095~d$.}
\label{bsplcUN}
\end{figure}

Table \ref{tabbsp1} contains an excerpt of the 3-dimensional parameter
space classified Cepheid catalog (the corresponding light curves are
shown in Fig. \ref{bsplcFM1}, Fig. \ref{bsplcFM2},
Fig. \ref{bsplcFM3}, Fig. \ref{bsplcFM4}, Fig. \ref{bsplcFO},
Fig. \ref{bsplcT2} and Fig. \ref{bsplcUN}). Table \ref{tabbsp1} is
available in its entirety in machine-readable format in the online
journal. A portion is shown here for guidance regarding its form and
content.

\begin{deluxetable}{c@{\extracolsep{\fill}}ccccccccccc}
 \tablecaption{Excerpt of the 3-dimensional parameter space
    classified Cepheid catalog. Only a portion of the complete
      table is shown here.  The table in its entirety is available in
      machine-readable format in the online journal. The columns
      contain (from left to right): identifier, Cepheid type, RA
      (J2000.0), DEC (J2000.0), significance (sig), period (determined
      from the \rps~band light curve), mean magnitude in the
      \rps~band, mean magnitude in the \ips~band, Fourier coefficient
      \fA, Fourier coefficient \fP, Wesenheit index and the
      decline/rise factor (c.f. Fig. \ref{Fig.risedecline})}
  \tablehead{\colhead{identifier} & \colhead{type} & \colhead{RA} &
    \colhead{DEC} & \colhead{sig} & \colhead{P} &\colhead{$r_{P1}$} &
    \colhead{$i_{P1}$} & \colhead{$\fA$} & \colhead{$\fP$} &
    \colhead{$W$} & \colhead{decl./rise} }
 
  \startdata
    103-27590 & FM &    11.1608 &    42.1585 &    16.40 &    4.398 &    21.34 $\pm$     0.08 &    21.20 $\pm$     0.08 &     0.37 $\pm$     0.03 &     4.37 $\pm$     0.11 &    20.80  $\pm$     0.38 &     2.80\\ 
    040-12157 & FM &     9.8918 &    40.4165 &    16.09 &    6.304 &    21.10 $\pm$     0.05 &    20.92 $\pm$     0.05 &     0.41 $\pm$     0.03 &     4.66 $\pm$     0.10 &    20.40  $\pm$     0.25 &     2.24\\ 
    102-03621 & FM &    11.0967 &    41.9573 &    19.07 &   10.224 &    20.04 $\pm$     0.02 &    19.80 $\pm$     0.05 &     0.19 $\pm$     0.01 &     4.87 $\pm$     0.07 &    19.09  $\pm$     0.18 &     2.14\\ 
    103-12114 & FM &    11.3482 &    42.0301 &    18.53 &   10.370 &    20.73 $\pm$     0.10 &    20.40 $\pm$     0.10 &     0.13 $\pm$     0.02 &     4.32 $\pm$     0.15 &    19.47  $\pm$     0.48 &     1.15\\ 
    064-30826 & FM &    10.1540 &    41.2511 &    32.34 &   16.563 &    19.84 $\pm$     0.03 &    19.63 $\pm$     0.03 &     0.32 $\pm$     0.01 &     4.78 $\pm$     0.02 &    19.02  $\pm$     0.14 &     1.94\\ 
    053-14189 & FM &    10.5470 &    40.8116 &    28.07 &   27.591 &    19.46 $\pm$     0.02 &    19.10 $\pm$     0.02 &     0.44 $\pm$     0.01 &     4.80 $\pm$     0.02 &    18.06  $\pm$     0.11 &     4.38\\ 
    065-11258 & FM &    10.2662 &    41.1512 &    33.25 &   45.740 &    18.78 $\pm$     0.02 &    18.48 $\pm$     0.04 &     0.43 $\pm$     0.00 &     4.89 $\pm$     0.01 &    17.61  $\pm$     0.15 &     3.62\\ 
    090-30644 & FM &    11.0918 &    41.9226 &    18.76 &   66.429 &    18.55 $\pm$     0.02 &    18.20 $\pm$     0.03 &     0.19 $\pm$     0.01 &     4.91 $\pm$     0.04 &    17.21  $\pm$     0.12 &     1.82\\ 
    051-15274 & FO &     9.7241 &    40.7097 &    33.30 &    2.730 &    21.39 $\pm$     0.08 &    21.33 $\pm$     0.08 &     0.08 $\pm$     0.04 &     4.85 $\pm$     0.44 &    21.17  $\pm$     0.39 &     1.14\\ 
    090-07109 & FO &    10.6804 &    41.6543 &    30.35 &    3.683 &    21.23 $\pm$     0.03 &    20.91 $\pm$     0.05 &     0.13 $\pm$     0.04 &     4.35 $\pm$     0.34 &    20.02  $\pm$     0.20 &     1.60\\ 
    041-31081 & T2 &    10.5598 &    40.5848 &    29.91 &   21.514 &    21.47 $\pm$     0.03 &    21.23 $\pm$     0.04 &     0.33 $\pm$     0.03 &     5.40 $\pm$     0.09 &    20.56  $\pm$     0.20 &     3.19\\ 
    078-12906 & T2 &    10.9476 &    41.4886 &    27.59 &   33.832 &    20.92 $\pm$     0.05 &    20.79 $\pm$     0.07 &     0.44 $\pm$     0.03 &     5.56 $\pm$     0.07 &    20.41  $\pm$     0.33 &     2.01\\ 
    103-01636 & UN &    11.4924 &    41.9421 &    16.31 &    5.251 &    20.39 $\pm$     0.03 &    20.08 $\pm$     0.05 &     0.37 $\pm$     0.03 &     4.14 $\pm$     0.09 &    19.22  $\pm$     0.19 &     3.38\\ 
    077-16136 & UN &    10.5815 &    41.4385 &    28.22 &   26.096 &    20.47 $\pm$     0.21 &    20.08 $\pm$     0.21 &     0.42 $\pm$     0.01 &     4.81 $\pm$     0.04 &    18.98  $\pm$     1.02 &     4.68\\ 
    \enddata
    \label{tabbsp1}
\end{deluxetable}
 
\clearpage

\begin{figure}[h!]
  \epsscale{0.5}
  \plotone{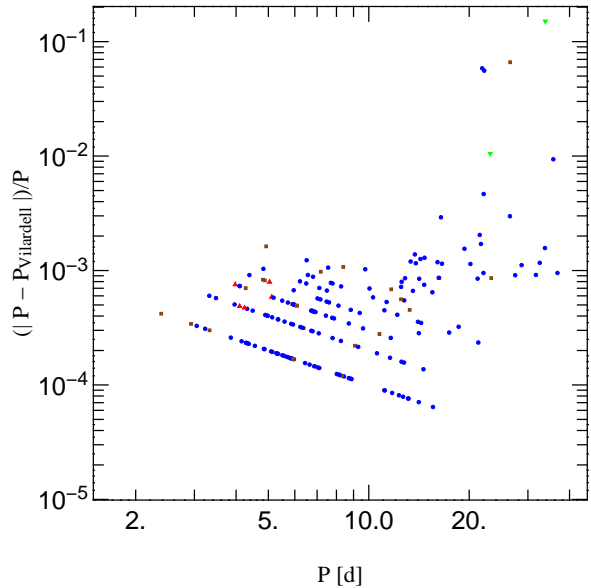}
  \caption{Relative difference of our periods with the 225 (187 FM
    (circles), 9 FO (upwards pointing triangles), 2 Type II (downward
    pointing triangles) and 27 UN Cepheids (squares)) matched Cepheids
    from \cite{2007A&A...473..847V}. The light curve of the Cepheid
    with the largest difference in the period is shown in the right
    panel of Fig. \ref{bsplcT2}. The cascaded form of the relative
    difference is due to the precision of the compared values.}
  \label{Fig.compVil}
\end{figure}

We match the position of the Cepheids in our sample with the 416
Cepheids from \cite{2007A&A...473..847V} and found a coincidence for
225 Cepheids (187 FM, 9 FO, 2 Type II and 27 UN Cepheids) with a
matching radius of 1 arcsec. The periods of both samples show a good
agreement (Fig. \ref{Fig.compVil}). The light curve of the Cepheid
with the largest difference in the period (5.101 days) is shown in the
right panel of Fig. \ref{bsplcT2}. The reference image shows no other
close source and as can be seen from the right panel of
Fig. \ref{bsplcT2} the period determination is reasonable, which
suggests a slightly wrong period in the Vilardell sample. The periods
for the other 5 light curves with a relative difference larger than
0.5\% we were able to confirm by visual inspection, but in some cases
we can not rule out that there is another close source within the 1
arcsec matching radius and that the difference is caused by that
source.

We also matched our Cepheids with the Cepheids of the DIRECT project
\citep{2004ASPC..310...33M} and the WeCAPP project
\citep{2006A&A...445..423F} with the matching radius of 1 arcsec. We
found matches for 216 Cepheids from 332 DIRECT Cepheids (187 FM, 3 FO,
0 Type II and 26 UN Cepheids) and 26 matched Cepheids from 126 WeCAPP
Cepheids (15 FM, 0 FO, 4 Type II and 7 UN Cepheids). Taking also the
\cite{2007A&A...473..847V} sample into account we matched 354 unique
Cepheids.

\begin{figure}[h!]
  \epsscale{1.0}
  \plottwo{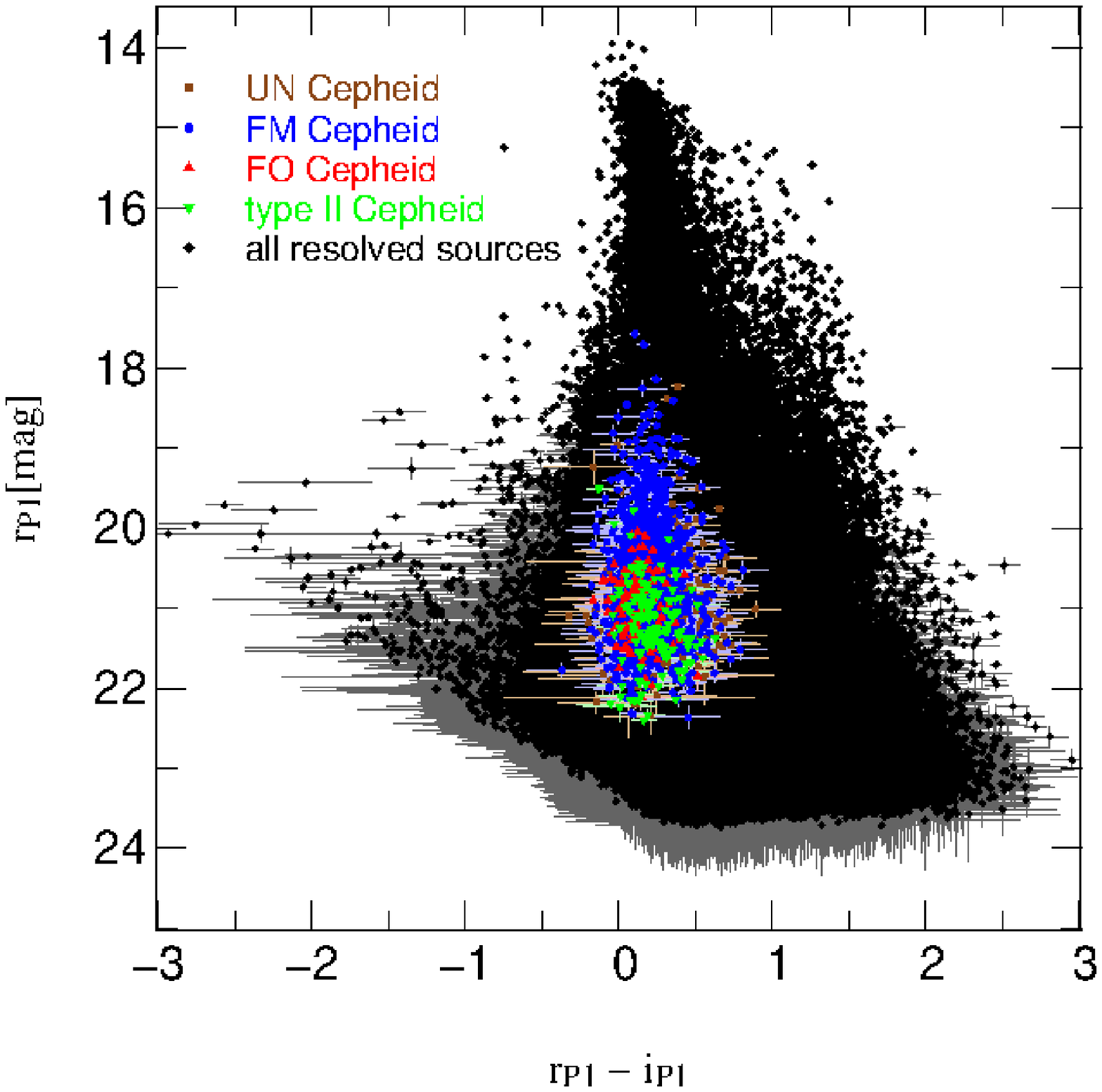}{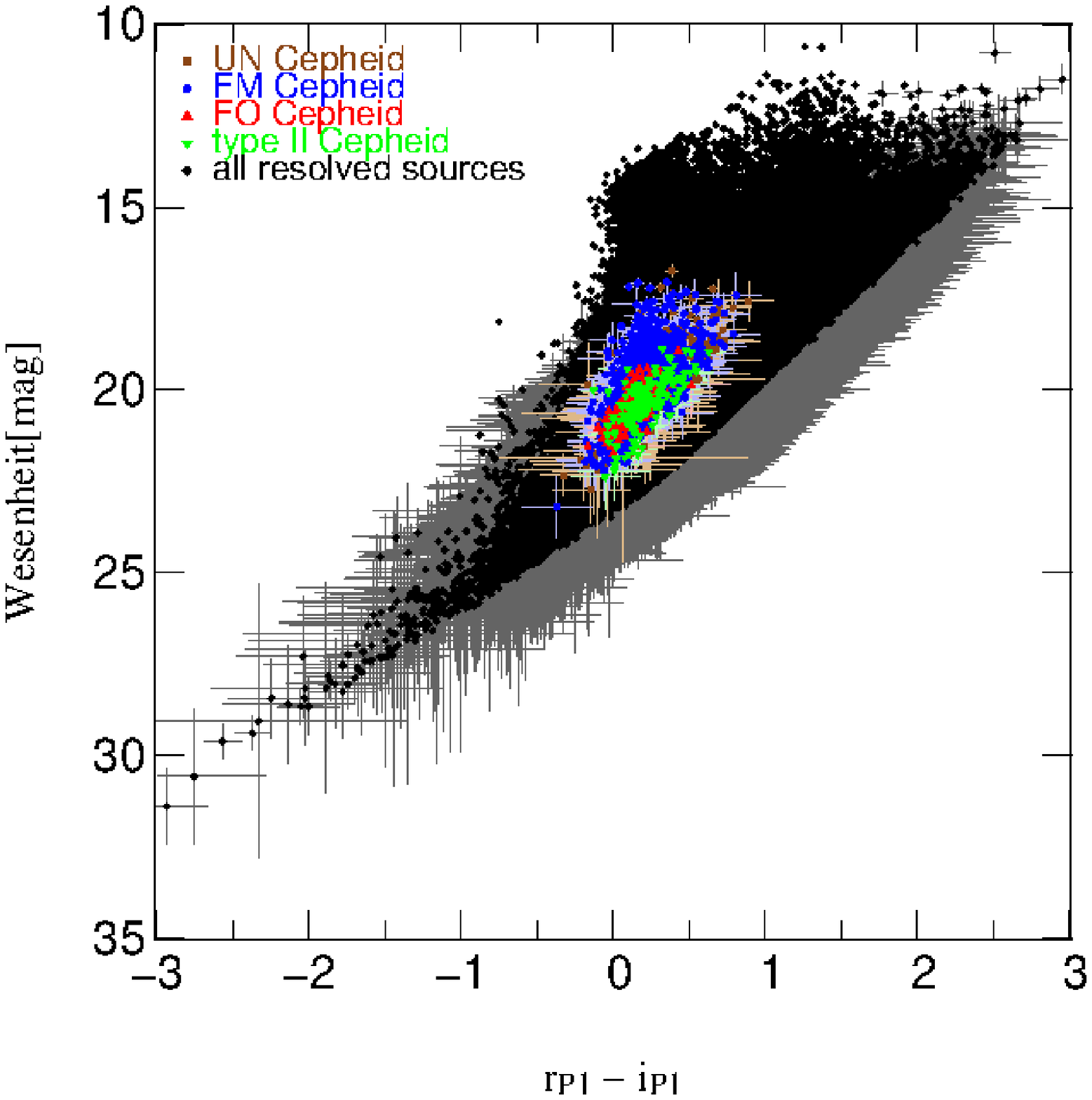}
  \caption{Color Wesenheit Diagram and Color Magnitude Diagram for all
    the 724894 point sources that were detected in the $\rps$
    band. Cepheid variable stars are color coded. Left panel: Color
    Magnitude Diagram. Right panel: Color Wesenheit Diagram.}
  \label{Fig.colall}
\end{figure}

\begin{figure}[h!]
  \epsscale{1.0}
  \plottwo{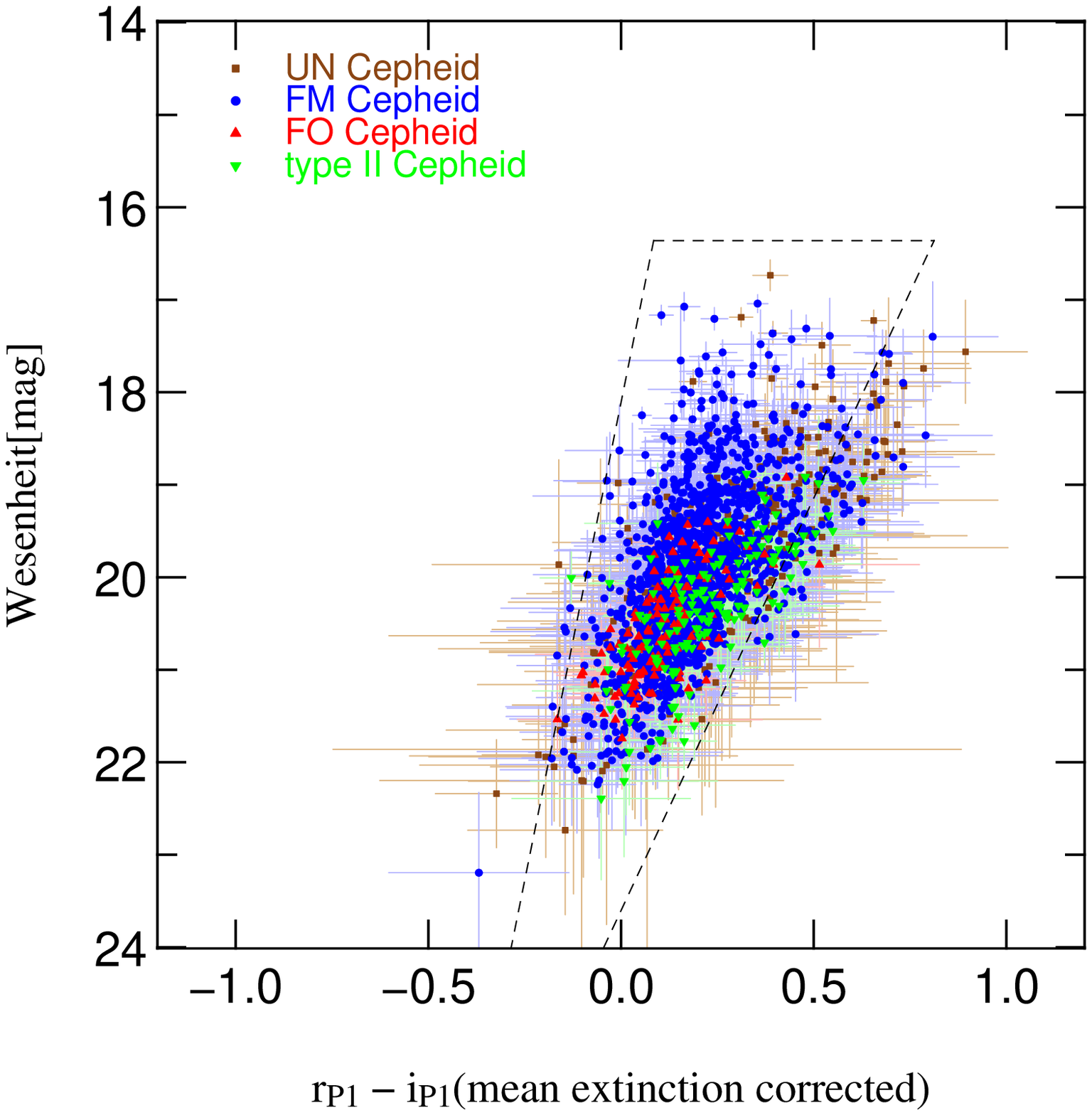}{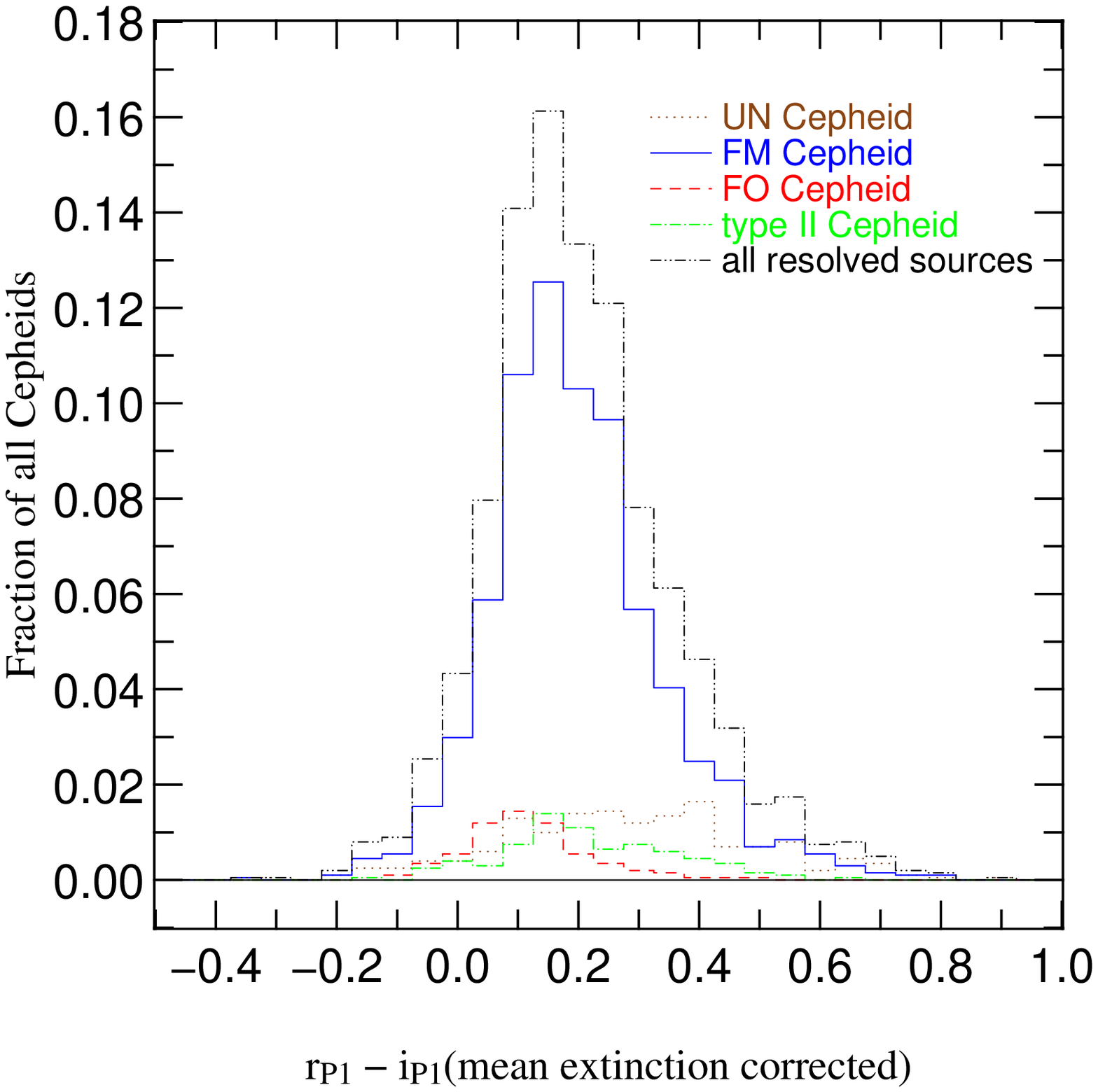}
  \caption{Left panel: Color Wesenheit Diagram with the domain (black dashed
    lines) we require the data points to be in, consistent with their
    margin of error (c.f. Section \ref{2.3}). Right panel: Histogram of the color distribution.}
  \label{Fig.colWhist}
\end{figure}

Fig. \ref{Fig.colall} shows the color magnitude diagram and the color
Wesenheit diagram for all resolved sources in the \rps~band and the
Cepheids of the 3-dimensional parameter space classified Cepheid
catalog. Left panel of Fig. \ref{Fig.colWhist} shows the color
Wesenheit diagram for the Cepheids alone. We require that the Cepheids
lie within the domain enclosed by the dashed line which results from
theoretical predictions of the instability strip (see Section
\ref{2.3}).

Using a simple color cut for the selection of Cepheids would either
yield too many interlopers (wide cut) or cut too many Cepheids
(narrow cut) as a simple color cut can not account for the tilt of the
instability strip in the CMDs (c.f. right panel of
\ref{Fig.colWhist}). The use of the instability strip allows for a
strict sample cut and yields a color distribution dominated by Cepheid
stars (c.f. left panel of Fig. \ref{Fig.colWhist}).

\begin{figure}[h!]
  \epsscale{0.5}
  \plotone{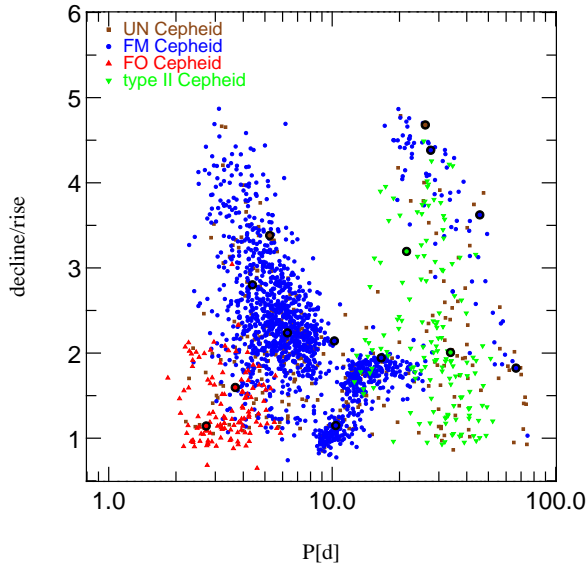}
  \caption{Decline/rise factor for the 3-dimensional parameter space
    classified Cepheid catalog. This factor describes the ratio of the
    time from the minimal to the maximal magnitude (not flux!)
    compared to the time from the maximal to the minimal magnitude and
    it is a good indicator of the overall shape of the light curve. We
    observe different scopes that correspond to the position of the
    bump. The 14 Cepheids framed by a black circle are shown in
    Fig. \ref{bsplcFM1}, Fig. \ref{bsplcFM2}, Fig. \ref{bsplcFM3},
    Fig. \ref{bsplcFM4}, Fig. \ref{bsplcFO}, Fig. \ref{bsplcT2} and
    Fig. \ref{bsplcUN}.}
  \label{Fig.risedecline}
\end{figure}

To describe the overall shape of the Cepheid light curve and to trace
the bump progression (also known as Hertzsprung progression,
\citealt{2000A&A...360..245B}) we introduce the decline/rise
factor. This factor describes the ratio of the time from the minimal
to the maximal magnitude (not flux!) compared to the time from the
maximal to the minimal magnitude. Fig. \ref{Fig.risedecline} shows
that there are different scopes that correspond to the different
positions of the bump. Around a period of 10 days two bumps occur on
each side of the minimal magnitude and the light curve gets quite
symmetrical. This can be verified in our sample light curves in
Fig. \ref{bsplcFM2}.

%\clearpage

\section{Results}
\label{sec.results}

So far, the most complete catalog of Cepheid variable stars in M31 had
been presented by \cite{2007A&A...473..847V}. They found 416 Cepheids
of which they classified 281 as FM and 75 as FO objects. We here
present a dramatically larger catalog with 2009 candidates, almost 5
times the number of objects in the earlier catalog. The factor
$\sim$~4.8 difference has to be attributed to the different patrol
fields: \cite{2007A&A...473..847V} observed 0.32 square degree within
north-eastern lobe of the disk of M31 (and some contribution from
bulge and halo) while our Pan-STARRS 1 data cover the full disk, the
complete bulge and a large fraction of the M31 halo (only a small part
of the halo has been analyzed for this work).

At the distance of M31 (750 to 770 kpc), an arcsecond corresponds to
$\sim$3.7~pc. Therefore, ground-based imaging of individual stars
within M31 is subject to blending and crowding, especially for young
stars like Type I Cepheids which are often located in young clusters
or associations.  Even at the resolution of the HST cameras, blending
is an issue. The impact of blending has been discussed in detail by
\cite{2007A&A...473..847V} for M31 and \cite{2012AJ....144..113C} for
M33.  \cite{2012AJ....144..113C} - comparing ground based images to
HST images - find severe effects ($\ge$ 10 \% level) for more than 30
\% of the stars in $V$ and $I$.

\subsection{FO-to-FM ratio}

Interestingly, the relative amount of FO stars as compared to the FM
pulsators differs between Vilardell et al. and our findings, namely,
0.27 vs. 0.09.  Both values have their largest uncertainties in the
number of unclassified objects as it is rather uncertain to which of
the two pulsational modes these objects belong. Incomplete detection
can further add uncertainties. As the Vilardell et al. data were
obtained with a slightly larger telescope, and with $B,V$ filter
bands, they might have a better completeness for the on average bluer
and fainter FO stars. Therefore, it is even more puzzling that the
FO-to-FM number ratio based on the David Dunlap Observatory Sample
\citep{1995IBVS.4148....1F} of the Milky Way ($\sim$~0.06) and of the
General Calatog of Variable Stars \citep{GCVS2012} within the Milky
Way ($\sim$~0.13) matches nicely our ratio.

We note that the FO-to-FM ratios as derived from the OGLE surveys of
the lower metallicity stars forming dwarf galaxies LMC, SMC, and IC
1613 yield ratios in the range 0.4 to 0.7 (see
e.g. \citealt{1999AcA....49..437U} and references therein, and the
OGLE project web pages for the catalogs). Comparing these values to
the one of the two spiral galaxies might be indicative for a
metallicity trend, despite the rather large scatter of the data in
this possible relation. FO Cepheids preferentially populate the blue
boundary of the instability strip. As isochrones easily demonstrate
(see e.g. Fig.~1 in \citealt{1977ApJ...218..633B}), lower
metallicity objects have blue loops in post main sequence evolutions
which extend far to the blue and cross the complete strip while higher
metalicity shrinks the extent more to the red and the strip is no more
fully crossed. Therefore, a trend between host metallicity and
FO-to-FM ratio is not surprising.

\subsection{Amplitude ratio}

\begin{figure}[h!]
    \epsscale{0.5}
    \plotone{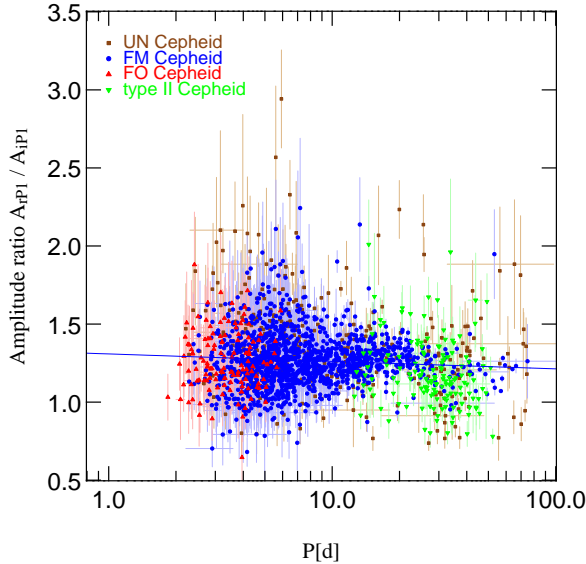}
    \caption{Amplitude ratio for the 3-dimensional parameter space
      classified Cepheid catalog. We observe a slight slope
      ($A_{\rps}/A_{\ips} = a +b \cdot \log(P)$) with $a = 1.309 \pm
      0.006$, $b = -0.0478 \pm 0.0014$ and a dispersion of 0.1680.}
    \label{Fig.ampratio}
  \end{figure}

It is well known that the amplitude of light variations of Cepheids
decreases from blue to red wavebands. In several recent investigations
of extragalactic samples, the ratio between the amplitudes in two
bands has been used to select Cepheid candidates. We do not use the
amplitude ratio for selecting good Cepheid candidates, but derive a
ratio between the $\rps$ and $\ips$ of $\sim$~1.3 with a slight
dependence on the period (see Fig.~\ref{Fig.ampratio} for our M31
sample). This is in excellent agreement with the high quality
photometry of galactic Cepheids (FM and FO) by
\cite{1968CoLPL...7...57W}. From their light curves we find an average
ratio in $R$ and $I$ of 1.28~$\pm$~0.09, and their data points are
very well represented by our fit to the M31 sample data (including the
slope).

\subsection{Period distribution}
 
\begin{figure}[h!]
    \epsscale{0.5}
    \plotone{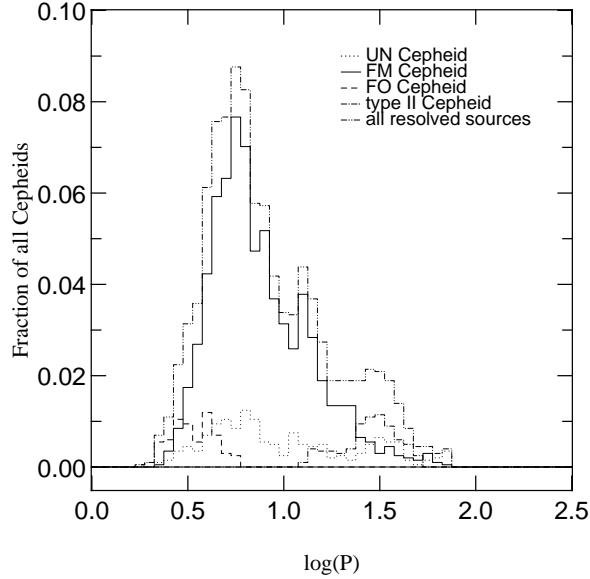}
    \caption{Normalized period distribution of the 3-dimensional
      parameter space classified Cepheid catalog. The period
      distribution of the FM Cepheids does show a secondary peak as
      described in \cite{2007A&A...473..847V}.}
    \label{Fig.per3d}
  \end{figure}

The FM period distribution of the presented catalog
(Fig. \ref{Fig.per3d}) shows a double peaked-distribution with peaks
at $\sim$~0.75 and 1.1 in log(P) ($\sim$ 5.5 days and $\sim$ 12.5
days), respectively. The peaks coincide with the ones found previously
by \cite{2007A&A...473..847V}. Also, the shape of the distribution
agrees quite well with the exception that our sample shows a steeper
cut-off to the shortest periods which probably is related to the
somewhat shallower PS1 data. Thus, our larger data set fully supports
the more general finding in the Milky Way and M31 that objects of
solar metallicity show a complex period distribution as compared to
e.g. the Magellanic Clouds and other low-metal star forming galaxies
(\citealt{1977ApJ...218..633B}, \citealt{2007A&A...473..847V},
\citealt{1999AJ....117..920A}).

We note that the period distributions of M~101 and NGC~4258, two other
large spirals with more than 800 and 300 detected Cepheids,
respectively, show similarly double peaked distribution with a local
minimum at $\sim$~10~d (\citealt{2011ApJ...733..124S},
\citealt{2004ASPC..310...33M}).  Thus, this shape of the distribution
seems to be a rather common feature of the Cepheid population of
large, solar metallicity spiral galaxies.  A full discussion of
this topic is beyond the scope of this paper.

\subsection{PL relation}
\label{chapterpl}
 
\begin{figure}[h!]
  \epsscale{1.0}
  \plottwo{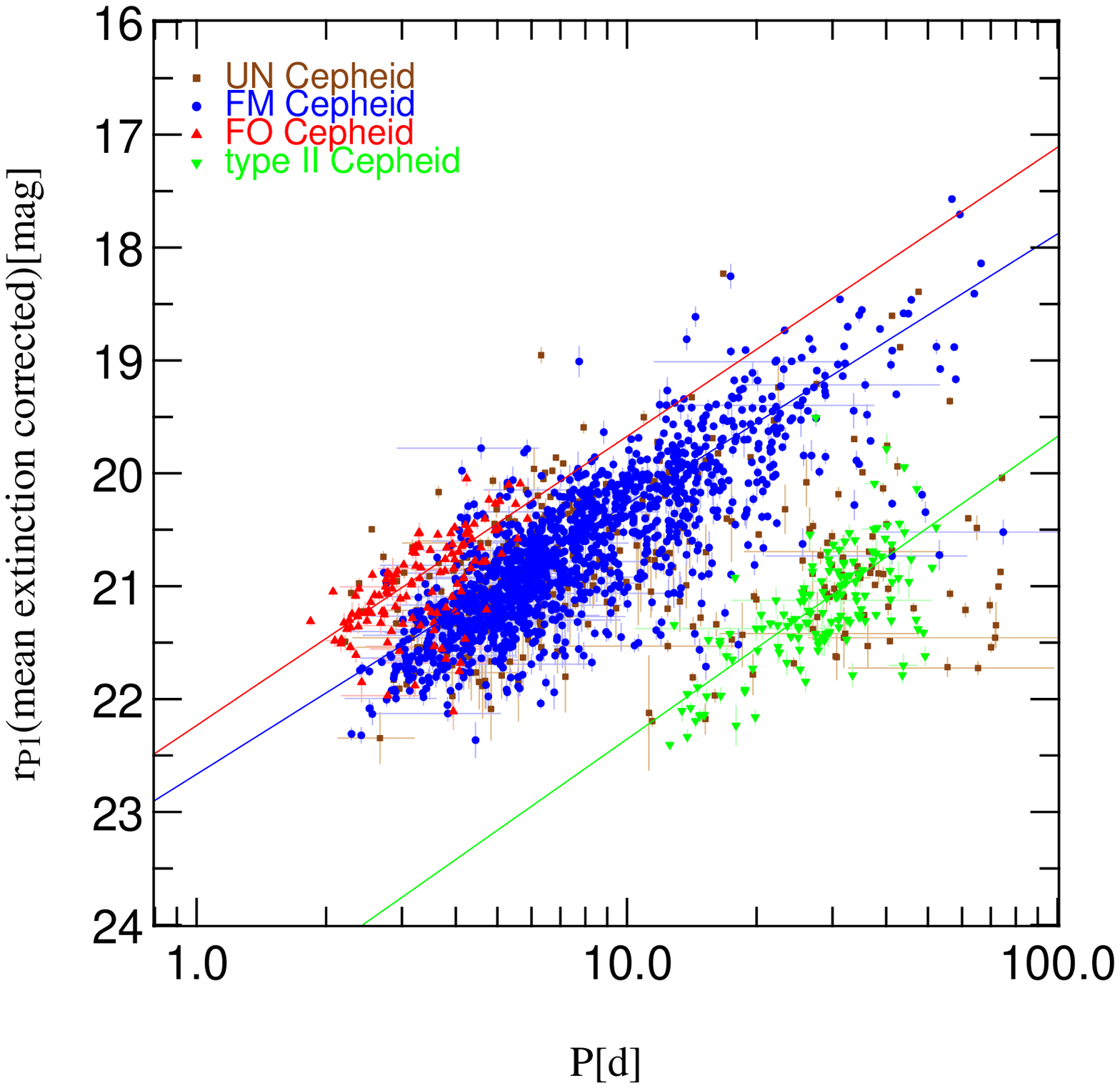}{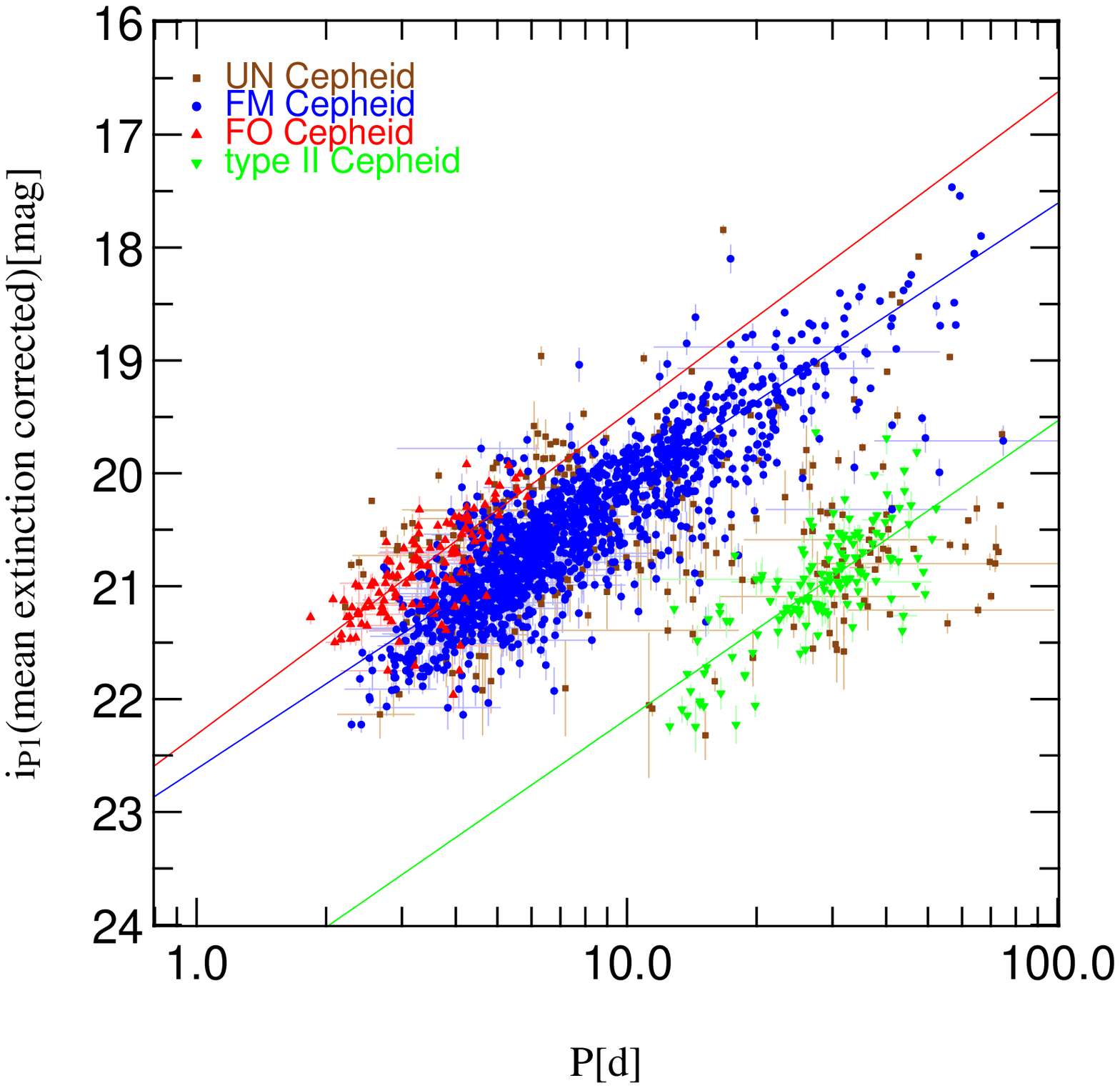}
  \caption{Period apparent magnitude relations for the 3-dimensional
    parameter space classified Cepheid catalog. The according fits
    shown as solid lines are given in table \ref{tabPL}. Left panel:
    Period apparent magnitude relation in the \rps~ band. Right panel:
    Period apparent magnitude relation in the \ips~band.}
  \label{Fig.PL3d}
\end{figure}

\begin{figure}[h!]
  \epsscale{1.0}
  \plottwo{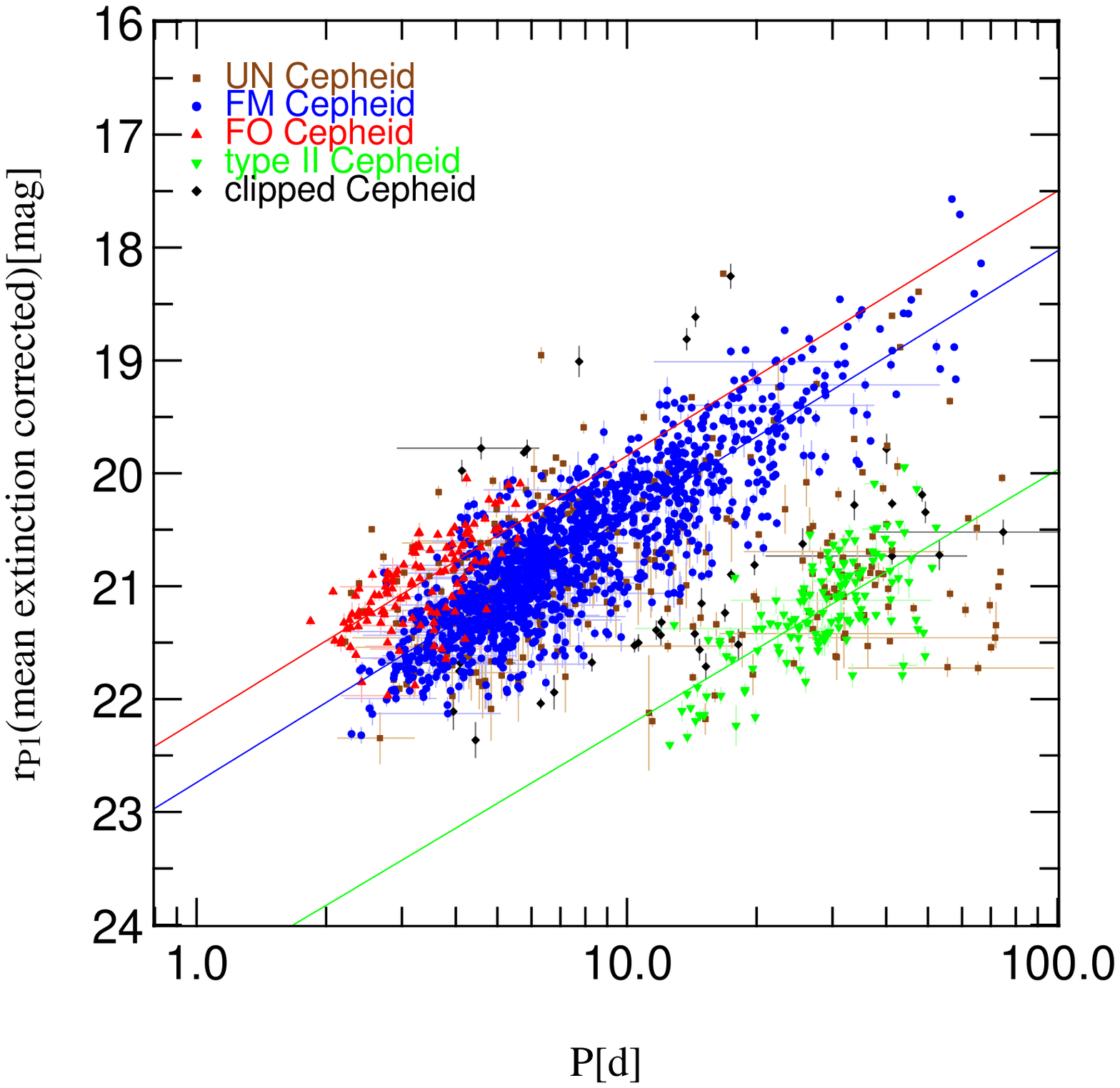}{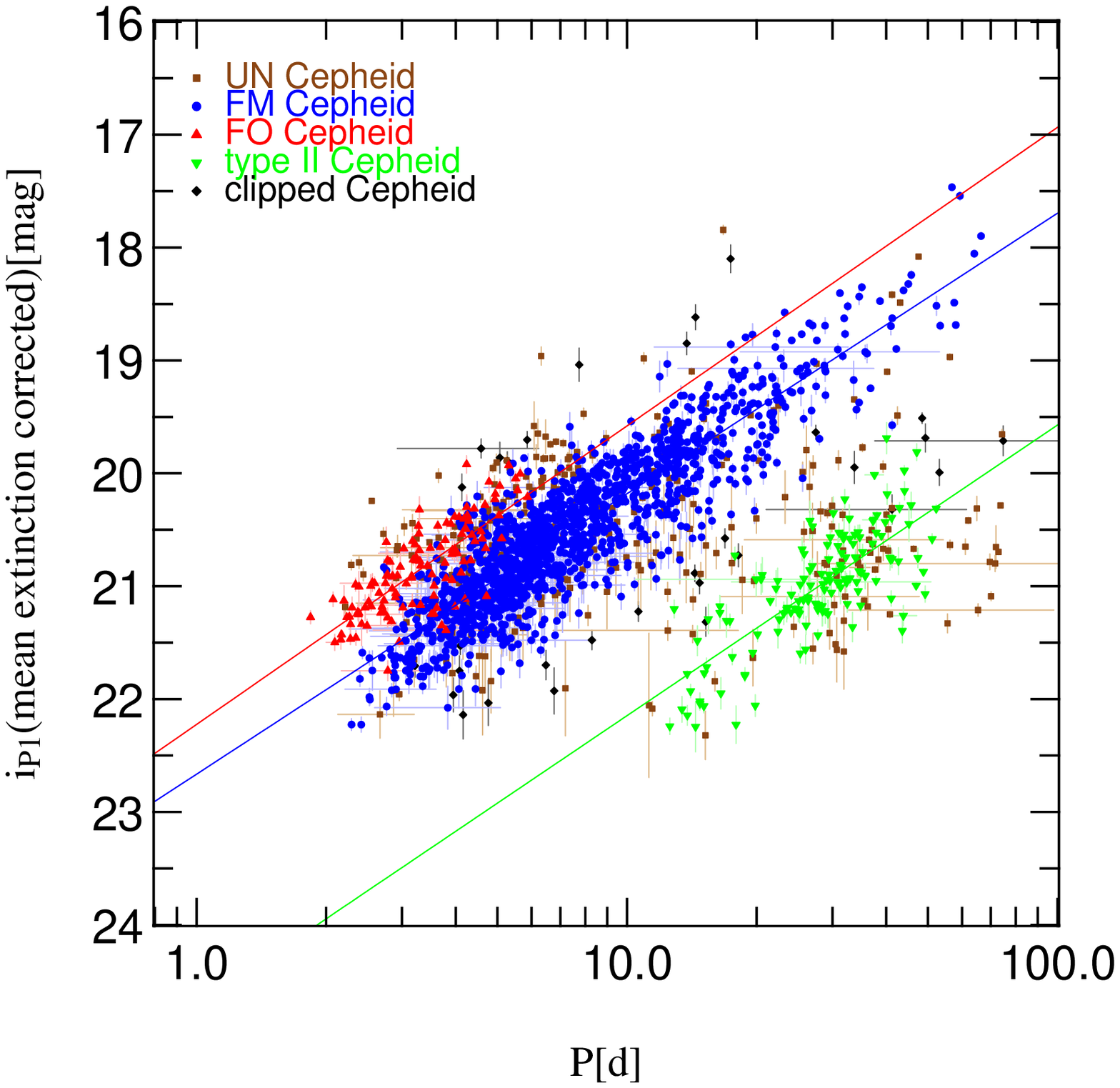}
  \caption{Period-apparent magnitude relations with iterative $3
    \sigma$ clipping. For the error bars in the $3 \sigma$ clipping we
    use the dispersion of the relation which we calculate for each
    iteration step. The according fits shown as solid lines are given
    in table \ref{tabPL}. Left panel: Period apparent magnitude
    relation in the $\rps$ band. Right panel: Period apparent
    magnitude relation in the $\ips$ band.}
  \label{Fig.PL3dClip}
\end{figure}

The Cepheids of the 3-dimensional parameter space classified Cepheid
catalog are used to determine the Period-apparent magnitude relations
(PL) for the different Cepheid types (Fig. \ref{Fig.PL3d}). To remove
systematic outliers that e.g. result from blending or the wrong
extinction correction\footnote{Note that as described in Section
  \ref{2.3} we only correct for the foreground extinction if there is
  no color excess given in the \cite{2009A&A...507..283M} map.}  we
additionally perform iterative $3 \sigma$ clipping
(Fig. \ref{Fig.PL3dClip}). But in contrast to the $3 \sigma$ clipping
that is used in Section \ref{2.6} for the Period-Wesenheit relation,
we use the dispersion of the data points relative to the best fit
relation (which we calculate for each iteration step) as an error for
each Cepheid magnitude. Otherwise the small errors would result in the
elimination of most of the Cepheids, instead of just clipping the
systematic outliers. In comparison the use of the dispersion is not
necessary for the Period-Wesenheit clipping since the errors in the
Wesenheit index are large. The slopes, zero points, the numerical fit
errors of these values and the dispersion of the different relations
are given in table \ref{tabPL} for a linear fit of the form:
\begin{equation}
  m = a \cdot \log(P) + b.
\end{equation}
For comparison we also added the period-Wesenheit relation from
Fig. \ref{Fig.PWesenheit3d} and the relations for the manually
classified Cepheid catalog.  The value of the slope $a$ and the values
for the intersection $b$ in the usually applied linear approximation
of the PLR have been a long standing matter of debate (see Section
\ref{chapterintro} and literature cited therein). Recently,
\cite{2011ApJ...734...46F} derived slope values from basic physical
principles taking into account the surface temperature and effective
surface area. These slope values depend on wavelength (expressed as
observing band dependence). The predictions of
\cite{2011ApJ...734...46F} agree rather well with empirical values of
\cite{2007A&A...476...73F} (Milky Way Cepheids) and of
\cite{2007ApJ...667...35N} (LMC) including the wavelength dependence.
\cite{2010ApJ...715..277B} obtained slope values from pulsation
models, but show an underlying relation with a weak additional
curvature term (see their Fig. 1). This quadratic behavior is well
approximated by a two-slope linear approach with different slope for
periods shorter and longer than 10 days
(\citealt{2005MNRAS.363..831N}, \citealt{2008A&A...477..621N}) and are
in good agreement with the broken slope proposed by
\cite{2009A&A...493..471S}. \cite{2010ApJ...715..277B} provide their
slope predictions as a function of metallicity and waveband for short,
long and overall period range.  \cite{2012arXiv1209.4090D} obtain
slope values similar to \cite{2010ApJ...715..277B} from pulsation
models for different period ranges in the SDSS filters. The obtained
slopes for $Z=0.02$ differ from those obtained by
\cite{2010ApJ...715..277B}.

  The slope values we derive are surprisingly shallow. In table
  \ref{tabPL} we also provide slopes for a subsample restricted to
  Cepheids with periods longer than 10 days. The slopes of the full
  sample fits in table \ref{tabPL} and those for the subsample with $P
  > 10 \, \mathrm{d}$ yield basically identical values. 

\begin{figure}[h!]
  \epsscale{0.5}
  \plotone{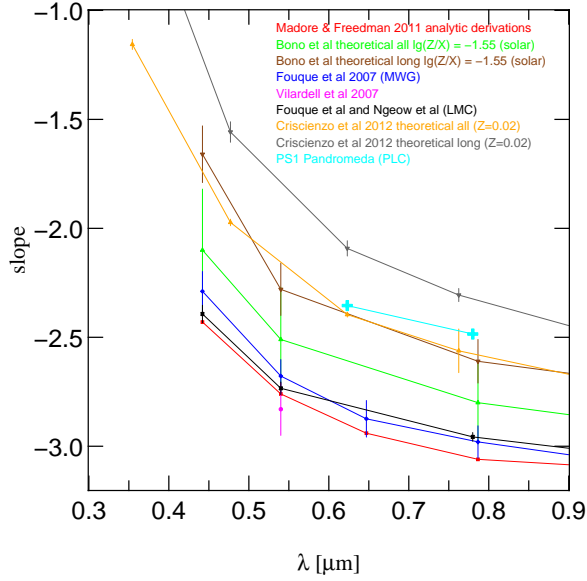}
  \caption{PLR slope values as function of wavelength. Our full
      sample and subsample with $P > 10 \, \mathrm{d}$ yield basically
      identical slopes (color: cyan). They agree with the theoretical
      predictions of \cite{2010ApJ...715..277B} for long periods
      (color: brown).}
  \label{Hopp}
\end{figure}

  As can be seen in Fig. \ref{Hopp} the shallow slope of the subsample
  with $P > 10 \, \mathrm{d}$ agrees with the prediction of
  \cite{2010ApJ...715..277B} for long periods. This is a warning for
  applying local PLR's to distant galaxies Cepheid samples which have
  even shallower (absolute) magnitude limits and are even more
  dominated by long period Cepheids: The slope values used by
  \cite{2012ApJ...745..156R} for M31 and by \cite{2011ApJ...730..119R}
  for 9 more distant hosts perfectly agree with our findings and with
  the predictions of \cite{2010ApJ...715..277B} for long
  periods. Thus, the effective period range available for a galaxy
  impacts the derived apparent distance values if the single slope
  solution for the PL is applied.

\begin{table}[h]
  \footnotesize
  \centering
  \caption{Summary of the results of the different period-apparent magnitude relations (PL),
    period-Wesenheit relations (PLW) and the period-apparent magnitude relations with iterative 
    $3 \sigma$ clipping. For the error bars in the $3 \sigma$ clipping (PLC) we use the dispersion
    of the relation which we calculate for each iteration step. Additionally a fit to a subsample with a period restriction (PLP) for FM Cepheids is provided.}
  \begin{tabular}{*{6}{|c}|}
    \tableline
    \tableline
    Relation & Catalog & Type & Slope [mag/log(d)]& Zero Point [mag] & Dispersion [mag] \\
    \tableline
    \tableline
    PL ($\rps$ band) & 3d & FM &   -2.39370 $\pm$    0.00110 &   22.66600 $\pm$    0.00380 &    0.42500\\ 
    \tableline
    PLC ($\rps$ band) & 3d & FM &   -2.35440 $\pm$    0.00930 &   22.73900 $\pm$    0.03390 &    0.35010\\ 
    \tableline
    PLP ($\rps$ band, $\mathrm{P} > 10 \, \mathrm{d}$) & 3d & FM &   -2.39060 $\pm$    0.00160 &   22.66420 $\pm$    0.00980 &    0.55550\\ 
    \tableline
    PL ($\rps$ band) & manual & FM &   -2.39800 $\pm$    0.00130 &   22.63410 $\pm$    0.00490 &    0.40700\\ 
    \tableline
    \tableline
    PL ($\rps$ band) & 3d & FO &   -2.56090 $\pm$    0.00370 &   22.23350 $\pm$    0.01800 &    0.35040\\ 
    \tableline
    PLC ($\rps$ band) & 3d & FO &   -2.34510 $\pm$    0.02650 &   22.18960 $\pm$    0.12520 &    0.29360\\ 
    \tableline
    PL ($\rps$ band) & manual & FO &   -2.46520 $\pm$    0.00680 &   22.14830 $\pm$    0.02920 &    0.23840\\ 
    \tableline
    \tableline
    PL ($\rps$ band) & 3d & T2 &   -2.68030 $\pm$    0.00380 &   25.03480 $\pm$    0.03850 &    0.38690\\ 
    \tableline
    PLC ($\rps$ band) & 3d & T2 &   -2.27150 $\pm$    0.02850 &   24.51130 $\pm$    0.28290 &    0.34340\\ 
    \tableline
    PL ($\rps$ band) & manual & T2 &   -2.40420 $\pm$    0.00720 &   24.61830 $\pm$    0.06830 &    0.30380\\ 
    \tableline
    \tableline
    PL ($\ips$ band) & 3d & FM &   -2.50490 $\pm$    0.00130 &   22.61890 $\pm$    0.00470 &    0.35710\\ 
    \tableline
    PLC ($\ips$ band) & 3d & FM &   -2.48490 $\pm$    0.00830 &   22.66520 $\pm$    0.03000 &    0.31090\\ 
    \tableline
    PLP ($\ips$ band, $\mathrm{P} > 10 \, \mathrm{d}$) & 3d & FM &   -2.43540 $\pm$    0.00190 &   22.52990 $\pm$    0.01240 &    0.43030\\ 
    \tableline
    PL ($\ips$ band) & manual & FM &   -2.46910 $\pm$    0.00160 &   22.55290 $\pm$    0.00610 &    0.33780\\ 
    \tableline
    \tableline
    PL ($\ips$ band) & 3d & FO &   -2.84470 $\pm$    0.00490 &   22.31370 $\pm$    0.02380 &    0.31750\\ 
    \tableline
    PLC ($\ips$ band) & 3d & FO &   -2.64490 $\pm$    0.02290 &   22.22470 $\pm$    0.10760 &    0.25230\\ 
    \tableline
    PL ($\ips$ band) & manual & FO &   -2.94140 $\pm$    0.00840 &   22.36100 $\pm$    0.04000 &    0.24290\\ 
    \tableline
    \tableline
    PL ($\ips$ band) & 3d & T2 &   -2.63900 $\pm$    0.00460 &   24.81500 $\pm$    0.04680 &    0.34530\\ 
    \tableline
    PLC ($\ips$ band) & 3d & T2 &   -2.57460 $\pm$    0.02710 &   24.72140 $\pm$    0.26840 &    0.32680\\ 
    \tableline
    PL ($\ips$ band) & manual & T2 &   -2.68790 $\pm$    0.00830 &   24.83940 $\pm$    0.07920 &    0.30570\\ 
    \tableline
    \tableline
    PLW & 3d & FM &   -3.00260 $\pm$    0.00620 &   22.57140 $\pm$    0.02180 &    0.44290\\ 
    \tableline
    PLW & manual & FM &   -3.00380 $\pm$    0.00760 &   22.57660 $\pm$    0.02830 &    0.40260\\ 
    \tableline
    \tableline
    PLW & 3d & FO &   -3.57730 $\pm$    0.02210 &   22.43020 $\pm$    0.10730 &    0.41600\\ 
    \tableline
    PLW & manual & FO &   -3.59930 $\pm$    0.03830 &   22.50680 $\pm$    0.17810 &    0.39570\\ 
    \tableline
    \tableline
    PLW & 3d & T2 &   -3.12580 $\pm$    0.02140 &   24.98560 $\pm$    0.21670 &    0.51350\\ 
    \tableline
    PLW & manual & T2 &   -2.89640 $\pm$    0.03900 &   24.60340 $\pm$    0.36910 &    0.49050\\ 
    \tableline
    \tableline 
  \end{tabular}
  \label{tabPL}
\end{table}

\begin{figure}[h!]
  \epsscale{0.5} \plotone{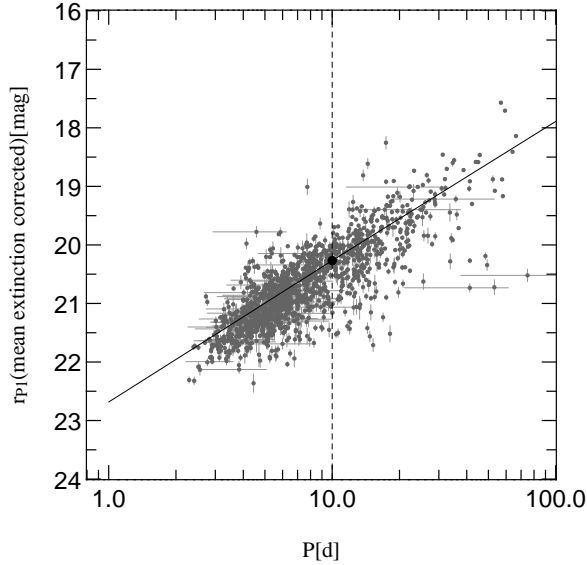}
  \caption{Common suspension point fit for FM Cepheids in the $\rps$
    band (c.f. equations \ref{susp1} and \ref{susp2}). The resulting
    slopes for the short period and long period subsample are very
    similar.}
  \label{FMknick}
\end{figure}

In a next step we want to check if our sample is dominated by long
period Cepheids. If our sample would be dominated by the long period
Cepheids, this would explain why the slopes of our full sample do not
agree with the slopes of \cite{2010ApJ...715..277B} for all Cepheids.
Therefore we want to compare the slopes of the short period ($P \leq
10 \, \mathrm{d}$) Cepheids to those of the long period
Cepheids. Simple linear fits to the short period samples provide
almost similar slopes to those of to the long period samples, but with
a different zero points. This means that the fits for the different
period ranges are not continuous at 10 days. Hence we introduce a
different procedure for the fit of the broken slope proposal. We fit
two linear slopes ($a_1$ and $a_2$) with a common suspension point
($y_{10}$) at 10 days ($x_{10}=10 \, \mathrm{d}$):
\begin{eqnarray}
  \label{susp1}
  y = a_1 \log (\frac{x}{x_{10}}) + y_{10} \quad \quad x \leq x_{10}\\
  \label{susp2}
  y = a_2 \log (\frac{x}{x_{10}}) + y_{10} \quad \quad x > x_{10}
\end{eqnarray}
For the FM Cepheids in the $\rps$ band this results in
\begin{eqnarray*}
  y_{10} = 20.26755 \pm 0.000004 \\
  a_1 = -2.41618 \pm 0.00007 \\
  a_2 = -2.37669 \pm 0.00004 
\end{eqnarray*}
with a dispersion of 0.36345 for the short periods and a dispersion of
0.55620 for the long periods. This fit is shown in
Fig. \ref{FMknick}. As can be seen also this fit with a common
suspension point does produce similar slopes for the short and long
period Cepheids.

\begin{figure}[h!]
  \epsscale{0.5} \plotone{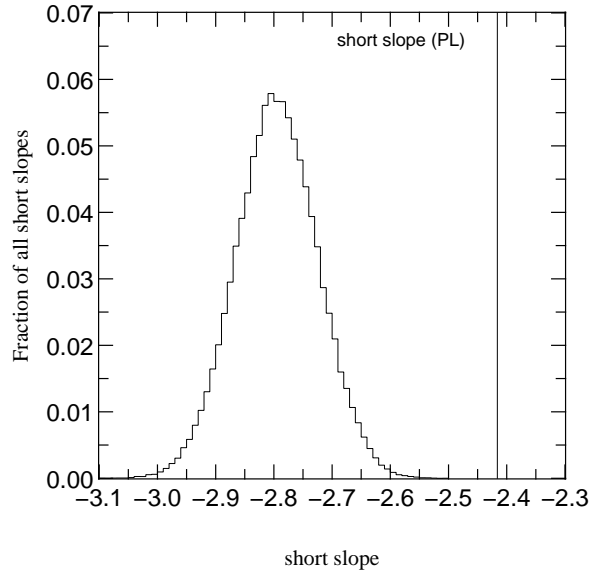}
  \caption{Distribution of the period subsample slopes for a Gaussian
    distribution around the slope -2.8 for the FM Cepheids in the
    $\rps$ band. The width of the Gaussian distribution is chosen to
    be the dispersion of the common suspension fit of the short period
    Cepheids (0.36345). A total of 100000 Gaussian distributions was
    used to determine the distribution of the short slopes. As can be
    seen the shallow slope from Fig. \ref{FMknick} cannot be explained
    by the dispersion of the PLR. }
  \label{Fig.gauss}
\end{figure}

To check if the dispersion of the PLR is concealing a steeper slope we
generate a simulated subsample for the short period Cepheids (in the
$\rps$ band) in such a way that the periods of the Cepheids stay the
same, but the magnitude is given by a Gaussian distribution around the
slope -2.8. The width of the Gaussian distribution is chosen to be the
dispersion of the common suspension fit of the short period
Cepheids. The subsample of the long period Cepheids stays the same. We
generate 100000 Gaussian distributions and perform the common
suspension point fit. The resulting distribution of the short period
slopes is shown in Fig. \ref{Fig.gauss}. As can be seen the shallow
slope of the short period subsample cannot be explained by the
dispersion of the PLR.

The most likely explanation for the shallow slope at short periods is
that we seem to select the brighter Cepheids at short periods due to
the Malmquist bias and that most short period Cepheids are too faint
to be detected, so that we cannot determine the slope for the short
periods correctly. To confirm this we would need to perform
completeness tests that are beyond the scope of this work.

Nevertheless the good agreement of our empirical values for the long
period Cepheids with the predictions of \cite{2010ApJ...715..277B}
supports the finding of \cite{2009A&A...493..471S} of a different
slope for long and short period Cepheids, depending on the
observational band, and the more general prediction of a curvature
term in the PLR.

\subsection{Spatial Cepheid distribution}

\begin{figure}[h!]
  \epsscale{1.0}
  \plottwo{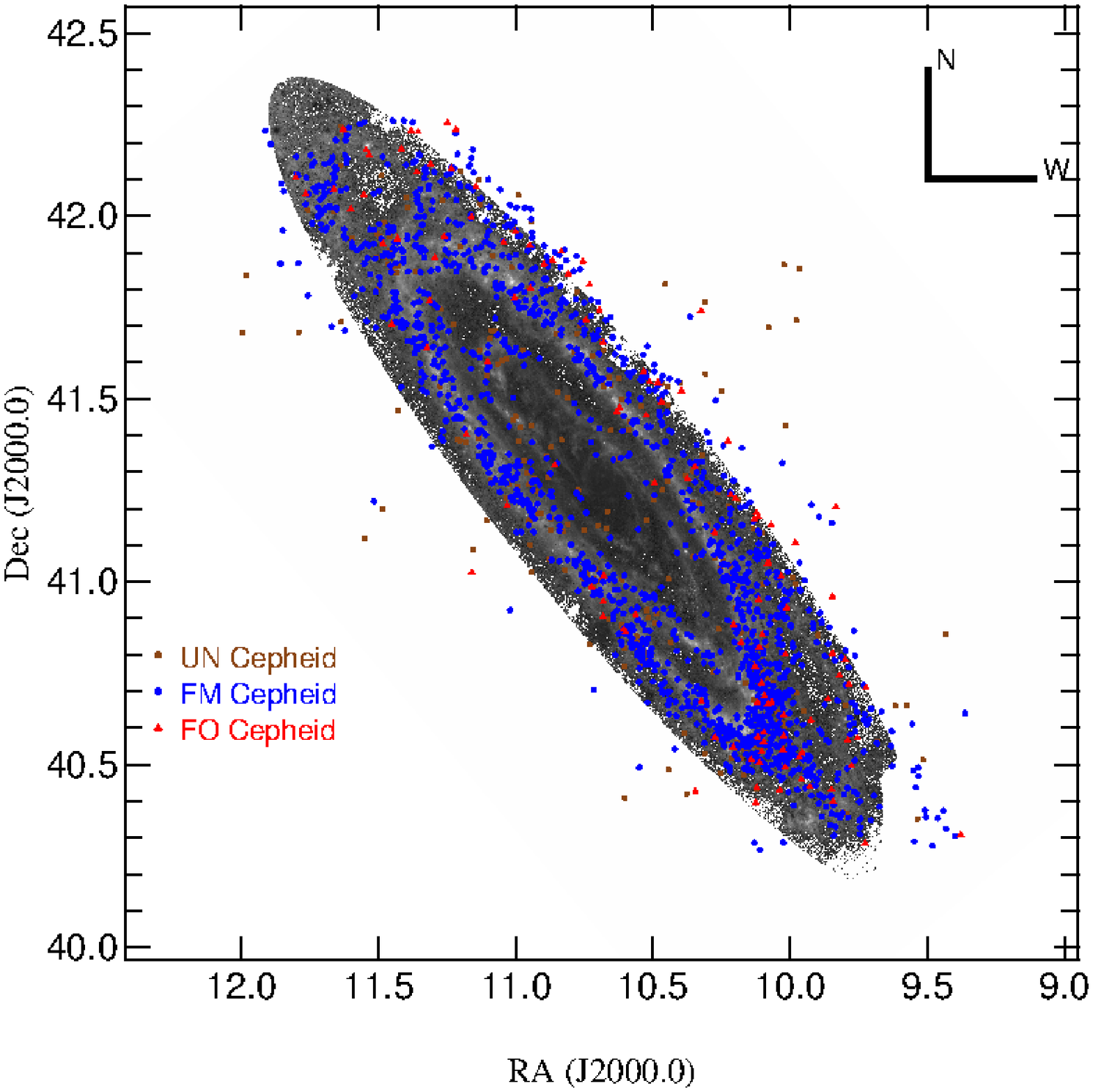}{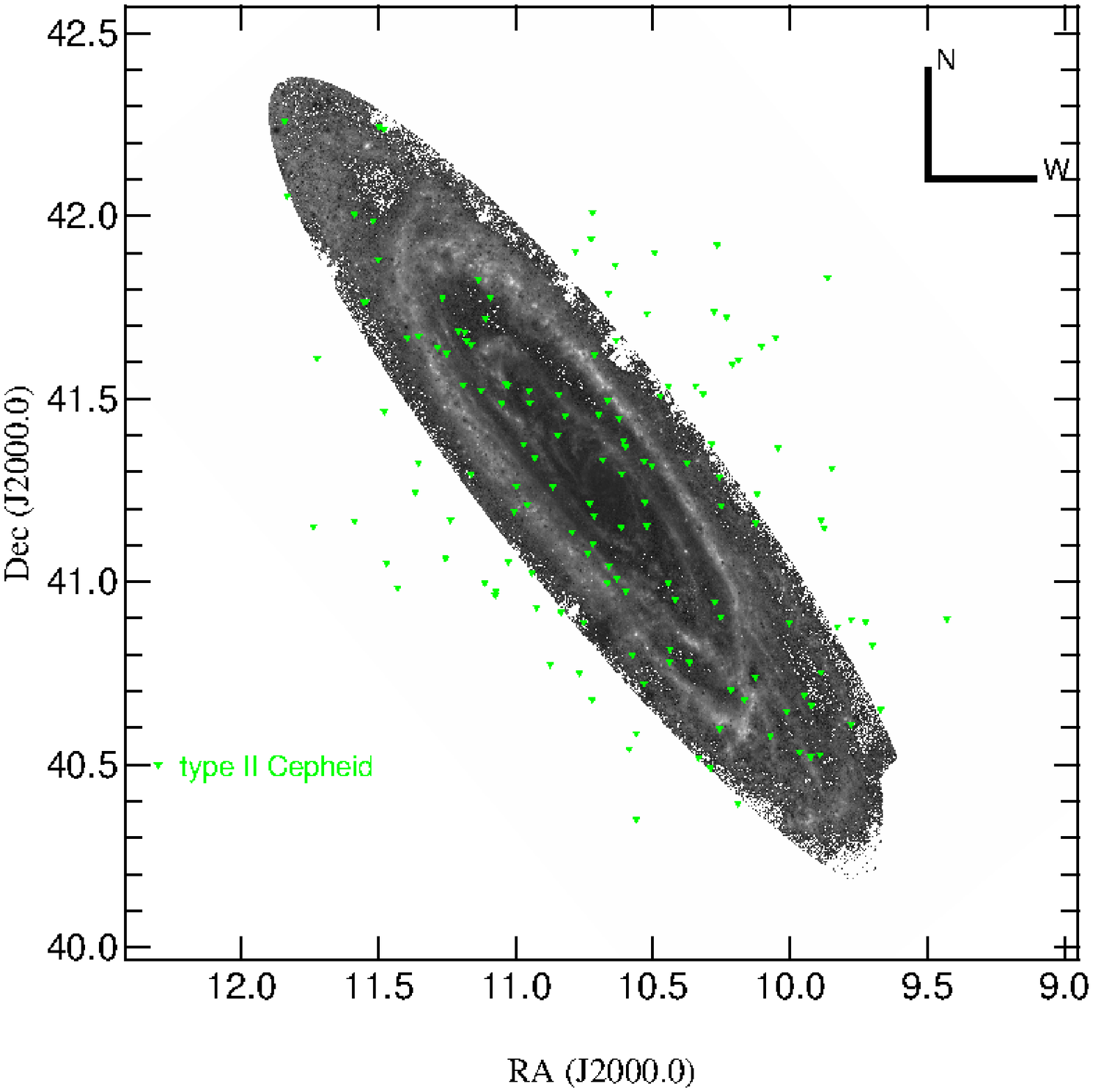}
  \caption{RA-Dec of the 3-dimensional parameter space classified
    Cepheid catalog, plotted over the E(B-V) map of
    \cite{2009A&A...507..283M}. It can clearly be seen that the FM and
    FO Cepheids trace the spiral arms, while the Type II Cepheids
    trace the M31 halo. Left panel: FM and FO Cepheid
    distribution. Right panel: Type II Cepheid distribution.}
\label{Fig.RADEC3d}
\end{figure}

\begin{figure}[h!]
  \epsscale{0.5}
  \plotone{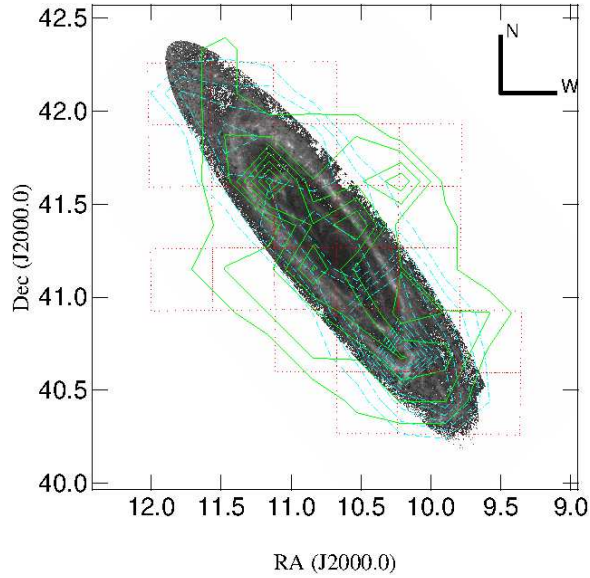}
  \caption{Contour plot of the RA-Dec distribution for the
    3-dimensional parameter space classified Cepheid catalog, plotted
    over the E(B-V) map of \cite{2009A&A...507..283M}. It can be seen
    that the FM and FO distribution (cyan dashed lines) follows the
    spiral arms, while the Type II distribution (green solid lines)
    traces the M31 halo. The Type II distribution is limited by the
    skycells (red dotted lines) that were analyzed until now. The ten
    contour line levels represent 1\% - 10\% of the respective Cepheid
    type number. For the FM and FO distribution this number is the
    total number of FM and FO combined.}
  \label{Fig.RADEC3dcont}
\end{figure}

We now investigate the location of the Cepheids in the M31
plane. Fig. \ref{Fig.RADEC3d} shows the position of the Cepheids
plotted over the E(B-V) map of \cite{2009A&A...507..283M}. The FM and
FO Cepheids are concentrated towards the disk and clearly show a ring
like structure (left panel in Fig. \ref{Fig.RADEC3d}), while the older
Type II Cepheids trace the halo of M31 (right panel in
Fig. \ref{Fig.RADEC3d}). Fig. \ref{Fig.RADEC3dcont} shows a contour
plot of the distribution of the Cepheids. While we can see the same
behavior of the FM and FO Cepheids as in Fig. \ref{Fig.RADEC3d}, we
observe that the Type II distribution is limited by the area of the
skycells that were analyzed. For the next data release we will
increase this area so as to better trace the halo. We do not discuss
the spatial distribution of Cepheids around NGC 206 but refer the
reader to \cite{1997A&A...326..442M}.

\subsection{Spatial age distribution}

\begin{figure}[h!]
  \epsscale{1.0}
  \plottwo{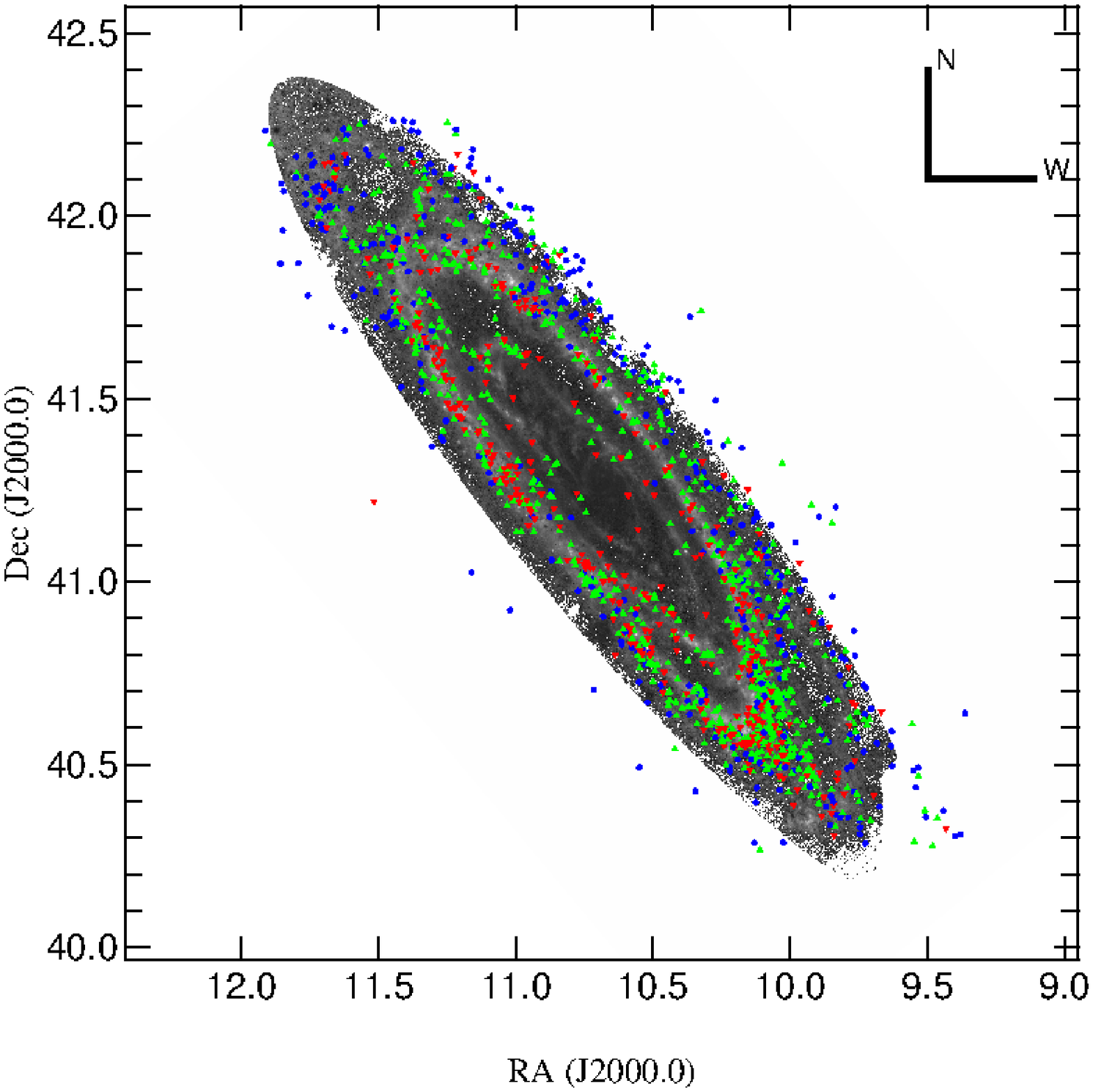}{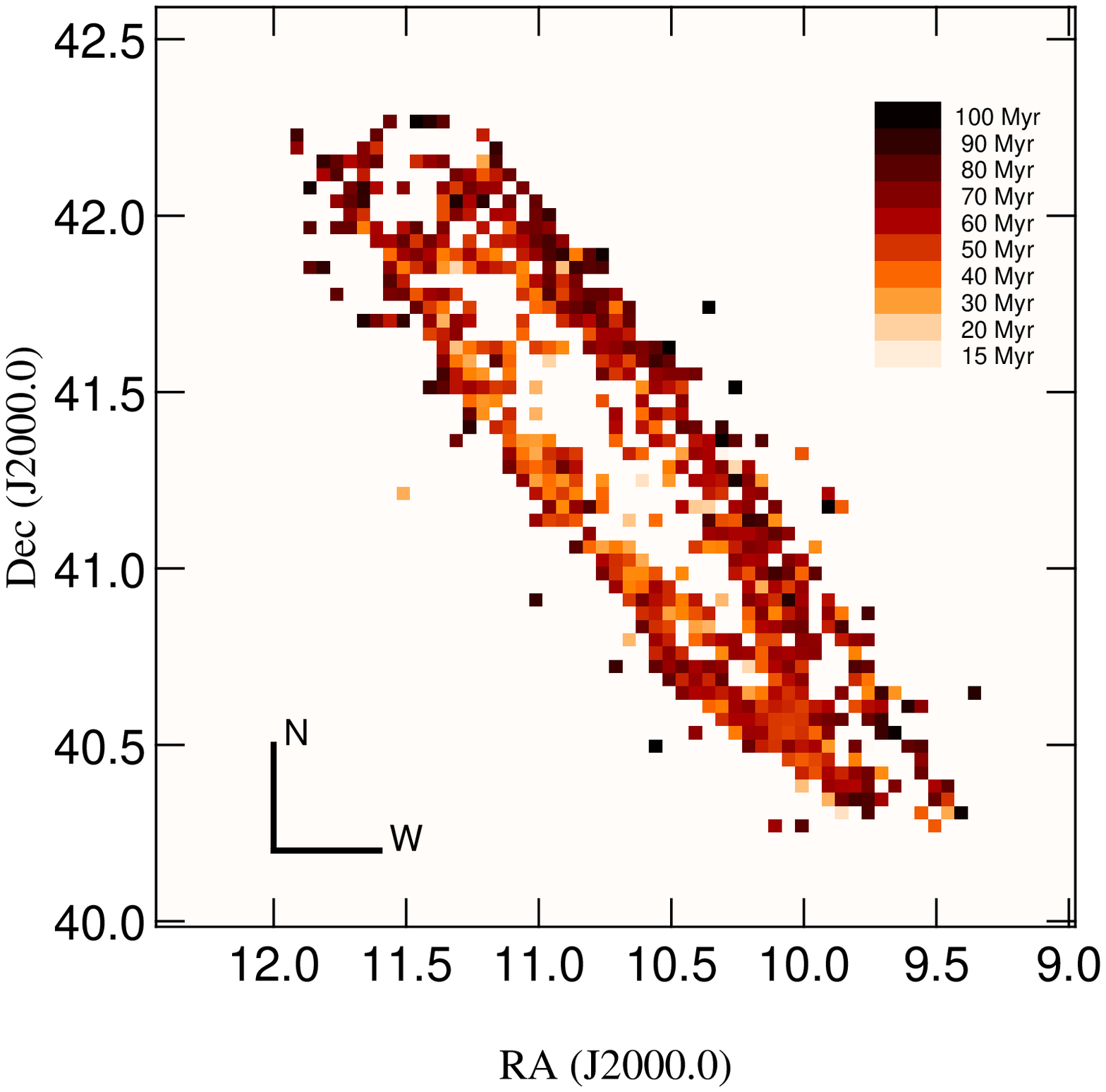}
  \caption{Spatial age distribution for the 3-dimensional parameter
    space classified Cepheid catalog for different Cepheid ages,
    plotted over the E(B-V) map of \cite{2009A&A...507..283M}. To
    calculate the age we use the period-age relations from
    \cite{2005ApJ...621..966B}. The spatial age distribution shows
    that the star formation in last $\sim 100~Myr$ was concentrated in
    a ring, which is in good agreement with
    \cite{2012ApJ...751...74D}. Blue points: FM and FO Cepheids with
    $t \geq 70~Myr$; Green upward pointing triangle: FM and FO
    Cepheids with $70~Myr > t \geq 40~Myr$; Red downward pointing
    triangle: $t < 40~Myr$. Left panel: RA-Dec distribution. Right
    panel: RA-Dec distribution with the median age for each bin.}
  \label{Fig.age}
\end{figure}

The spatial distribution of FM and FO Cepheids can be used to obtain a
spatial age distribution for M31. We use the period-age relation for
FM and FO Cepheids given by \cite{2005ApJ...621..966B} to determine
the age of each Cepheid.  The spatial age distribution shown in
Fig. \ref{Fig.age} indicates that star formation in the last $\sim
100~\mathrm{Myr}$ was concentrated in a ring which is in good
agreement with \cite{2012ApJ...751...74D}.  Fig. \ref{Fig.age} also
hints a correlation between the Cepheid age and the distance to the
center of M31.

\begin{figure}[h!]
  \epsscale{1.0}
  \plottwo{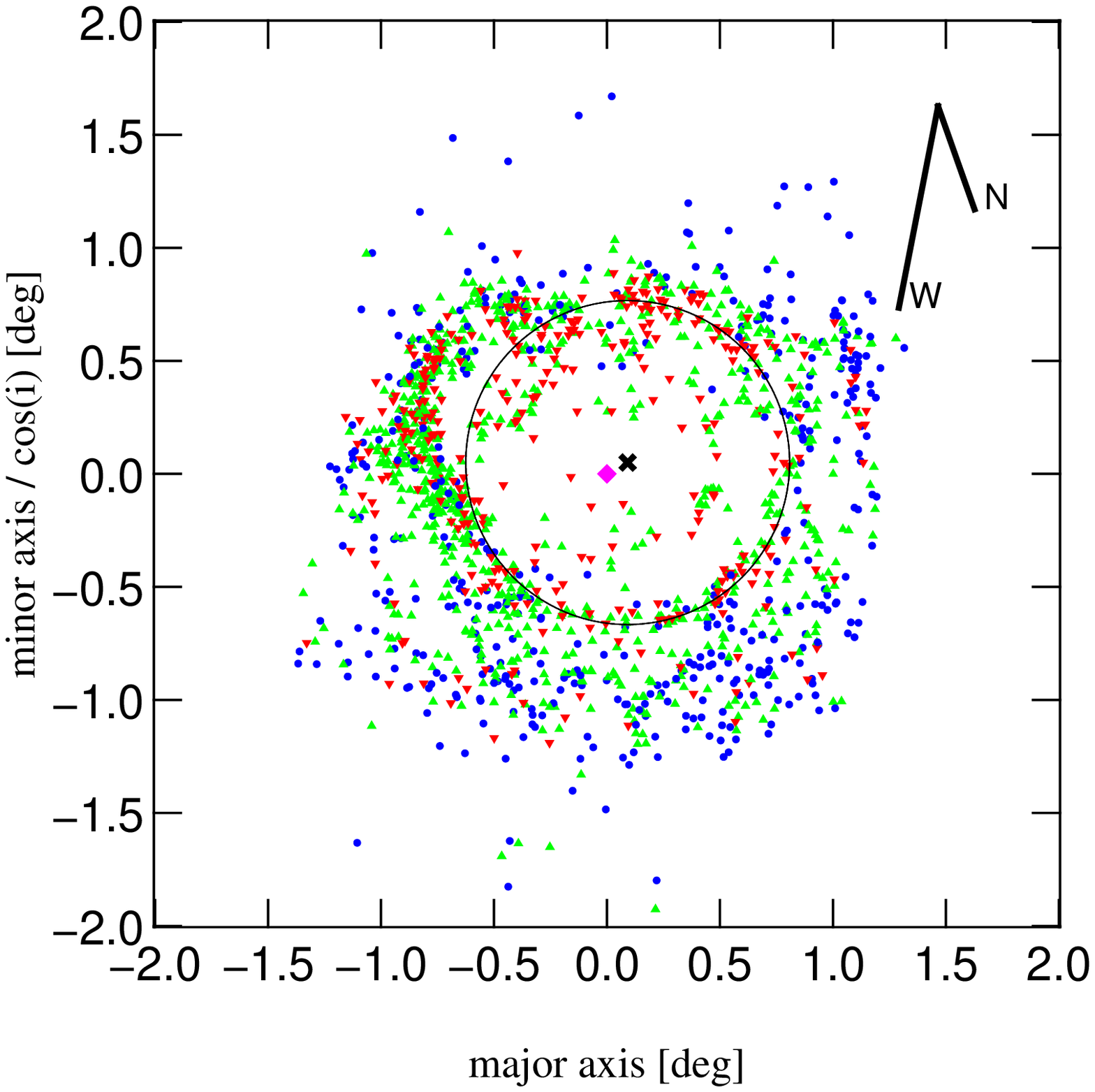}{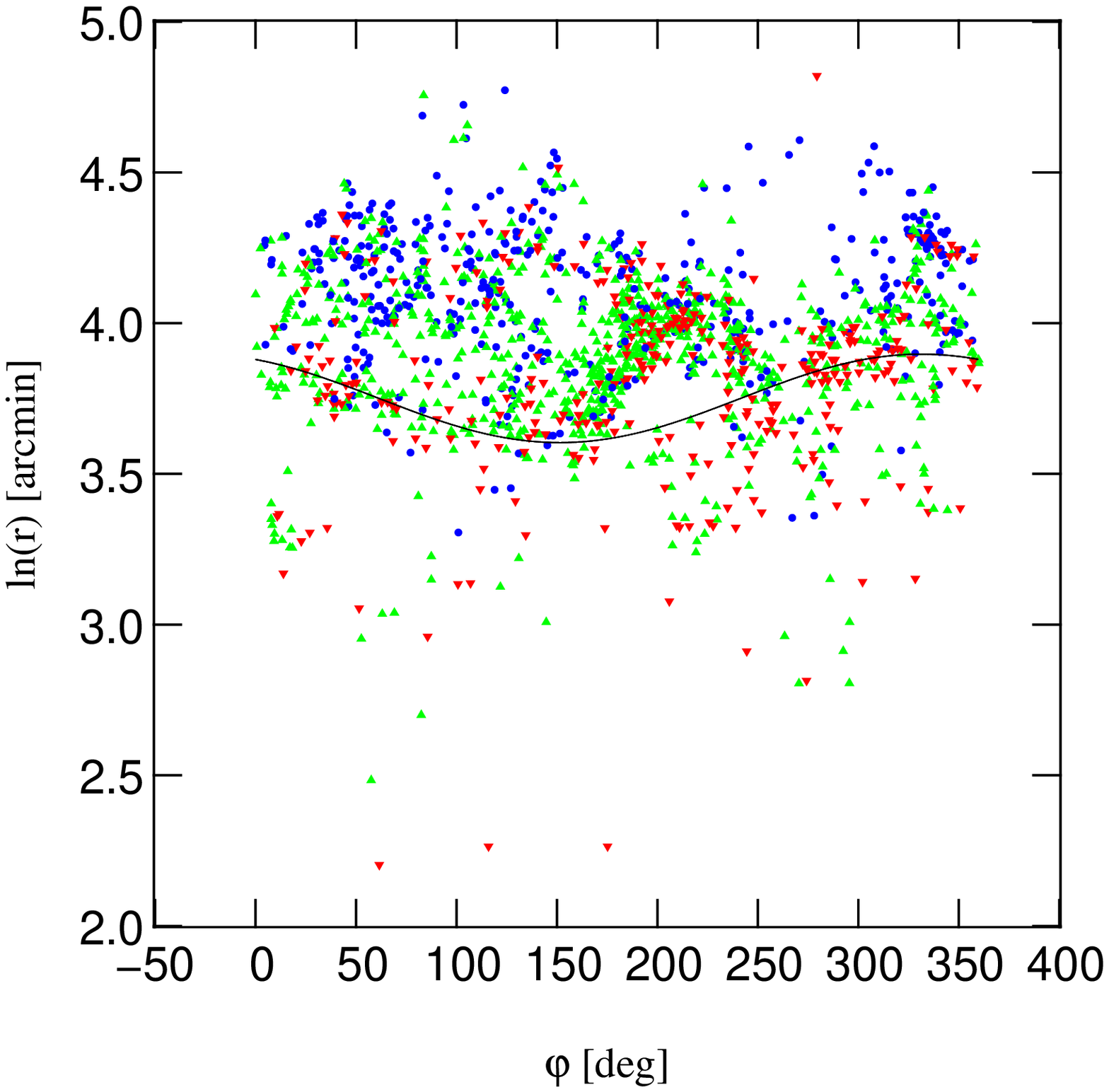}
  \caption{Deprojected spatial age distribution for the 3-dimensional
    parameter space classified Cepheid catalog. To calculate the age
    we use the period-age relations from
    \cite{2005ApJ...621..966B}. The ring and the splitting of the ring
    near M32 have been discussed in \cite{2006ApJ...638L..87G}.It can
    be seen that the star formation is clustered at the 10 kpc ring
    \citep{2012ApJ...751...74D}. Blue points: FM and FO Cepheids with
    $t \geq 70~Myr$; Green upward pointing triangle: FM and FO
    Cepheids with $70~Myr > t \geq 40~Myr$; Red downward pointing
    triangle: $t < 40~Myr$, black solid line: 10 kpc ring from
    \cite{2006ApJ...638L..87G} (the center of the ring is the black
    cross and the magenta point the center of M31). Left panel:
    Deprojected age map. Right panel: Deprojected age map in polar
    coordinates.}
\label{Fig.gordondeproj}
\end{figure}

\begin{figure}[h!]
  \epsscale{0.5}
  \plotone{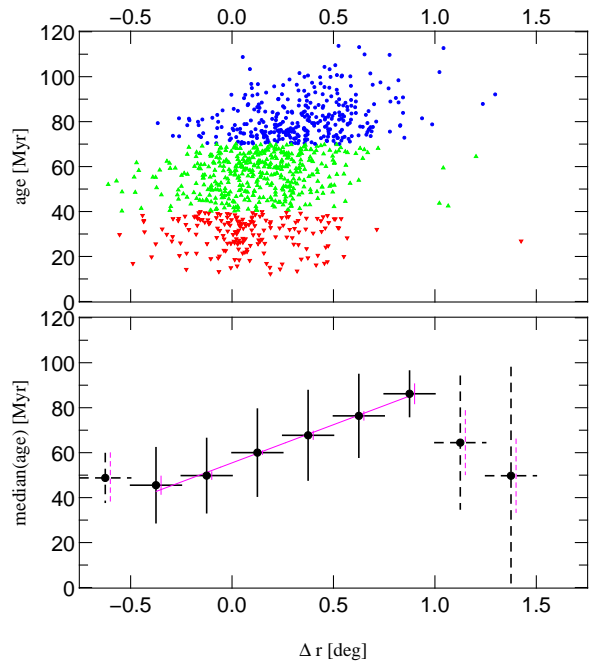}
  \caption{Age distribution as a function of distance to the 10 kpc
    ring. The definition of the symbols and colors is the same as in
    Fig. \ref{Fig.gordondeproj}. The splitted part of the ring ($150
    \leq \varphi \leq 260$, c.f. right panel in
    Fig. \ref{Fig.gordondeproj}) was excluded. In the top panel the
    age distribution for the FM and FO Cepheids can be seen. The
    bottom panel shows the median age for bins with width of $0.5$
    deg. The first and the last two bins contain less than 10 Cepheids
    and thus can be neglected. The errors of the medians have been
    determined with the bootstrap method (c.f. Section
    \ref{sec.period}).  The errors determined with the bootstrap
    method are shown in magenta with an offset of 0.025 deg for better
    visibility. The dispersion of each bin is shown in black. We
    observe an age gradient and the fit to the median age ($y = a + b
    \cdot x$) is shown as a solid magenta line. The age gradient
    suggests that the star formation related to the interaction to M32
    moved inwards.  No Cepheids can be found in the top left corner of
    the upper panel (the center of M31). We hardly detect old (faint)
    Cepheids in this region, most likely due to the low signal to
    noise ratio in the center of M31. The lack of young Cepheids in
    the outer region of the ring is no selection bias, since we detect
    old (faint) Cepheids in the seventh bin, but no young Cepheids,
    although they should be more easily detectable than old
    Cepheids. This leaves us with the conclusion that the age gradient
    is real.}
  \label{Fig.gradient}
\end{figure}

In order to check for this correlation we deproject the spatial
distribution. We use $75^\circ$ for the inclination and $37^\circ$ for
the tilt angle and the radius and offset of the 10 kpc ring given by
\cite{2006ApJ...638L..87G}. The deprojected age map is shown in
Fig. \ref{Fig.gordondeproj}. We found the splitting of the ring as
previously described in
\cite{2006ApJ...638L..87G}. \cite{2006ApJ...638L..87G} attributes the
star formation of the ring and its splitting to a passage of M32
through the M31 disk. In a next step we analyze the age of the
Cepheids as a function of distance to the 10 kpc ring. For this
analysis we exclude all Cepheids in the splitted part of the ring
($150 \leq \varphi \leq 260$, c.f. right panel in
Fig. \ref{Fig.gordondeproj}). Fig. \ref{Fig.gradient} shows the median
Cepheid age as a function of the distance to the 10 kpc ring in bins
of $0.5$ deg. The first and the last two bins contain less than 10
Cepheids and can therefore be neglected.  The errors in each bin have
been determined with the bootstrap method (c.f. Section
\ref{sec.period}) and are rather small compared to the dispersion in
the bins we consider.  The upper left corner of the upper panel of
Fig. \ref{Fig.gradient} contains no Cepheids. In this region (the
center of M31) the signal to noise ratio is low, so that this is
likely the reason that we hardly detect faint Cepheids (i.e. Cepheids
with small periods; the study of the detection efficiency for Cepheids
will be subject of a further work) and thus we hardly detect old
Cepheids in this region. In the seventh bin we detect old (faint)
Cepheids but no young Cepheids, although they should be more easily
detectable than old Cepheids. Therefore the lack of young Cepheids in
the outer region of the ring is no selection bias. This leaves us with
the conclusion that the age gradient is real. To measure the strength
of the age gradient we perform a fit to the median age with $y = a + b
\cdot x$:
\begin{eqnarray*}
  a = 55.4941 \pm 0.7692 ~ [Myr] \\
  b = 33.8517 \pm 0.6100 ~ [Myr/deg]
\end{eqnarray*}
with a $\frac{\Delta \chi^2}{\mathrm{d.o.f.}} = 0.36$. We therefore
conclude that the median age of the stellar population decreases by
$\sim 34~Myr/deg$ inwards. A possible interpretation for this is that
the star formation is related to the interaction with M32 moving
inwards.
 
%\clearpage

\section{Conclusion and Outlook}
\label{sec.conclusion}

We present a large sample of Cepheids in M31 in the $\rps$ and $\ips$
filters. We develop an automatic Cepheid detection and classification
scheme based on the Fourier parameters $P$, $\fA$ and $\fP$ and the
location of the instability strip in the Wesenheit-color plane. This
makes the Cepheid detection and classification less biased than
traditional methods and facilitates Cepheid detection in large
surveys.

We find 1440 fundamental mode (FM) Cepheids, 126 Cepheids in the first
overtone mode (FO), 147 belonging to the Population II types and 296
Cepheids that could not be classified. 354 of the 2009 Cepheids could
be found in other surveys (\citealt{2007A&A...473..847V},
\citealt{2006A&A...445..423F} and the DIRECT project,
\citealt{2004ASPC..310...33M}) with a matching radius of 1 arcsec. The
matching tables are provided in electronical form.

The spatial distribution of FM and FO Cepheids follows the 10 kpc ring
in M31, while the Type II Cepheids are distributed throughout the halo
of M31. The spatial age distribution shows a positive age gradient
with increasing distance from the 10 kpc ring, which implies that the
star formation in the 10 kpc ring moved inwards.

The analysis of our period luminosity relations (PLR) indicates that
the PLR is slightly curved. This has ramifications on the
determination of extra galactic distances, since most of the commonly
used calibrations of the PLR are dominated by the short period
Cepheids and the typical Cepheids that are used for the extra galactic
distance determination are long period Cepheids.

The next data release will cover more area of the M31 halo, thus we
will be able to trace the halo better. With another year of
observations we will be able to extend the sample to larger
periods. Overlapping skycells will improve the calibration and enable
completeness tests.

The complete catalog of the 2009 Cepheids can be found electronically
in the CDS database.

\acknowledgments 

We acknowledge the careful reading and helpful comments by the
anonymous referee. The Pan-STARRS1 Surveys (PS1) have been made
possible through contributions of the Institute for Astronomy, the
University of Hawaii, the Pan-STARRS Project Office, the Max-Planck
Society and its participating institutes, the Max Planck Institute for
Astronomy, Heidelberg and the Max Planck Institute for
Extraterrestrial Physics, Garching, The Johns Hopkins University,
Durham University, the University of Edinburgh, Queen's University
Belfast, the Harvard-Smithsonian Center for Astrophysics, the Las
Cumbres Observatory Global Telescope Network Incorporated, the
National Central University of Taiwan, the Space Telescope Science
Institute, and the National Aeronautics and Space Administration under
Grant No. NNX08AR22G issued through the Planetary Science Division of
the NASA Science Mission Directorate.  This research was supported by
the DFG cluster of excellence ‘Origin and Structure of the Universe’
(www.universe-cluster.de).

\bibliographystyle{apj}

\end{document}